\documentclass[onecolumn,sort&compress,numbers]{els-mrw}

\usepackage{amssymb}
\usepackage{amsmath}
\usepackage{txfonts}
\usepackage{helvet}
\usepackage[T1]{fontenc}
\usepackage[colorlinks,pdfusetitle]{hyperref}
\hypersetup{colorlinks=true,allcolors=[rgb]{1,0.56,0},linktocpage=true}

\usepackage{ifpdf}
\usepackage{feynmp}
\ifpdf
\fi

\newcommand{\td}[2]{\frac{d #1}{d #2}}
\newcommand{\ket}[1]{|#1\rangle}
\newcommand{\e}[1]{\times10^{#1}}

\begin{document}
\setlength{\unitlength}{1mm}
\begin{fmffile}{foo}

\chapter{Neutrino Oscillations in the Three Flavor Paradigm}
\author[1]{Peter B.~Denton}
\address[1]{\orgname{Brookhaven National Lab}, \orgdiv{High Energy Theory Group, Physics Department}, \orgaddress{Upton, NY 11973, USA}}
\date{\today}

\maketitle

\begin{abstract}[Abstract]
The three-flavor neutrino oscillation model describes the well-studied phenomenon of neutrinos produced in association with one charged lepton: electron, muon, or tau, and then later detected in association with a possibly different charged lepton.
While somewhat surprising, the firm experimental discovery of the phenomenon in the late 1990s and early 2000s has lead to a revolution in particle physics as the nature of neutrinos has been explored with heightened vigor ever since.
At the core of the phenomenon are the six neutrino oscillation parameters.
These parameters are fundamental and not predicted from anything else in our model of particle physics.
At the time of writing this chapter, many of them have been measured, but several key questions remain that are to be answered by neutrino oscillations themselves.
These questions have motivated some of the largest and most involved particle physics experiments built to date.
This chapter will develop the basics of neutrino oscillation theory and build intuition for the role of the oscillation parameters and how they are measured, as well as the important role of the matter effect in neutrino oscillations.
\end{abstract}

\begin{keywords}
Neutrino Oscillations\sep Beyond the Standard Model\sep CP Violation\sep Matter Effect\sep Solar Neutrinos\sep Atmospheric Neutrinos\sep Reactor Neutrinos\sep Accelerator Neutrinos
\end{keywords}

\let\oldaddcontentsline\addcontentsline
\renewcommand{\addcontentsline}[3]{}
\section*{Key Points}
\let\addcontentsline\oldaddcontentsline
\begin{itemize}
\item The discovery that neutrinos oscillate represents the only particle physics evidence we have of physics beyond the Standard Model, and indicates a new scale at $<1$ eV.
\item Neutrino oscillations requires 2 to 3 neutrinos to have masses, that the masses are all different, and that there is mixing among those neutrinos.
\item There are six parameters governing standard three-flavor oscillations; a rough picture exists but several key questions exist.
\end{itemize}

\section{Introduction}
Neutrinos are often produced and detected in association with a given charged lepton -- electron, muon, or tau -- which can be easily identified.
The fact that a neutrino produced in association with one charged lepton later may be less likely to interact with the same charged lepton and more likely to interaction with another is known as the phenomenon of neutrino oscillations.

The idea of neutrino oscillations traces back many decades, although typically initial hints of neutrino oscillations appear in attempts to address other physics problems.
For example Pontecorvo discussed $\mu^+e^-\leftrightarrow\mu^-e^+$ mixing in 1957 \cite{Pontecorvo:1957cp}, and several earlier papers have some physics components of neutrino oscillation theory.
In 1962 Maki, Nakagawa, and Sakata worked out much of the basic physics of two-flavor neutrino oscillations \cite{Maki:1962mu}.
The unitary lepton mixing matrix relating the neutrino mass eigenstates to the charged-lepton flavor eigenstates is called the PMNS matrix $U_{\rm PMNS}$ referring to these two papers.

The first experimental evidence for neutrino oscillations came from Ray Davis's Homestake Mine solar neutrino experiment through the 1960s to 1980s \cite{Davis:1968cp,Cleveland:1998nv} in association with John Bahcall's theory prediction for the solar neutrino flux \cite{Bahcall:1968hc,Bahcall:2000nu}.
This showed that the flux of neutrinos from the Sun above a certain threshold energy was about a factor of about three\footnote{Note that the fact that there are three neutrinos and the fact that solar flux is lower than expected by about a factor of three is seemingly unrelated; see section \ref{sec:flavor} below.} less than expected.
Nonetheless, due to the challenging experiment as well as the complexities of the theory prediction, the result was not interpreted as evidence of new fundamental particle physics until decades later.
Another issue was the unclear physics landscape that could conceivably describe the effect\footnote{In addition, a deficit was also seen in the lower energy part of the solar neutrino spectrum where the predictions were better \cite{Bahcall:1968xb}, but the measurements were even more challenging \cite{GALLEX:1998kcz,SAGE:1999nng,Gavrin:2001sz,GNO:2000avz}.}.

The landscaped evolved in 1998 when Super-Kamiokande reported high significance evidence that neutrinos produced in the Earth's atmosphere change flavors depending on how far through the Earth they propagated \cite{Super-Kamiokande:1998kpq}.
Given the energy and zenith angle (which corresponds to distance) distributions, this clearly put neutrino oscillations as a new physics phenomenon to be taken seriously.

Shortly thereafter, in a pair of papers in 2001 and 2002, SNO reported confirmation of both the Homestake measurement and the Bahcall prediction of solar neutrinos \cite{SNO:2001kpb,SNO:2002tuh}.
This also confirms that neutrinos do change flavors and, in combination with lower energy solar neutrino data from the $pp$ process by SAGE, GALLEX, and GNO \cite{GALLEX:1998kcz,SAGE:1999nng,Gavrin:2001sz,GNO:2000avz}, confirms that the solar neutrino flavor changing depends on the neutrino energy.
Interpretation of solar neutrino in an oscillation framework also requires an understanding of how neutrinos interact with matter (see section \ref{sec:matter effect} below) \cite{Wolfenstein:1977ue} and, specifically, in an environment with changing density (see section \ref{sec:msw} below) \cite{Mikheyev:1985zog}.

While there has been a leading candidate, known as the Higgs mechanism, for the mass generation of all the other massive particles for many years \cite{Higgs:1964pj,Englert:1964et} which was confirmed in 2012 \cite{ATLAS:2012yve,CMS:2012qbp}, no such clear picture exists for neutrinos even today.
\textit{Understanding the underlying mass generation mechanism of neutrinos is one of the biggest open questions in particle physics.}
The most promising approach is neutrinoless double beta decay experiments.
Other options include measuring the cosmic neutrino background or measurements at the LHC and other high energy colliders.
Unfortunately, it is unlikely that neutrino oscillation experiments will provide insight into answering this question; as such we will focus on what neutrino oscillations can tell us about neutrinos in this chapter.

Since the turn of the millennium, the field of neutrino oscillations has exploded in terms of experiment, phenomenology, and theory.
The experimental landscape includes a veritable alphabet soup of experiments covering a many orders of magnitude in neutrino energy and baseline.
Phenomenonologists have clarified many interesting and non-trivial relationships among the oscillation parameters in the context of realistic experiments.
Finally, theorists have continued to put the oscillation framework on increasingly solid ground and have developed many new physics models to test the standard three-flavor framework.
In the last several decades considerable progress has been made improving our understanding of the phenomenon of neutrino oscillations quite a bit.
The underlying physics model that governs the existence of neutrino mass, however, remains elusive.

In this chapter, we will discuss the connection between the neutrino oscillation phenomenon and the rest of our understanding of particle physics, noting primarily how it does not fit in to the Standard Model, in section \ref{sec:nu osc SM}.
We will then investigate in section \ref{sec:nu osc probs} neutrino oscillation probabilities: the key component relating the underlying physics parameters to the physical observables.
In terms of physics research, these probabilities play a similar role as cross sections in other areas of particle physics.
This section will contain the bulk of the physics relevant for this chapter.
The potential ability to predict the parameters in the neutrino sector, often in connection with the parameters in the quark sector, is also still an open question in physics and will be discussed in section \ref{sec:flavor}.
We will then discuss a number of alternative means of probing the same parameters that neutrino oscillations probe in section \ref{sec:non osc} and summarize this chapter in section \ref{sec:summary}.

This chapter follows and, significantly expands upon, some of \cite{Denton:2022een}.
The target audience of this chapter is any scientist from bachelor student to senior faculty who is aiming to begin their education on neutrino oscillation physics and has an understanding of quantum mechanics.

\section{Neutrino Oscillations and the Standard Model of Particle Physics}
\label{sec:nu osc SM}
It is widely held that neutrinos do not have mass in the Standard Model (SM) of particle physics.
For example, Steven Weinberg, who is responsible for our modern understanding of how leptons fit together in the SM \cite{Weinberg:1967tq}, famously said that neutrinos do not have mass in his SM \cite{NYT_Weinberg}.
While the SM is not immutable and it can evolve, the decision of what is in it does not rest with any individual.

The key in deciding what is in the SM is in the name: ``Standard.''
Because there is no leading ``standard'' means of adding mass to neutrinos in our model of particle physics, the notion of massive neutrinos remains beyond the SM for the time being.
This is because there are at least two equally compelling means of adding a mass term for neutrinos, and neither has clear advantages or disadvantages over the other, other than personal taste.

On the one hand, neutrinos may have a Dirac mass term due to the Higgs mechanism \cite{Higgs:1964pj,Englert:1964et} along with new right handed neutrino states that are SM gauge singlets.
The mass generation mechanism follows the same approach as the quarks or the charged leptons and nothing, a priori, would seem to forbid such a mass term from existing.
Many physicists, however, feel uncomfortable with the notion of very small $\lesssim10^{-12}$ Yukawa couplings.
That is, many suspect that dimensionless numbers should be within a few orders of magnitude of their largest possible value.
Notably the top quark Yukawa coupling to the Higgs field is $\mathcal O(1)$, but the rest of the fermions are broadly distributed down the electron Yukawa coupling to the Higgs field which is almost one million times smaller.
The neutrino Yukawa couplings would necessarily be even smaller, by at least another factor of about a million.
While for some, this is motivation to look for alternative solutions, in reality, nothing forbids such small couplings from existing (see also the strong CP problem).

An alternative solution is that neutrinos have a new kind of mass term that no other particles in the SM have (or could have) called a Majorana mass term \cite{Majorana:1932ga}.
This is only possible if neutrinos are fundamentally different from the other fermions (charged leptons and quarks) and is only conceivable for neutrinos in the first place because they do not have electric charge.
In this case, there are only left-handed neutrinos and right-handed antineutrinos which, due to the presence of a mass term, can transform into each other at rate that is suppressed at large momentum.
In fact, for $p_\nu\gg m_\nu$, which is almost always the case in current experiments, the two fundamental natures of neutrinos behave identically with corrections that scale like $(p_\nu/m_\nu)^2$, see the Dirac-Majorana Confusion Theorem \cite{Kayser:1982br,Shrock:1982sc}.
Since $m_\nu\lesssim0.1$ eV and $p_\nu\gtrsim1$ MeV in neutrino oscillation experiments, the difference appears at the $\sim10^{-14}$ level or smaller which is undetectable.

If there is a Majorana mass term, there may well also be a Dirac mass term.
The presence of two mass terms requires a diagonalization to find that physical mass state.
Depending on the hierarchies involved, this may lead to a fairly elegant solution typically called type-I seesaw \cite{Minkowski:1977sc}, see also \cite{Yanagida:1979as,Gell-Mann:1979vob,Yanagida:1980xy,Mohapatra:1979ia}.

In any of the above cases, some choice must always be made.
Specifically, if a Dirac mass term is the only one assumed, then the Majorana mass term must be excluded somehow, typically by imposing lepton number conservation.
If the Majorana case (or a seesaw case) is assumed then a completely novel kind of mass term must exist as would lepton number violating processes.
In any scenario, new fields must be added, but their behavior is quite different from that of known fields in all the different scenarios.
It is for these reasons that neutrino masses are not in the SM because there is no obvious ``standard'' choice for how to do so.
Thus, neutrino oscillations is physics beyond the Standard Model (BSM) and investigations for further new physics in oscillations is sometimes called BSM physics, but also sometimes called New Physics to represent the fact that even though neutrino oscillations is already BSM physics, there may also be additional phenomenon beyond the three-flavor physics such as sterile neutrinos, non-standard neutrino interactions, faster than predicted neutrino decay, or many other new physics scenarios \cite{Arguelles:2019xgp}.

For better or for worse, the rich structure of neutrino mass generation models have no impact on existing and upcoming neutrino oscillation experiments.
This makes it somewhat easier to extract the oscillation parameters described below, but also makes it more challenging to learn about the fundamental nature of neutrinos.
Thus the discussion in the rest of this chapter is agnostic about the nature of neutrino mass which is best probed in non-oscillation experiments.

\section{Neutrino Oscillation Probabilities}
\label{sec:nu osc probs}
The fundamental observable that describes the nature of neutrino oscillations is known as a neutrino oscillation probability.
These probabilities depend on the initial and final flavor, the neutrino energy, the distance traveled, and the fundamental neutrino parameters.
In terms of the measured spectrum, this can be qualitatively expressed as
\begin{equation}
\td{N_{\nu_\beta}}{E_\nu}(L)=P(\nu_\alpha\to\nu_\beta;L,E)\td{N_{\nu_\alpha}}{E_\nu}(L=0)\,,
\end{equation}
where $\td NE(L)$ is the spectrum of neutrinos measured at distance $L$, $\alpha$ and $\beta$ are neutrino flavors $\{e,\mu,\tau\}$, and $P$ is the neutrino oscillation probability function.
That is, the spectrum of neutrinos at some distance $L$ from the source is the spectrum of neutrinos at the production point times the probability of transforming from one state to another over a distance $L$ and with a neutrino energy $E$.
The probabilities may also depend on the presence of matter between production and detection.

The neutrino oscillation mechanism can be understood with three steps.
First, when a neutrino is produced in association with a charged lepton\footnote{Neutrinos can also be produced in $Z$ decays, but as this process is independent of flavor, no oscillation effect occurs.} via the weak interaction, it is really the superposition or linear combination of three different neutrino mass states.
This is expressed as
\begin{equation}
\ket{\nu_\alpha}=\sum_{i=1}^3U_{\alpha i}^*\ket{\nu_i}\,,
\label{eq:mass2flavor}
\end{equation}
where $U$ is a complex $3\times3$ unitary matrix called the Pontecorvo-Maki-Nakagawa-Sakata (PMNS) \cite{Pontecorvo:1957cp,Maki:1962mu} matrix.
The PMNS matrix can be parameterized in many ways, but a useful one that is widely used across the field \cite{Denton:2020igp} is as the product of three rotations with a single complex phase:
\begin{equation}
U=
\begin{pmatrix}
1\\
&c_{23}&s_{23}\\
&-s_{23}&c_{23}
\end{pmatrix}
\begin{pmatrix}
c_{13}&&s_{13}e^{-i\delta}\\
&1\\
-s_{13}e^{i\delta}&&c_{13}
\end{pmatrix}
\begin{pmatrix}
c_{12}&s_{12}\\
-s_{12}&c_{12}\\
&&1
\end{pmatrix}\,,
\label{eq:pmns}
\end{equation}
where the common shorthand $s_{ij}=\sin\theta_{ij}$ and $c_{ij}=\cos\theta_{ij}$ is used.
In addition, we note that Greek letters $\alpha$, $\beta$, \dots\ typically denote flavor states and mid-alphabet Latin letters $i$, $j$, \dots\ typically denote mass states.

Second, since the mass states are those that actually propagate; they accumulate a phase during propagation due to the solution to the Schr\"odinger equation.
That is,
\begin{equation}
\ket{\nu_i(t)}=e^{-iE_it}\ket{\nu_i(t=0)}\,,
\label{eq:phase accumulation E_i}
\end{equation}
where $E_i$ is the energy of mass state $i$ and $t$ is the propagation time.
We use the fact that in all oscillation experiments neutrinos are known to be ultrarelativistic to expand the energy as
\begin{equation}
E_i\simeq p_i+\frac{m_i^2}{2p_i}\,.
\end{equation}
We note that in quantum mechanical superposition, only the interference of phases can be measured, thus a constant term can be subtracted out.
This allows us to rewrite the phase accumulation in eq.~\ref{eq:phase accumulation E_i} as
\begin{equation}
\ket{\nu_i(t)}=e^{-im_i^2t/2p_i}\ket{\nu_i(t=0)}\,.
\label{eq:phase accumulation msq}
\end{equation}
Now we note that to an excellent approximation (although see section \ref{sec:wave packet} below), we can take the time $t$ as the baseline $L$ and the momentum $p_i$ as the energy $E$ the same for all mass states.
This allows us to write the phase accumulation in its final form as
\begin{equation}
\ket{\nu_i(t)}=e^{-im_i^2L/2E}\ket{\nu_i(t=0)}\,.
\label{eq:phase accumulation final}
\end{equation}

The third and final step is projecting the mass states back to the flavor state for an interaction in the detector which an associated charged lepton.
By unitarity, this follows from the same equation as the opposite projection shown in eq.~\ref{eq:mass2flavor},
\begin{equation}
\ket{\nu_i}=\sum_{\alpha\in\{e,\mu,\tau\}}U_{\alpha i}\ket{\nu_\alpha}\,.
\end{equation}

We now combine these three parts to write the transition amplitude
\begin{equation}
\mathcal A(\nu_\alpha\to\nu_\beta;L,E)=\sum_{i=1}^3U_{\alpha i}^*e^{-im_i^2L/2E} U_{\beta i}\,,
\label{eq:amplitude}
\end{equation}
and the neutrino oscillation probability is
\begin{equation}
P(\nu_\alpha\to\nu_\beta;L,E)=|\mathcal A(\nu_\alpha\to\nu_\beta;L,E)|^2\,.
\label{eq:P from A}
\end{equation}

This computation can also be usefully expressed in the Hamiltonian framework.
Such a Hamiltonian in the neutrino flavor basis can be written as
\begin{equation}
H=\frac1{2E}U
\begin{pmatrix}
0\\&\Delta m^2_{21}\\&&\Delta m^2_{31}
\end{pmatrix}
U^\dagger\,,
\label{eq:Hvac}
\end{equation}
where $\Delta m^2_{ij}=m^2_i-m^2_j$ and we have again used the fact that we can subtract off a term from all states, as is evident by eqs.~\ref{eq:amplitude}-\ref{eq:P from A}.
Given the Hamiltonian, the transition amplitude can be simply written as
\begin{equation}
\mathcal A(\nu_\alpha\to\nu_\beta;L,E)=\left[\exp\left(-iHL\right)\right]_{\alpha\beta}\,.
\end{equation}
This approach is beneficial as we will modify the energy levels due to the presence of matter in subsections \ref{sec:matter effect} and \ref{sec:msw} below.

These amplitudes can then be expanded out in various ways by applying trigonometric identities, the unitarity of $U$, and the assumption of CPT invariance.
A useful generic way of writing the full probability is
\begin{align}
P(\nu_\alpha\to\nu_\beta;L,E)={}&\delta_{\alpha\beta}
-4\sum_{i>j}\Re(U_{\alpha i}U_{\beta i}^*U_{\alpha j}^*U_{\beta j})\sin^2\left(\frac{\Delta m^2_{ij}L}{4E}\right)\nonumber\\
&\pm8J\sin\left(\frac{\Delta m^2_{21}L}{4E}\right)\sin\left(\frac{\Delta m^2_{31}L}{4E}\right)\sin\left(\frac{\Delta m^2_{32}L}{4E}\right)\,,
\label{eq:P general}
\end{align}
where $\delta_{\alpha\beta}$ is one if $\alpha=\beta$ and zero otherwise, and 
\begin{equation}
J=\Im(U_{e1}U_{e2}^*U_{\mu1}^*U_{\mu2})=s_{12}c_{12}s_{13}c_{13}^2s_{23}c_{23}\sin\delta\,,
\label{eq:J}
\end{equation}
is an invariant pointed out by Cecilia Jarlskog \cite{Jarlskog:1985ht}, see also \cite{Krastev:1988yu}.
Note that $J$ is independent of which $2\times2$ subset of the PMNS matrix is used to calculate it, up to an overall sign.
The final triple sign term in eq.~\ref{eq:P general} proportional to the Jarlskog invariant $J$ is only non-zero for appearance channels ($\alpha\neq\beta$), as is evident by the definition in eq.~\ref{eq:J}.
In addition, the sign in front of the Jarlskog invariant $J$ is positive for $\nu_e\to\nu_\mu$ and changes sign if $\alpha$ and $\beta$ are swapped, one of $\alpha$ or $\beta$ is changed to a new flavor (that is not the same as the other), or if neutrinos are changed to antineutrinos.

These equations allow one to compute neutrino oscillation probabilities in vacuum, but do not offer any insights into the relationships between what can realistically be measured and the underlying oscillation parameters.
The rest of this section will be devoted to understanding the behavior of the six neutrino oscillation parameters: the two mass squared differences $\Delta m^2_{21}$ and $\Delta m^2_{31}$, the three mixing angles $\theta_{12}$, $\theta_{23}$, and $\theta_{12}$, and the one complex phase $\delta$.

While the full three-flavor picture of neutrino oscillations involves many subtle aspects, we will begin here with a brief discussion in the context of two-flavor oscillations before expanding to the complete three-flavor picture.
Neutrino probabilities are typically classified into disappearance and appearance probabilities.
Disappearance channels are simpler than appearance channels in several ways.
Typically if an experiment can produce one flavor, it can measure it, but it may or may not be able to easily measure other flavors (e.g.~reactor neutrino experiments cannot easily detect $\nu_\mu$ or $\nu_\tau$ because the energies are not high enough to produce the associated charged leptons).
In addition, the probabilities in disappearance tend to depend on fewer oscillation parameters due to the way we typically construct the PMNS matrix (see eq.~\ref{eq:pmns}) making it easier to extract those that they do depend on and also makes it easier to design such an experiment because fewer external parameters influence the experimental design.
Finally, given the known nature of the oscillation parameters and neutrino cross sections also tends to suppress appearance probabilities, although it need not do so generically.
We will now discuss these two classes of channels in more detail.

\subsection{Disappearance}
\subsubsection{Theory and Phenomenology}
A disappearance neutrino oscillation probability channel is the detection of one flavor of neutrino in a source that is dominantly the same flavor.
In the event that only one $\Delta m^2$ and one mixing angle dominates, it is easy to approximate the full probability as
\begin{equation}
P(\nu_\alpha\to\nu_\alpha;L,E)_{\rm 2-flav}=1-\sin^2(2\theta)\sin^2\left(\frac{\Delta m^2L}{4E}\right)\,.
\end{equation}
We immediately note that $\theta=45^\circ$ leads to maximal mixing where the probability becomes zero at $\frac LE=2\pi(2n+1)$ for $n\in\mathbb Z$.

The full three flavor probability is shown in eq.~\ref{eq:P general} without the final $J$ term.
We note that for disappearance the $\Re(U_{\alpha i}U_{\beta i}^*U_{\alpha j}^*U_{\beta j})$ term becomes just $|U_{\alpha i}|^2|U_{\alpha j}|^2$.
We also note that the way that PMNS matrix is typically parameterized (see eq.~\ref{eq:pmns}) is designed to extract the underlying oscillation parameters as simply as possibly \cite{Denton:2020igp}.
That is, medium baseline ($L\sim1$ km) reactor disappearance experiments can measure $4|U_{e3}|^2(1-|U_{e3}|^2)=\sin^2(2\theta_{13})$ which depends only on $\theta_{13}$.
Solar experiments measure $|U_{e2}|^2=s_{12}^2c_{13}^2$ which allows for the easy extraction of $\theta_{12}$ and long-baseline reactor neutrino experiments ($L\gtrsim50$ km) which measure $4|U_{e1}|^2|U_{e2}|^2=\sin^2(2\theta_{12})c_{13}^4$.
Similarly, atmospheric and accelerator disappearance measurements measure $4|U_{\mu3}|^2(1-|U_{\mu3}|^2)\approx\sin^2(2\theta_{23})$ (given that $s_{13}^2$ is small) which also allows for a fairly straightforward extraction of $\theta_{23}$.

Disappearance experiments tend to be easier than appearance experiments for various reasons and the vast majority of neutrino oscillation measurements are predominantly disappearance measurements.
One reason is that since the probabilities are 1 at high energies and tend to return to close to one; there is not a suppression in the rate due to the probability.
Another reason is that many experiments have enough energy to produce electron neutrinos, but not enough to produce a muon (or tau) thus even if the neutrinos would be detected in the $\nu_\mu$ state, since a muon cannot be produced the cross section is zero.
Finally, given the known oscillation parameters, the largest potentially practical appearance probabilities tend to be $\nu_\mu\to\nu_\tau$, but similarly the $\tau$ production threshold is slowly rising and the cross section is generally suppressed while the $\nu_\mu\to\nu_e$ appearance probability tends to be at the $\sim5-10\%$ level.

\subsubsection{Experimental Efforts}
\label{sec:dis exp}
Technically the first disappearance measurements made used solar neutrinos at the Homestake experiment \cite{Davis:1968cp} which detected $\nu_e$ while the solar neutrino flux is predominantly $\nu_e$ at production.
Since solar neutrinos behave quite differently from most oscillation experiments and generally do not particularly experience oscillations (see \ref{sec:msw} below) this is not a typical disappearance measurement.

The next example is with atmospheric neutrinos as measured by Super-Kamiokande \cite{Super-Kamiokande:1998kpq,Super-Kamiokande:2004orf,Super-Kamiokande:2005mbp,Super-Kamiokande:2006jvq,Super-Kamiokande:2010orq,Super-Kamiokande:2017yvm,Super-Kamiokande:2023ahc} and more recently by IceCube \cite{IceCube:2013pav,IceCube:2014flw,IceCube:2017lak,IceCubeCollaboration:2023wtb,IceCubeCollaboration:2024ssx}.
Atmospheric neutrinos contain multiple flavors at production from the full pion decay chain, as well as from some kaons, but the neutrino flux is dominantly muon neutrinos with subleading contributions from electron neutrinos and muon antineutrinos.
Given what we know of the oscillation parameters, atmospheric neutrinos are well described by
\begin{equation}
P(\nu_\mu\to\nu_\mu;L,E)\approx1-\sin^2(2\theta_{23})\sin^2\left(\frac{\Delta m^2_{\mu\mu}L}{4E}\right)\,,
\label{eq:2 flavor disappearance muon}
\end{equation}
where the effective frequency is the $\nu_\mu$ weighted average of $\Delta m^2_{31}$ and $\Delta m^2_{32}$ \cite{Nunokawa:2005nx,Parke:2024xre}:
\begin{align}
\Delta m^2_{\mu\mu}&=m_3^2-\frac{|U_{\mu1}|^2m_1^2+|U_{\mu2}|^2m_2^2}{|U_{\mu1}|^2+|U_{\mu2}|^2}\\
&=s_{12}^2\Delta m^2_{31}+c_{12}^2\Delta m^2_{32}+\cos\delta s_{13}\sin(2\theta_{12})\tan\theta_{23}\Delta m^2_{21}\\
&\approx
s_{12}^2\Delta m^2_{31}+c_{12}^2\Delta m^2_{32}\,.
\end{align}
The approximations made are that the $\Delta m^2_{21}$ oscillation frequency is too slow to contribute here: $\Delta m^2_{21}/|\Delta m^2_{\mu\mu}|\ll1$ which is satisfied as that ratio is known to be $\sim3\%$.
The two remaining frequencies (i.e.~$\Delta m^2$'s) are similar in size, but since $\theta_{12}<45^\circ$ the probability is slightly more sensitive to $\Delta m^2_{32}$ than $\Delta m^2_{31}$.
The further corrections to eq.~\ref{eq:2 flavor disappearance muon} due to the small but nonzero value of $\theta_{13}$ are small enough to ignore in many cases.
Finally, the matter effect (discussed below in \ref{sec:matter effect}) is somewhat surprisingly small in this channel due to a cancellation \cite{Denton:2024thm}.

The next example of disappearance experiments is with reactor electron antineutrinos. 
There are two classes of experiments that measure oscillations in the three-flavor framework: medium-baseline with $L\sim1$ km and long-baseline with $L\gtrsim50$ km.
The medium baseline experiments measure an effective probability
\begin{equation}
\left.P(\bar\nu_e\to\bar\nu_e;L,E)\right|_{\rm MBL}\approx1-\sin^2(2\theta_{13})\sin^2\left(\frac{\Delta m^2_{ee}L}{4E}\right)\,,
\end{equation}
where the effective frequency is the $\nu_e$ weighted average of $\Delta m^2_{31}$ and $\Delta m^2_{32}$ \cite{Nunokawa:2005nx,Parke:2016joa,Parke:2024xre}\footnote{See also \cite{DayaBay:2013yxg,DayaBay:2015ayh} for slightly different definitions of this quantity.}:
\begin{align}
\Delta m^2_{ee}&=m_3^2-\frac{|U_{e1}|^2m_1^2+|U_{e2}|^2m_2^2}{|U_{e1}|^2+|U_{e2}|^2}\\
&=c_{12}^2\Delta m^2_{31}+s_{12}^2\Delta m^2_{32}\,.
\end{align}
The matter effect (see section \ref{sec:matter effect} below) also tends to be small in these experiments due to the low energy of these experiments traveling through the crust, although there is a subleading effect, see \cite{Li:2016txk,Khan:2019doq}.

The long-baseline reactor experiments measure at a leading level
\begin{equation}
\left.P(\bar\nu_e\to\bar\nu_e;L,E)\right|_{\rm LBL}\approx1-\sin^2(2\theta_{12})\sin^2\left(\frac{\Delta m^2_{21}L}{4E}\right)\,,
\end{equation}
although a sufficiently precise measurement, which is anticipated to be achieved in the 2030s, will measure all the relevant terms in the full probability which is
\begin{align}
P(\bar\nu_e\to\bar\nu_e;L,E)=
1-c_{13}^4\sin^2(2\theta_{12})\sin^2\left(\frac{\Delta m^2_{21}L}{4E}\right)&\nonumber\\
-c_{12}^2\sin^2(2\theta_{13})\sin^2\left(\frac{\Delta m^2_{31}L}{4E}\right)&\\
-s_{12}^2\sin^2(2\theta_{13})\sin^2\left(\frac{\Delta m^2_{32}L}{4E}\right)&\nonumber\,.
\end{align}

The final examples of disappearance experiments are those from accelerator experiments which will measure a similar probability as that in atmospheric disappearance channels.
These require the most sophisticated experimental configurations as well as some knowledge about the oscillation parameters to even design them.
The first with evidence of disappearance in accelerator neutrinos was K2K \cite{K2K:2004iot,K2K:2006yov} followed by 
MINOS \cite{MINOS:2007ixr,MINOS:2008kxu,MINOS:2011neo,MINOS:2011qho,MINOS:2012dbe,MINOS:2013utc,MINOS:2014rjg}, T2K \cite{T2K:2012qoq,T2K:2013bzi,T2K:2014ghj,T2K:2015sqm,T2K:2017krm,T2K:2020nqo,T2K:2023mcm}
and NOvA \cite{NOvA:2016vij,NOvA:2017abs,NOvA:2018gge,NOvA:2019cyt,NOvA:2021nfi}.
All of these benefits use a controlled beam to reduce backgrounds due to direction and timing information as well as a combination of near and far detectors placed at specific locations on the Earth to see one oscillationo given the $\Delta m^2$'s and the typical neutrino energies in their particular beam.
The near detector measures the largely unoscillated flux to determine the flux times cross section while the far detector measures the same thing with the addition of neutrino oscillations.
Naively a ``far over near'' ratio would yield the probability, after accounting for exposure effects, and thus easy extraction of the oscillation parameters.
In reality numerous issues complicate this requiring some modeling of the flux and cross section; improving our understanding of these issues is an active area of research.
While K2K and MINOS were both on-axis with broad band beams, NOvA and T2K employ an off-axis approach to reduce the energy spread and flux uncertainty at the cost of a reduced flux; this is especially useful for appearance measurements.

There have also been numerous anomalous neutrino results which have been interpreted as disappearance with $\Delta m^2$'s inconsistent with those that are well established, such as the often discussed LSND \cite{LSND:2001aii} and MiniBooNE \cite{MiniBooNE:2020pnu} results which seem to prefer anomalous oscillation frequencies at $\gtrsim5\sigma$.
These results are in the appearance channel and are thus subject to constraints from both relevant disappearance channels and are in considerable tension with data from $\nu_\mu$ disappearance \cite{IceCube:2020phf,MINOS:2017cae} and comparisons with other similar experiments such as MicroBooNE \cite{MicroBooNE:2022sdp}; global analyses of oscillation data fail to find a good solution \cite{Dentler:2018sju,Diaz:2019fwt,Boser:2019rta}.
Theree is one main result that has not yet been directly ruled outand is at $>5\sigma$ which is in gallium experiments \cite{SAGE:2009eeu,Kaether:2010ag,Barinov:2021asz}, although this is in some tension from short-baseline reactor neutrinos.
In addition, all hints for new oscillation frequencies are in significant tension with cosmological data \cite{Hagstotz:2020ukm}.
Additional neutrino oscillation frequencies, if confirmed, would imply at least one additional mass state.
Measurements from $Z$ boson decays at LEP confirm that only three neutrinos lighter than $M_Z/2=46$ GeV exist \cite{ALEPH:2005ab}; thus any new state would have to be sterile and not experience the weak interaction.
For the remainder of this chapter, we will focus on the standard oscillation frequencies among the dominantly active neutrino states.

In the late 2020s and the 2030s, three new experiments plan to significantly improve our knowledge of the oscillation parameters.
These are JUNO \cite{JUNO:2015zny}, which has the primary goal of measuring long-baseline reactor neutrinos with additional capabilities to measure atmospheric and solar neutrinos and has recently presented their first oscillation results \cite{JUNO:2025gmd}.
The other two are Hyper-Kamiokande \cite{Abe:2011ts} and DUNE \cite{DUNE:2020ypp} which will both measure accelerator, atmospheric, and solar neutrino oscillations.

\subsection{Appearance}
\label{sec:appearance}
\subsubsection{Theory and Phenomenology}
As mentioned above, appearance experiments tend to be more challenging than disappearance experiments for a variety of reasons, and this is represented in the significantly later measurements of appearance and the lower precision on these measurements.
Without appearance measurements, however, some oscillation parameters are impossible to fully determine and others are vastly harder.

The appearance probability has components similar to the disappearance probability, notably three $\sin^2(\Delta m^2_{ij}L/4E)$ terms.
The initial 1 is not present because as $L\to0$ the appearance probability should return to 0, unlike for disappearance where it should return to 1.
Finally, an additional CP (or T) violating term is also expected to exist, see eq.~\ref{eq:P general} above.

Appearance probabilities depend considerably on all the oscillation parameters including the most challenging oscillation parameters to determine.
Specifically, the sign of $\Delta m^2_{31}$ (the same as the sign of $\Delta m^2_{32}$) which is known as the atmospheric mass ordering\footnote{The sign of $\Delta m^2_{21}$ known as the solar mass ordering has been clear back to early solar neutrino data.}, whether $\theta_{23}$ is above or below $45^\circ$ known as the octant problem, and the value of $\delta$ which quantifies how much CP is violated.
The sensitivity to the mass ordering in particular is driven by the matter effect (see section \ref{sec:matter effect} below) which is particularly relevant for appearance through the crust with neutrino energies $\gtrsim1$ GeV.

\subsubsection{Experimental Efforts}
Appearance measurements are predominantly accelerator neutrinos where electron neutrinos are detected in beams of largely muon neutrinos which provide the first clear measurements of appearance.
The main appearance data sets come from measurements by T2K \cite{T2K:2011ypd,T2K:2013bqz,T2K:2013ppw,T2K:2015sqm,T2K:2017hed,T2K:2017rgv,T2K:2018rhz,T2K:2019ird,T2K:2019bcf,T2K:2023smv} and NOvA \cite{NOvA:2016kwd,NOvA:2017abs,NOvA:2018gge,NOvA:2019cyt,NOvA:2021nfi} which have detected appearance in both neutrino beams and antineutrino beams.
These experiments both leverage a near detector and far detector to constrain flux and cross section uncertainties and have measured $\mathcal O(100)$ appearance events each and are both statistics limited.
These accelerator appearance experiments provide the dominant constraints on the several parameters that are particularly challenging to probe otherwise.
Although they do not currently have sufficient sensitivity to answer key oscillation questions, they do provide competitive, and soon world leading, precision measurements on $\theta_{23}$ and $|\Delta m^2_{\mu\mu}|$ (and thus $|\Delta m^2_{31}|$).

Additionally, several $\nu_\tau$ appearance measurements exist \cite{MammenAbraham:2022xoc}.
Notably OPERA recorded about 8 $\nu_\tau$ appearance events leveraging a high resolution emulsion detector \cite{OPERA:2010pne,OPERA:2013tlg,OPERA:2014fax,OPERA:2015wbl,OPERA:2018nar}.
Atmospheric neutrino experiments have also detected tau neutrinos coming dominantly from the $\nu_\mu\to\nu_\tau$ appearance channel at Super-Kamiokande \cite{Super-Kamiokande:2017edb} and IceCube \cite{IceCube:2019dqi} detecting $\mathcal O(100-1000)$ events with significances $\sim3-4\sigma$.
Finally, a handful $\nu_\tau$ events have been detected from astrophysical neutrinos \cite{IceCube:2020fpi,IceCube:2024nhk} which is presumably appearance since tau neutrinos are not expected to be produced astrophysically.

Another source of appearance is in solar neutrinos which are all produced as $\nu_e$, notably the electron elastic scattering channel (ES) and the neutral current channel (NC).
The ES cross section is dominantly $\nu_e$ but with considerable ($\gtrsim10\%$) $\nu_\mu$ and $\nu_\tau$ contributions.
The NC cross section is flavor blind.
Thus SNO reported a measurement of solar neutrinos with both ES \cite{SNO:2001kpb} and NC \cite{SNO:2002tuh} at levels incompatible with a $\nu_e$ flux only at high significance.
While this is a measurement of appearance, it does not provide information about CP violation because the $\nu_\mu$ and $\nu_\tau$ cross sections are the same\footnote{The ES cross section is slightly different for $\nu_\mu$ and $\nu_\tau$ neutrinos at the one-loop level \cite{Sarantakos:1982bp} which does lead to a small and likely undetectable signal of CP violation \cite{Brdar:2023ttb,Kelly:2024tvh}.} in both channels, thus by unitarity they effectively only measure $P(\nu_e\to\nu_\mu)+P(\nu_e\to\nu_\tau)=1-P(\nu_e\to\nu_e)$ which is CP even.

We briefly comment that not all oscillation experiments are either disappearance or appearance experiments.
Notably, atmospheric neutrinos are dominantly $\nu_\mu$ at the production point, but have significant $\nu_e$ contribution due to kaons and heavier mesons at all energies and muon decays at lower energies \cite{Honda:2015fha}, making them somewhat harder to classify and requiring more complicated procedures to extract precision oscillation data.

\subsection{Wave-Packet Picture}
\label{sec:wave packet}
The above description works within a simple quantum mechanics framework, with two key assumptions.
The first is that the time from production to detection, which is used in the Schr\"odinger equation, is approximately the same as the distance, which is measured experimentally: $t\simeq L$.
The second is that the three neutrino mass states are produced with fixed momentum: $E_i\simeq p+\frac{m_i^2}{2p}\simeq p+\frac{m_i^2}{2E}$.
It would seem that each of these approximations introduce additional corrections $\mathcal O(m^2/2E)$, which is exactly the same size as the effect that we are measuring.

A much more careful treatment, however, finds the simple picture presented in section \ref{sec:nu osc probs} gives the correct answer, see e.g.~\cite{Lipkin:2005kg} and also \cite{Akhmedov:2009rb,Akhmedov:2017mcc,Akhmedov:2022bjs}.
This requires a careful treatment of the above effects as well as the energy and spatial resolution of the production and detection and finds that the separate neutrino mass eigenstates remain coherent in all conceivable oscillation experiments with the exception of astrophysical neutrinos (including solar neutrinos) which are known to be fully decohered.
There are also phenomenological searches for evidence of decoherence, typically theoretically ascribed to quantum gravity effects; no such evidence has been found \cite{DayaBay:2016ouy,Stuttard:2020qfv,deGouvea:2020hfl,deGouvea:2021uvg,DeRomeri:2023dht,deGouvea:2024syg}.

\subsection{Matter Effect}
\label{sec:matter effect}
A key effect in neutrino oscillation experiments is the presence of matter in the path of neutrinos.
While absorption and deflection in available materials (the Earth or the Sun) are not relevant until neutrinos reach high energies $>$TeV which are far beyond where oscillations can be observed, the presence of a background field of electrons does affect neutrino oscillations at measurable levels.
Wolfenstein pointed out that the presence of electrons in most matter, and not positrons or muons or taus, contributes to the effective energy of the electron neutrino state during propagation \cite{Wolfenstein:1977ue}, even though neutrinos typically do not propagate in this state\footnote{Note that the matter effect is often erroneously called the MSW effect which refers to a different phenomenon described in sections \ref{sec:msw} and \ref{sec:sn}.} and this arises from the Feynman diagrams shown in fig.~\ref{fig:VCC}.

\begin{figure}
\centering
\begin{fmfgraph*}(30,20)
\fmfleftn{i}2
\fmfrightn{o}2
\fmflabel{$\nu_\alpha$}{i2}
\fmflabel{$f$}{i1}
\fmflabel{$\nu_\alpha$}{o2}
\fmflabel{$f$}{o1}
\fmf{fermion}{i1,v1,o1}
\fmf{fermion}{i2,v2,o2}
\fmf{photon,label=$Z$}{v1,v2}
\end{fmfgraph*}\hspace*{1.2in}
\begin{fmfgraph*}(30,20)
\fmfleftn{i}2
\fmfrightn{o}2
\fmflabel{$\nu_e$}{i2}
\fmflabel{$e^-$}{i1}
\fmflabel{$e^-$}{o2}
\fmflabel{$\nu_e$}{o1}
\fmf{fermion}{i1,v1,o1}
\fmf{fermion}{i2,v2,o2}
\fmf{photon,label=$W$}{v1,v2}
\end{fmfgraph*}
\vspace*{0.1in}
\caption{The two relevant forward elastic Feynman diagrams pointed out in \cite{Wolfenstein:1977ue}.
The left diagram is the neutral current diagram which affects all flavors equally.
The right diagram is the charged current diagram which only affects electron neutrinos.}
\label{fig:VCC}
\end{figure}

The impact of the matter effect can be best understood in the Hamiltonian framework (see eq.~\ref{eq:Hvac}) where we add in additional energy-like contributions to the neutrinos because the mediators are spin-1 bosons.
\begin{equation}
H=\frac1{2E}U
\begin{pmatrix}
0\\&\Delta m^2_{21}\\&&\Delta m^2_{31}
\end{pmatrix}
U^\dagger
+
\begin{pmatrix}
V_{\rm CC}+V_{\rm NC}\\&V_{\rm NC}\\&&V_{\rm NC}
\end{pmatrix}\,,
\label{eq:H matter}
\end{equation}
where the charged-current (CC) and neutral-current (NC) potentials are
\begin{align}
V_{\rm CC}&=\pm\sqrt2G_Fn_e\,,\\
V_{\rm NC}&=\mp\frac12\sqrt2G_Fn_n\,,
\end{align}
and where $G_F$ is Fermi's constant, $n_e$, $n_n$ are the electron and neutron number densities, respectively, and the upper (lower) signs are for neutrinos (antineutrinos).
Then one can solve the Schr\"odinger as described in section \ref{sec:nu osc probs} and, if the density varies slowly enough, one can again use the expression of eq.~\ref{eq:amplitude} to compute the oscillation probabilities; there may be faster or more clear ways of computing the probabilities in matter, however.

We note several key phenomenon related to the matter effect.
First, $V_{\rm NC}$ can be dropped from the above Hamiltonian\footnote{Note that this is not true for sterile neutrinos or certain other new physics models.} for the same reason that $m_1^2/2E$ can be subtracted out as well: neutrino oscillations are not sensitive to quantities proportional to the identity matrix, in any basis.
Second, the matter effect is relevant for many experiments in the Earth including the crust and in the Sun.
Third, the size of the effect relative to the $\Delta m^2$'s is given by the quantity $a$ which is defined as
\begin{equation}
a\equiv2EV_{\rm CC}\approx1.52588\e{-4}\left(\frac{Y_e\rho}{\rm g\cdot{\rm cm}^{-3}}\right)\left(\frac E{\rm GeV}\right){\rm eV}^2\,,
\end{equation}
where $Y_e$ is the electron fraction and is $\sim0.5$ in the Earth and is larger in the Sun and $\rho$ is the density which is $\sim3$ g/cm$^3$ in the Earth's crust and $\sim50-100$ g/cm$^3$ in the center of the Sun.

\begin{figure}
\centering
\includegraphics[width=0.49\textwidth]{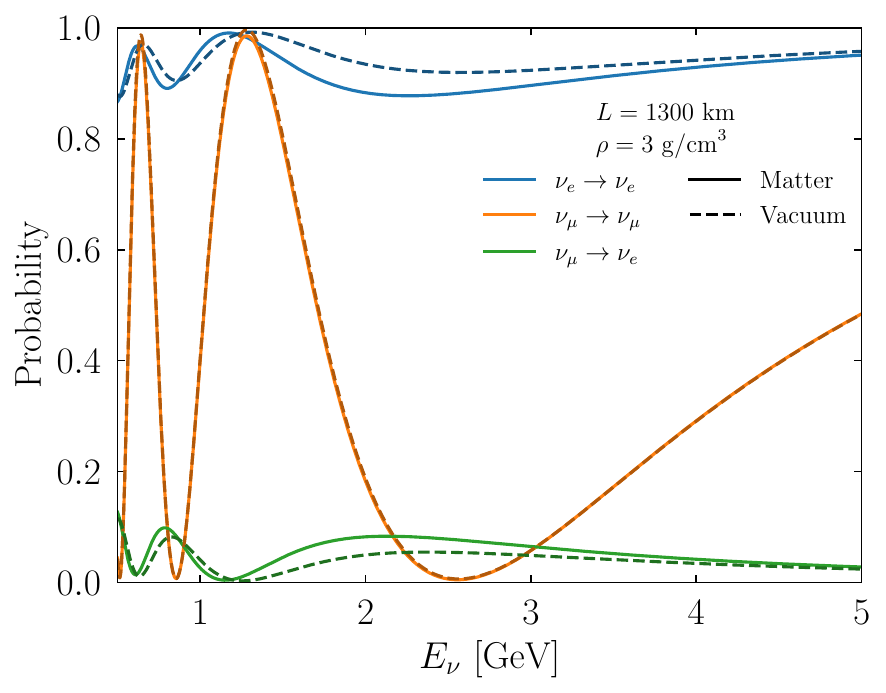}
\includegraphics[width=0.49\textwidth]{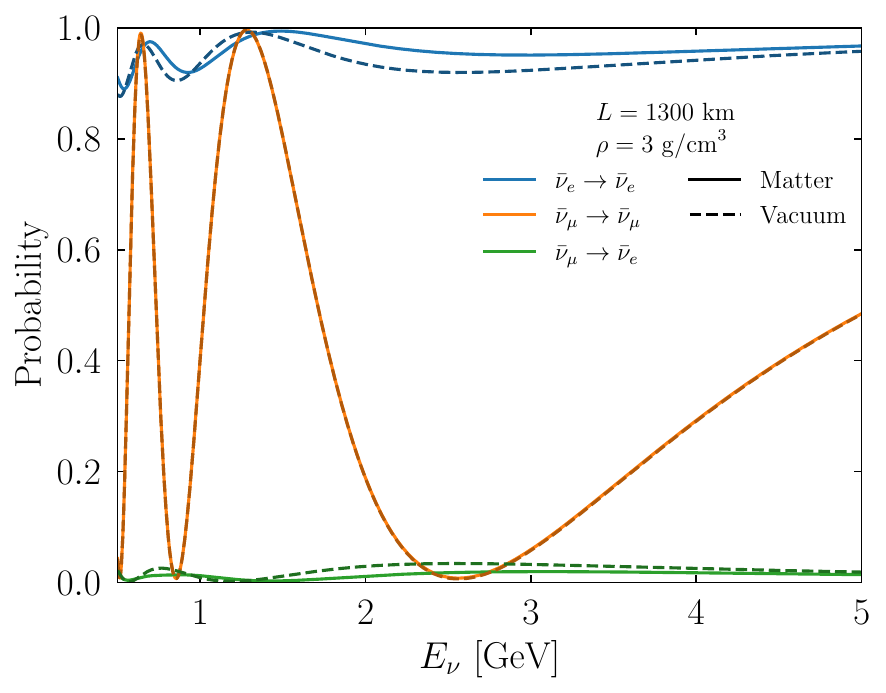}
\caption{Neutrino oscillation probabilities for $\nu_e$ disappearance (blue), $\nu_\mu$ (orange), and $\nu_\mu\to\nu_e$ appearance (green) through the Earth's crust at a baseline of 1300 km -- the baseline of DUNE -- at energies sensitive to the first atmospheric oscillation.
The dashed curves show the probabilities in vacuum.
The \textbf{left} panel is for neutrinos and the \textbf{right} panel is for antineutrinos.
The probabilities were calculated with \texttt{NuFast} \cite{Denton:2024pzc}.}
\label{fig:probabilities}
\end{figure}

A useful means of understanding how the matter effect impacts neutrino oscillations is to work in a framework where we leverage the intuition built about vacuum oscillations and then understand how the presence of matter modifies the effective oscillation parameters \cite{Minakata:2015gra,Denton:2016wmg,Denton:2018hal,Denton:2018fex,Denton:2018cpu,Denton:2019yiw,Denton:2019ovn,Denton:2019qzn,Agarwalla:2013tza}, for an overview of various techniques, see \cite{Barenboim:2019pfp}.
The matter effect is essential for breaking the degeneracy to determine the sign of the $\Delta m^2$'s.
We show various appearance and disappearance probabilities in vacuum and in matter through the Earth's crust in fig.~\ref{fig:probabilities}.

\subsection{Matter Effect in the Sun}
\label{sec:msw}
The matter effect in the Sun also modifies the neutrino propagation in a wide variety of ways depending on the true oscillation parameters.
The correct description, given the known oscillation parameters\footnote{For radically different oscillation parameters, additional corrections are necessary due to the presence of jump probabilities \cite{Parke:1986jy,Petcov:1987zj,Kuo:1988pn}.} is
\begin{equation}
P^\odot(\nu_e\to\nu_e)=\sum_{i=1}^3|\hat U_{ei}|^2|U_{ei}|^2\,,
\label{eq:P solar}
\end{equation}
where $\hat U$ is the matrix that diagonalizes the matter Hamiltonian (eq.~\ref{eq:H matter}) at the production point of the neutrinos in the Sun; in vacuum $\hat U\to U$.

Given the Sun's density, at low enough energies, $E\lesssim1$ MeV, we see that $\hat U\simeq U$ and the probability is
\begin{equation}
	P^{\odot}_{LE}(\nu_e\to\nu_e)=|U_{e1}|^4+|U_{e2}|^4+|U_{e3}|^4=c_{13}^4\left(1-\frac12\sin^22\theta_{12}\right)+s_{13}^4\approx1-\frac12\sin^22\theta_{12}\,,
\label{eq:P solar LE}
\end{equation}
where in the final step we have used the fact that $s_{13}^2\simeq0.02$ is small.
For high energies\footnote{For much higher energies above the atmospheric resonance $E\gtrsim10$ GeV additional effects are in play, see e.g.~\cite{Lehnert:2007fv}.
The solar neutrino flux is expected to end around 20 MeV.} $E\gtrsim10$ MeV the neutrinos produced in the Sun are dominantly $\nu_2$ and remain $\nu_2$ as they leave the Sun and propagate to the Earth, so the probability is
\begin{equation}
	P^\odot_{HE}(\nu_e\to\nu_e)=(|U_{e1}|^2+|U_{e2}|^2)|U_{e2}|^2+|U_{e3}|^4=c_{13}^4s_{12}^2+s_{13}^4\approx s_{12}^2\,.
	\label{eq:PsunHE}
\end{equation}
Eq.~\ref{eq:PsunHE} is commonly referred to as the MSW effect, referring to the paper that identified it by Mikheyev and Smirnov \cite{Mikheyev:1985zog} and the fact that it leverages the matter effect at all, which was identified in a different context earlier by Wolfenstein \cite{Wolfenstein:1977ue}.

\begin{figure}
\centering
\includegraphics[width=0.6\textwidth]{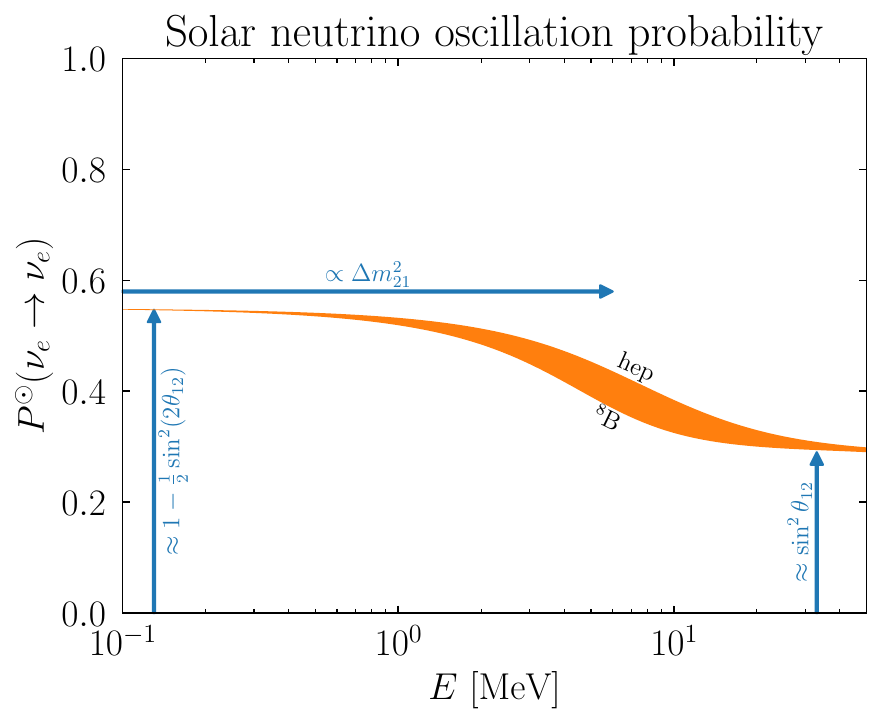}
\caption{The solar neutrino $\nu_e\to\nu_e$ probability as a function of neutrino energy across the relevant range.
The thickness of the band orange band represents the variation in the probability due to the different densities of the production regions within the Sun with hep flux produced at larger radii, the $^8$B flux produced closer to the center of the Sun.
The low energy (vacuum dominated) and high energy (MSW adiabatic region) probabilities are shown up to $\theta_{13}$ corrections.
Increasing $\Delta m^2_{21}$ increases the energy at which the transition happens.}
\label{fig:solar probability}
\end{figure}

While eq.~\ref{eq:P solar} describes the full probability including the transition region between the low energy and high energy regions described above, we can also approximate that region by modifying the vacuum probability in eq.~\ref{eq:P solar LE} to the two flavor solution in matter, and then include some three flavor corrections, to find that the probability is
\begin{align}
P^\odot(\nu_e\to\nu_e)&\approx\frac{c_{13}^4}2\left[1+\cos2\theta_{12}\frac{\cos2\theta_{12}-c_{13}^2a/\Delta m^2_{21}}{\sqrt{(\cos2\theta_{12}-c_{13}^2a/\Delta m^2_{21})^2+\sin^22\theta_{12}}}\right]+s_{13}^4\\
&\approx\frac12\left[1+\cos2\theta_{12}\frac{\cos2\theta_{12}-c_{13}^2a/\Delta m^2_{21}}{\sqrt{(\cos2\theta_{12}-c_{13}^2a/\Delta m^2_{21})^2+\sin^22\theta_{12}}}\right]\,,
\end{align}
which describes the physics in the transition region between the two cases very well.
It shows that larger values of $\theta_{12}$ (up to $45^\circ$) will decrease the probability at low energies and increase it at high energies, making it harder to identify the transition region, while smaller values of $\theta_{12}$ does the opposite in both regimes.
We also see that the transition region, defined by when $a_{\rm res}\approx\cos2\theta_{12}\Delta m^2_{21}/c_{13}^2$, increases in energy proportionally to $\Delta m^2_{21}$ and also depends on $\theta_{12}$.
The solar neutrino probability is shown in fig.~\ref{fig:solar probability} which depends on the density at the production region in the Sun.
As different nuclear processes happen at different radii and thus different densities, the probability also depends on the production process.
For example, $^8$B neutrinos are produced from close to the center of the Sun, while hep neutrinos (which have not yet been observed) are produced from larger radii, see e.g.~\cite{Denton:2025cbo,Zaidel:2025kdk}.
Other sources of neutrinos within the Sun have also been detected (see e.g.~\cite{BOREXINO:2014pcl}).

Additionally, when solar neutrinos are detected at night, they experience the additional matter effect of the Earth.
This leads to a small regeneration of electron neutrinos and a slight enhancement of the probability at the few \% level for energies above a few MeV; this is known as the day-night effect \cite{Bouchez:1986kb,Baltz:1986hn,Cribier:1986ak}.
It has not yet been significantly detected, but experiments like DUNE and Hyper-Kamiokande may be able to detect it in the future \cite{Capozzi:2018dat,DUNE:2024wvj,Hyper-Kamiokande:2018ofw,Barenboim:2023krl}.
This oscillation channel also dominantly depends on the same parameters: $\Delta m^2_{21}$ and $\theta_{12}$, albeit with rather different oscillation physics.

Solar neutrinos have been detected in a variety of experiments, dating back to the Homestake experiment from the 1960s to the 1980s \cite{Davis:1968cp,Cleveland:1998nv}.
This, combined with theory predictions \cite{Bahcall:1968hc,Bahcall:2000nu}, lead to the ``solar neutrino problem:'' why the flux of higher energy neutrinos seems to be a factor of $\sim3$ lower than the prediction.
The solar neutrino puzzle was conclusively resolved by SNO \cite{SNO:2001kpb,SNO:2002tuh} which confirmed via elastic scattering and neutral current measurements, which contain some $\nu_\mu$ and $\nu_\tau$ neutrinos, that both the Homestake measurement and the theory prediction were correct and neutrinos were changing flavor \cite{SNO:2002tuh}.
Since then, SNO has improved their measurements \cite{SNO:2011hxd} and Super-Kamiokande has also measured solar neutrinos \cite{Super-Kamiokande:1998qwk,Super-Kamiokande:1998zvz,Super-Kamiokande:1998oic,Super-Kamiokande:2001bfk,Super-Kamiokande:2002ujc,Super-Kamiokande:2005wtt,Super-Kamiokande:2008ecj,Super-Kamiokande:2010tar,Super-Kamiokande:2016yck,Super-Kamiokande:2023jbt}.
On the low energy side, gallium experiments \cite{GALLEX:1998kcz,SAGE:1999nng,Gavrin:2001sz,GNO:2000avz} have measured the sub-MeV solar neutrino flux, and Borexino has measured the solar neutrino flux across a broad range of energies \cite{BOREXINO:2014pcl}.

\subsection{Oscillations in Supernova}
\label{sec:sn}
Supernova (SN) are the most efficient neutrino sources in the universe converting about 99\% of the gravitational energy of large stars in neutrinos with energies in the 10s of MeV.
Numerous neutrino effects are present inside SN including neutrino-neutrino scattering, so-called fast-flavor conversions, and MSW conversions, as discussed above \cite{Raffelt:2007nv,Dighe:2008dq,Scholberg:2012id}.
These MSW conversions could lead to a discernible effect of the mass ordering in a galactic SN event, depending on the distance to the SN, the details of the SN model, and other effects \cite{Scholberg:2017czd,DeSalas:2018rby}\footnote{Note that the diffuse supernova neutrino background \cite{Suliga:2022ica}, containing the SN neutrinos from the entire universe, is beginning to be detected but depends fairly weakly on the atmospheric mass ordering \cite{Moller:2018kpn}.}.

\subsection{The Physics of Each Oscillation Parameter}
\begin{figure}
\centering
\includegraphics[width=0.6\textwidth]{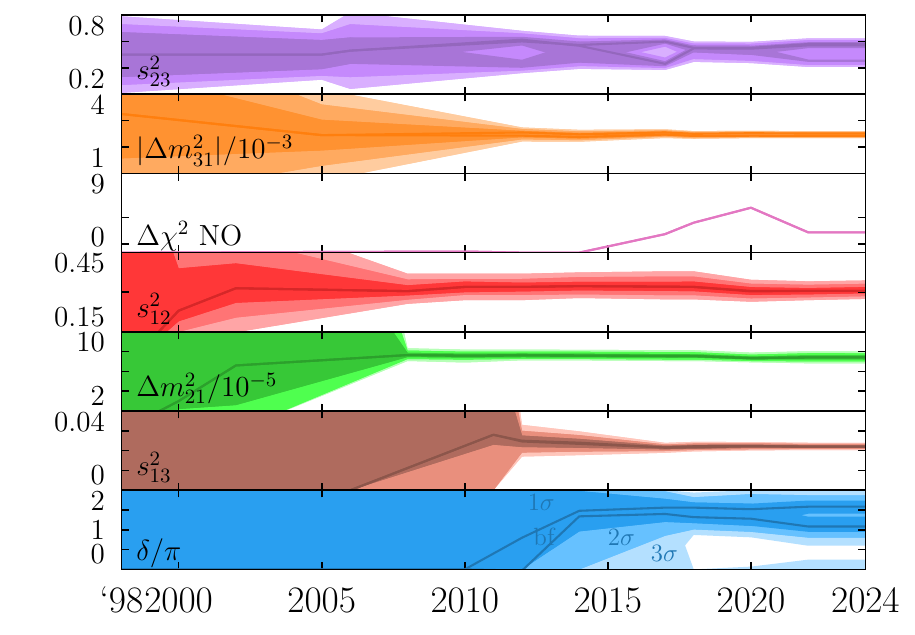}
\caption{The evolution of our understanding of the six oscillation parameters over the last $\sim2.5$ decades, see also \cite{Denton:2022een}.
``NO'' refers to the preference for the normal atmospheric mass ordering ($\Delta m^2_{31}>0$).
The best fit values are shown in solid lines and the shaded regions show the $1,2,3$ $\sigma$ preferred regions.
Data comes from \cite{Super-Kamiokande:1998kpq,Gonzalez-Garcia:2000opv,Maltoni:2002ni,Super-Kamiokande:2005mbp,Super-Kamiokande:2006jvq,Schwetz:2008er,Gonzalez-Garcia:2010zke,T2K:2011ypd,Forero:2012faj,Forero:2014bxa,deSalas:2017kay,Capozzi:2017ipn,Esteban:2020cvm}.}
\label{fig:parameter precision}
\end{figure}

We have discussed the physics of different neutrino oscillation channels in the other parts of this section including the impacts of some of the oscillation parameters.
Figure \ref{fig:parameter precision} shows the historical evolution of our understanding of the six oscillation parameters, along with a separate cutout for the sign of $\Delta m^2_{31}$, known as the atmosheric mass ordering.
Here we discuss the key effects of each of the six oscillation parameters.
Before we do that, we briefly discuss the choices of how one parameterizes the mixing matrix.

\subsubsection{Parameterizing the Mixing Matrix}
The neutrino oscillation mixing matrix, known as the PMNS matrix, can be, and has been, parameterized in numerous different ways.
The commonly used one in eq.~\ref{eq:pmns} is beneficial for several reasons, notably that it allows for many practical experiments to map their measurements fairly directly onto the actual oscillation parameters \cite{Denton:2020igp}.
Alternative parameterizations do not change the physics and could have been useful if Nature selected different regions of parameter space or the experimental landscape were different.

Once the PMNS matrix is defined, one must define which mass eigenstate is which.
If we better understood the relationship between the masses, or the mass splittings, and the mixings, this would be straightforward.
As they are not, alternative definitions must be employed and the literature is somewhat inconsistent on this point.
We recommend the definition that is, at the moment, the most robust \cite{Denton:2020exu,Denton:2021vtf}:
\begin{equation}
|U_{e1}|^2>|U_{e2}|^2>|U_{e3}|^2\,.
\end{equation}
That is, we know that the electron neutrino is mostly $\nu_1$ ($|U_{e1}|^2\sim2/3$), somewhat $\nu_2$ ($|U_{e2}|^2\sim1/3$), and a tiny bit $\nu_3$ ($|U_{e3}|^2=0.02$).
Thus the statement that high energy solar neutrinos (e.g.~$^8$B) have a lower disappearance probability than low energy solar neutrinos (e.g.~pp) implies that $\Delta m^2_{21}>0$, see also \cite{deGouvea:2000pqg}.

\subsubsection{\texorpdfstring{$\Delta m^2_{31}$}{Dmsq31}}
The most robustly measured neutrino oscillation parameter is the magnitude of $\Delta m^2_{31}$, also known as the atmospheric mass splitting or the atmospheric frequency, and is known to be $|\Delta m^2_{31}|\sim2.5\e{-3}$ eV$^2$.
It has been measured by eight different experiments including reactor-, long-baseline accelerator-, and atmospheric-experiments spanning four orders of magnitude in energy and baseline with an impressive level of agreement among them all.
$\Delta m^2_{31}$ is approximately the same $\Delta m^2_{32}$ up to a $\sim\Delta m^2_{21}/\Delta m^2_{31}\sim3\%$ correction.
This mass splitting can be thought of as a frequency and it dictates when ($L$) and at what energy ($E$) neutrino oscillations happen.
Future measurements will achieve sub percent precision on this parameter.

The sign of this parameter, known as the atmospheric mass ordering question, is one of the big unknowns in neutrino physics.
While current data is inconclusive on its sign \cite{Esteban:2024eli,Capozzi:2025wyn,Esteban:2026phq}, it is widely expected to be measured by numerous different techniques with the next generation of detectors.
The most robust approach is directly leveraging the matter effect, with which DUNE will rapidly reach high significance $>5\sigma$ \cite{DUNE:2022aul}.
Additionally, by leveraging knowledge of the solar mass ordering and high resolution measurement of the $\Delta m^2_{31}$ and $\Delta m^2_{32}$ frequencies, JUNO will have some sensitivity to the atmospheric mass ordering at the $\sim3\sigma$ level \cite{JUNO:2015zny}.
By comparing $\nu_e$ disappearance and $\nu_\mu$ disappearance \cite{Nunokawa:2005nx,Parke:2024xre}, combinations of next generation experiments will provide some information on the mass ordering.
Finally, a galactic supernova could provide some atmospheric mass ordering information as well \cite{Scholberg:2017czd}.

The mass ordering has important implications for other areas of physics.
Notably, cosmological data sets constrain the sum of the neutrino masses which are currently preferring parameters in some tension with the lower limit from the inverted ordering (and also in some tension with the lower limit from the normal ordering) \cite{eBOSS:2020yzd,DES:2021wwk,DESI:2024hhd}.
It also plays an important role in the expected neutrinoless double beta decay rates with easier to detect rates in the inverted ordering.
Finally, it affects measurements of the cosmic neutrino background, again with easier detections in the case of the inverted ordering \cite{Long:2014zva}.

\subsubsection{\texorpdfstring{$\Delta m^2_{21}$}{Dmsq21}}
While there are three mass squared differences, only two are independent.
The hierarchically different one from $\Delta m^2_{31}$ is $\Delta m^2_{21}$, known as the solar mass splitting, and is known to be $\Delta m^2_{21}\approx+7.5\e{-5}$ eV$^2$.
Then the third mass splitting is related to the other two by the sum rule $\Delta m^2_{32}=\Delta m^2_{31}-\Delta m^2_{21}$.
This mass splitting, $\Delta m^2_{21}$, affects solar neutrinos, but is not particularly well measured there due to few measurements in the transition region discussed in section \ref{sec:msw}, see also fig.~\ref{fig:solar probability}, which happens when the disappearance probability changes from $\sim0.55$ for $E\lesssim1$ MeV to $\sim0.3$ for $E\gtrsim5$ MeV.
Solar neutrinos experiencing the matter effect through the Earth, so-called nighttime solar neutrinos, are also sensitive to $\Delta m^2_{21}$.
Although this effect has not yet be definitively detected, it still does provide some information\footnote{The Earth's matter effect in solar neutrinos also provides information about $\theta_{12}$, but not more than the information in neutrinos directly from the Sun.} about $\Delta m^2_{21}$.
The best measurement comes from reactor neutrinos by JUNO \cite{JUNO:2025gmd}, previously it was by KamLAND \cite{KamLAND:2013rgu}.
Future measurements from JUNO will achieve sub percent precision \cite{JUNO:2022mxj}.
Measuring the complex phase $\delta$ depends on the true value of this parameter \cite{Denton:2023zwa,Denton:2023qmd}.

\subsubsection{\texorpdfstring{$\theta_{23}$}{theta23}}
In the usual parameterization, the largest mixing angle by value is $\theta_{23}$ and is called the atmospheric mixing angle.
It is known to be close to maximal: $\sim45^\circ$.
This means that the $\nu_\mu$ disappearance probability has been experimentally determined to be nearly zero at the first (highest energy) oscillation maximum.
This parameter was first measured by Super-Kamiokande \cite{Super-Kamiokande:1998kpq} and is currently best determined by long-baseline accelerator experiments MINOS \cite{MINOS:2013utc}, NOvA \cite{NOvA:2021nfi}, and T2K \cite{T2K:2023mcm} as well as atmospheric experiments IceCube \cite{IceCube:2019dqi} and Super-Kamiokande \cite{Super-Kamiokande:2023ahc}.

Disappearance measurements in the $\nu_\mu$ channels for these experiments constrain $\sin^2(2\theta_{23})$ which seems to be near one.
This leaves open an approximate degeneracy whereby it is not easy to tell if $\theta_{23}$ is greater than or less than $45^\circ$; this is known as the octant problem.
The octant problem is whether or not $\nu_3$ (the neutrino state that is least $\nu_e$) is dominantly $\nu_\tau$: the lower octant (or more ``normal'' choice, see e.g.~\cite{Denton:2020exu}) or is dominantly $\nu_\mu$: the upper octant.
The best means of probing this is via measuring an appearance probability which depends on $\sin^2\theta_{23}$.
This is challenging for the reasons outlined in subsection \ref{sec:appearance}, especially given the fact that many oscillation effects affect the probability at similar levels.

Current data is indecisive on the octant question and measuring this is a major goal of upcoming experiments DUNE and Hyper-Kamiokande which each have good sensitivity provided that $\theta_{23}$ is not too close to maximal.

\subsubsection{\texorpdfstring{$\theta_{12}$}{theta12}}
The next largest mixing angle in the usual parameterization is $\theta_{12}$ which is usually called the solar mixing angle and is $\sim34^\circ$.
This indicates that the long-baseline reactor neutrino disappearance probability drops to about 15\% at the first (highest energy) oscillation maximum.
This parameter is determined in three separate ways.

Historically, the first indication of this parameter comes from high energy (dominantly $^8$B) solar neutrinos which are sensitive to $\sin^2\theta_{12}$ from a combination of the Homestake measurement \cite{Cleveland:1998nv} and the Bahcall theory prediction \cite{Bahcall:1968hc} and is measured today by SNO \cite{SNO:2011hxd}, Borexino \cite{BOREXINO:2014pcl}, and Super-Kamiokande \cite{Super-Kamiokande:2016yck}.
Next, the low energy pp solar neutrino flux was determined which depends on $1-\frac12\sin^2(2\theta_{12})$ and was measured by SAGE, GALLEX, GNO, and Borexino \cite{GALLEX:1998kcz,SAGE:1999nng,Gavrin:2001sz,GNO:2000avz,BOREXINO:2018ohr}.
Finally, long-baseline ($\sim50$ km) reactor neutrinos measure a spectrum that also provides information on $\sin^2(2\theta_{12})$ as measured by KamLAND and JUNO \cite{KamLAND:2013rgu,JUNO:2025gmd}.
All approaches are consistent.
Improved measurements of solar neutrinos by DUNE \cite{Capozzi:2018dat} will improve the solar measurements somewhat, and improved long-baseline reactor neutrino measurements from JUNO will dramatically improve measurements of this parameter \cite{JUNO:2022mxj}.

\subsubsection{\texorpdfstring{$\theta_{13}$}{theta13}}
The final mixing angle, called $\theta_{13}$ or the reactor mixing angle, was thought to be quite small \cite{Albright:2006cw}, generally $\lesssim1^\circ$ by theoretical arguments.
Measurements of medium-baseline ($\sim1$ km) reactor neutrinos indicate that it is $\sim8.5^\circ$ as measured by Daya Bay \cite{DayaBay:2018yms}, RENO \cite{RENO:2018dro}, and Double Chooz \cite{DoubleChooz:2019qbj} with some earlier hints coming from various combinations of data sets \cite{Balantekin:2008zm,Fogli:2011qn}.
It is now one of the best measured oscillation parameters and future experiments are unlikely to significantly improve it.

It plays a subleading role in long-baseline accelerator appearance measurements as the variation in the probability due to varying $\delta$ also depends on $\theta_{13}$.
It also affects solar neutrinos slightly as it governs the size of the non-oscillating contribution to the flux (which is quite small) and a small correction to the effective size of the matter effect \cite{Denton:2019yiw}.

The precise measurement of this parameter, as well as where it sits in parameter space, plays a key role in flavor model predictions, see section \ref{sec:flavor}.

\subsubsection{\texorpdfstring{$\delta$}{delta} and CP Violation}
The final oscillation parameter in the three-flavor oscillation picture is called $\delta$ which is the one guaranteed physical complex phase.
When computing oscillation probabilities, the sign of this phase changes when changing between neutrinos and antineutrinos.
The value of this parameter is largely undetermined.
Currently NOvA and T2K have some sensitivity to this parameter at the $\sim2\sigma$ level, but their measurements are not obviously consistent.
Measuring this parameter is one of the primary physics goals of upcoming experiments DUNE \cite{DUNE:2020ypp} and Hyper-Kamiokande \cite{Hyper-Kamiokande:2018ofw}.

While $\delta$ is related to CP violation and parametrically describes one of three known numbers in particle physics that governs the amount of CP violation (the other two are the similar phase in the quark matrix where there is CP violation and in the gluon sector where CP seems to be conserved), it is a parameterization dependent quantity.
A better means of quantifying the amount of CP violation in the neutrino sector is via the Jarlskog invariant; see eq.~\ref{eq:J}.
Physical observables often scale with this parameter.
For example, ignoring matter effects, the difference between neutrino and antineutrino appearance at the first oscillation maximum for long-baseline accelerator experiments is approximately given by
\begin{equation}
P(\nu_\mu\to\nu_e)-P(\bar\nu_\mu\to\bar\nu_e)\approx8\pi J\frac{\Delta m^2_{21}}{\Delta m^2_{31}}\,.
\label{eq:Pme-Pbarme}
\end{equation}
The largest that this value can be at $J_{\max}=\frac1{6\sqrt3}\approx0.096$ is $\sim0.72$.
Largely due to the measurement of $\theta_{13}$, the maximum allowed value of $J$, given oscillation data, is 0.033 \cite{Denton:2020uda,Esteban:2024eli}, and thus the largest eq.~\ref{eq:Pme-Pbarme} can be is $\sim0.25$.

\section{Flavor Models}
\label{sec:flavor}
While the parameters governing the neutrino sector are treated as free parameters, as are the masses of the charged leptons and quarks (i.e.~the Yukawa couplings of the charged leptons and quarks to the Higgs boson) and the quark mixing matrix, it may be possible that some or all of these parameters are predicted from an underlying theoretical argument.
These flavor models are often related to some symmetry structure applied to some aspect of the fermion sector.
As the neutrino masses and mixings are the last known remaining parameters, they play a key role in assessing the predictivity of these models.
A lot of the attention in the field is focused on non-Abelian finite groups \cite{Kaplan:1993ej,Ishimori:2010au,Feruglio:2019ybq,Almumin:2022rml,Chauhan:2023faf,Denton:2023hkx}, although other classifications exist as well.
Flavor models can be built up from the low scale observables or generated at high scales and run down to the observables.

As an example, one scenario that was popular before $\theta_{13}$ was measured to be non-zero is called tri-bimaximal (TBM) \cite{Harrison:2002er,Harrison:2002kp,Harrison:2003aw} where the mixing matrix is
\begin{equation}
U^{\rm TBM}=
\begin{pmatrix}
\sqrt{\frac23}&\frac1{\sqrt3}&0\\
-\frac1{\sqrt6}&\frac1{\sqrt3}&-\frac1{\sqrt2}\\
-\frac1{\sqrt6}&\frac1{\sqrt3}&\frac1{\sqrt2}
\end{pmatrix}\,,
\end{equation}
which maps on to $\theta_{12}\approx35^\circ$, $\theta_{23}=45^\circ$, and $\theta_{13}=0$.
The prediction was very successful for the two larger mixing angles, but is quite wrong for $\theta_{13}$.
To address this, it is common to consider a scenario where the charged lepton mixing matrix is non-diagonal.
Then the mass matrix which is the product of the neutrino matrix and charged lepton matrix may be well fit to the data.
The corrections in the charged lepton matrix may be taken as related to the Cabibbo angle from the quark mixing matrix \cite{King:2007pr,Pakvasa:2007zj,King:2012vj} see also \cite{Giunti:2002ye,Minakata:2004xt,Datta:2005ci,Everett:2005ku}, for reviews see \cite{King:2013eh,King:2014nza,Petcov:2017ggy}.
These models get close to the measured values, but are at considerable tension largely due to the precision with which $\theta_{13}$ is measured.

Another example is texture zeros where certain elements of the Majorana mass matrix are zero \cite{Merle:2006du,Lashin:2011dn,Dev:2006qe,Verma:2020gpl,Xing:2003jf,Xing:2003ic,Grimus:2004hf,BenTov:2011tj,Liu:2012axa,Frampton:2002yf,Guo:2002ei,Xing:2002ta,Desai:2002sz,Xing:2002ap,Dev:2006if,Dev:2006xu,Fritzsch:2011qv,Honda:2003pg,Hollik:2017get}.
Models with three or more zeros have long since been ruled out \cite{Frampton:2002yf}.
Most two-texture zero scenarios are ruled out given the latest oscillation and cosmological data, only those with $M_{ee}=M_{e\mu}=0$ and $M_{ee}=M_{e\tau}=0$ are allowed \cite{Denton:2023hkx}.
Scenarios with one-texture zero are also considered.

\begin{figure}
\centering
\includegraphics[width=0.6\textwidth]{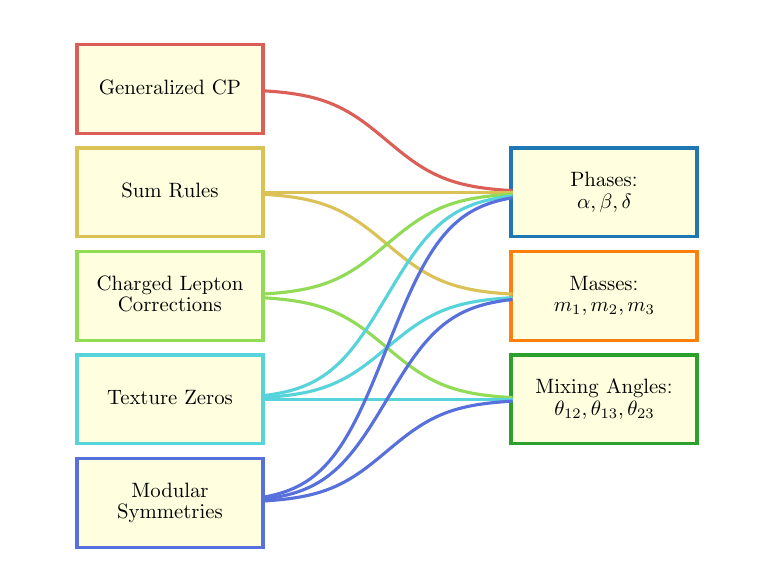}
\caption{Several flavor model prediction classifications (left) and the oscillation parameters affects (right).
Also shown are the so-called Majorana phases: $\alpha$ and $\beta$ which are physical if and only if neutrinos have a Majorana mass term.
Figure adapted from \cite{Denton:2023hkx}.}
\label{fig:flavor model schematic}
\end{figure}

Many other classes of models are also considered in the literature such as modular symmetries \cite{Feruglio:2017spp,Penedo:2018nmg,Kobayashi:2018vbk,Gehrlein:2020jnr,Kikuchi:2021ogn,Liu:2021gwa,Kobayashi:2023zzc}, generalized CP \cite{Wolfenstein:1981rk,Kayser:1984ge,Bilenky:1984fg,Branco:1986gr,Feruglio:2012cw,Holthausen:2012dk,Ding:2013nsa,Ding:2014hva,King:2014rwa,King:2014rwa,Ding:2014ssa,Hagedorn:2014wha,Ding:2014ora,Ding:2015rwa,Penedo:2018kpc}, and models with mass sum rules \cite{Barry:2010yk,Barry:2010yk,Bazzocchi:2009da,Ding:2010pc,Ma:2005sha,Ma:2006wm,Kang:2015xfa,Honda:2008rs,Brahmachari:2008fn,Altarelli:2005yx,Chen:2009um,Chen:2009gy,Cooper:2012bd,Altarelli:2009kr,Altarelli:2008bg,Hirsch:2008rp,Bazzocchi:2009pv,Everett:2008et,Boucenna:2012qb,Mohapatra:2012tb,Altarelli:2005yp,Altarelli:2006kg,Ma:2006vq,Bazzocchi:2007na,Bazzocchi:2007au,Lin:2008aj,Ma:2009wi,Ciafaloni:2009qs,Bazzocchi:2008ej,Feruglio:2013hia,Chen:2007afa,Ding:2008rj,Chen:2009gf,Feruglio:2007uu,Merlo:2011hw,Luhn:2012bc,Fukuyama:2010mz,Ding:2013eca,Lindner:2010wr,Hashimoto:2011tn,Ding:2011cm,Morisi:2007ft,Adhikary:2008au,Lin:2009bw,Csaki:2008qq,Hagedorn:2009jy,Burrows:2009pi,Ding:2009gh,Mitra:2009jj,delAguila:2010vg,Burrows:2010wz,Ahn:2014zja,Karmakar:2014dva,Ahn:2014gva,He:2006dk,Berger:2009tt,Kadosh:2010rm,Lavoura:2012cv,King:2012in,Adulpravitchai:2009gi,Dorame:2011eb,Dorame:2012zv,King:2013psa,Gehrlein:2017ryu,Gehrlein:2020jnr}, among others.
Some models in one class can also be partially or fully categorized in other classes.
These classifications are somewhat useful and they indicate what aspect of the neutrino mass matrix they predict, as shown in fig.~\ref{fig:flavor model schematic}.
Some approaches do not directly fall into these classification schemes such as Froggat-Nielsen \cite{Froggatt:1978nt} where $U(1)$ charges are applied to each state.

\section{Connection to Non-Oscillation Physics}
\label{sec:non osc}
While oscillations are generally the best means of probing the six parameters discussed above, there are other physical processes that connect to neutrino oscillations through an overlap in the parameters probed.
Some of them are sensitive to different combinations of the oscillation parameters, but at lower sensitivity or in a more model dependent fashion, while others probe a combination of the above six oscillation parameters and other neutrino parameters such the absolute neutrino mass scale or the Majorana phases, should they be physical.

\subsection{Cosmology}
\subsubsection{Large Scale Structure}
Cosmological measurements of large scale structure of matter combined with information about the standard model of cosmology from the cosmic microwave background provide the most promising means of probing the absolute neutrino mass scale.
These data sets provide constraints on the sum of the neutrino masses: $m_1+m_2+m_3$ and generally tend to constrain it to be small and consistent with zero: $\sum_{i=1}^3m_i\lesssim0.1$ eV, depending on the data sets used \cite{Planck:2018vyg,DESI:2024mwx}.
This should be compared to the lower limits for the normal and inverted orderings of 0.06 eV and 0.1 eV.
In fact, the data is actually pushing this number well below that allowed by even the normal ordering and actually somewhat in a region of parameter space where the effect on cosmology goes in the opposite direction expected from neutrino masses \cite{Craig:2024tky,Jiang:2024viw,Loverde:2024nfi} with a tension between cosmology and the terrestrial neutrino oscillation experiments at the $\sim$2-3 $\sigma$ level.
The masses of neutrinos are measured by determining the amount of power suppression at the redshift when neutrinos from the cosmic neutrino background become non-relativistic\footnote{At least two neutrino species are non-relativistic today; the lightest state may still be relativistic if it is light enough.}.

\subsubsection{Cosmic Neutrino Background}
The cosmic neutrino background (C$\nu$B), like the cosmic microwave background, is a relic from the hot, dense big bang and provides key information about the history of our universe.
If the mass ordering is normal then the heavier neutrinos are less electron neutrinos than if the mass ordering is inverted.
For this reason, a detection of the C$\nu$B depends on the mass ordering.
It also depends on whether neutrinos are Majorana or Dirac as if they are Dirac the non-relativistic left-handed neutrinos will now be half in right-helical states reducing the event rate relative to the Majorana case \cite{Long:2014zva}.

Detecting the C$\nu$B is challenging.
The primary effort is known as PTOLEMY \cite{PTOLEMY:2018jst,PTOLEMY:2019hkd} which uses neutrino capture on a very large amount of tritium but does, however, face some non-trivial theoretical issues with the physical possibility of achieving the necessary energy resolution \cite{Cheipesh:2021fmg,PTOLEMY:2022ldz}.
In any case, there are additional uncertainties in the local fluctuations in the density of the C$\nu$B relative to the average value across the universe due to gravitational clustering, and other effects \cite{Bauer:2022lri}.

\subsection{Neutrinoless Double Beta Decay}
Neutrinoless double beta decay (0$\nu\beta\beta$) is an unobserved physical process that would imply that lepton number is violated \cite{Furry:1939qr,Schechter:1981bd}.
If this rate is measured, it would be possible to extract a quantity
\begin{equation}
|m_{\beta\beta}|=\left|c_{12}^2c_{13}^2e^{i\alpha}m_1+s_{12}^2c_{13}^2e^{i\beta}m_2+s_{13}^2m_3\right|\,,
\end{equation}
where $\alpha$ and $\beta$ are the Majorana phases.
Such a measurement, when combined with oscillation data including the mass ordering, would provide one piece of information about a combination of the absolute mass scale (e.g.~$m_1$), $\alpha$, and $\beta$.
With an independent measurement of the mass scale from beta decay end points or cosmology (see below), this then maps on to one piece of information about $\alpha$ and $\beta$.
Determining each Majorana phase independently is not likely possible.

As in the inverted ordering there are two heavier states in the lightest scenario, compared with one heavier state in the normal ordering, the 0$\nu\beta\beta$ rate is generally enhanced in the inverted ordering.
Experiments are beginning to push into the inverted ordering parameter space lead by KamLAND-Zen \cite{KamLAND-Zen:2022tow} and aim to cover the inverted ordering in the coming years.
A non-detection at that point, if combined with a determination that the mass ordering was inverted, would imply that either neutrinos are not Majorana or that a more complicated scenario were at play.

\subsection{Beta Decay End Point}
A straightforward and relatively model independent approach of measuring the absolute neutrino mass scale is via precisely measuring the end point of the beta decay spectrum and identifying deviations in the spectrum due to the finite mass of neutrinos.
These are sensitive to the quantity
\begin{equation}
m_{\nu_\beta}=\sqrt{c_{12}^2c_{13}^2m_1^2+s_{12}^2c_{13}^2m_2^2+s_{13}^2m_3^2}\,.
\end{equation}
The best constraint comes from KATRIN using tritium and a massive spectrometer at $m_{\nu_\beta}<0.45$ eV at 90\% CL \cite{Katrin:2024cdt}.
They have a target of $m_{\nu_\beta}\sim0.2$ eV \cite{KATRIN:2021dfa,KATRIN:2001ttj}.
Project 8 uses a different technique that also uses tritium and measures the cyclotron radiation from the electron and aims to achieve a sensitivity of 0.04 eV \cite{Project8:2017nal}.
ECHo uses $^{163}$Ho and aims to achieve sub eV precision of the neutrino mass scale \cite{Gastaldo:2013wha} via electron capture.

\subsection{Supernova}
A future detection of neutrinos from a nearby supernova (SN) would provide a wealth of information about many areas of physics including neutrino physics.
As these neutrinos are not too high of energy with a spectrum peaking at $\sim$10-20 MeV and they travel considerable distances $\sim$10 kpc, the lower energy neutrinos of the heavier mass states will arrive somewhat later than the lighter mass states \cite{Zatsepin:1968kt}, see also \cite{Loredo:2001rx,Nardi:2003pr,Nardi:2004zg,Pagliaroli:2010ik,Lu:2014zma,Hyper-Kamiokande:2018ofw,Hansen:2019giq,Pompa:2022cxc,Pitik:2022jjh,Brdar:2022vfr,Denton:2024mlb}.
This effect requires a sharp feature in the signal such as the neutronization burst which is likely to be present in the SN \cite{Mirizzi:2015eza}, the chance that the SN forms a black hole \cite{Sekiguchi:2010ja,Gullin:2021hfv} which happens for some subset of SN explosions, and lastly a QCD phase transition in the SN \cite{Sagert:2008ka,Fischer:2010wp,Fischer:2017lag,Zha:2021fbi,Fischer:2021tvv,Kuroda:2021eiv,Bauswein:2022vtq,Lin:2022lck,Pitik:2022jjh} which may exist.

The constraints derived from SN1987A is that $m_\nu\lesssim$6-9 eV \cite{Loredo:2001rx}.
In the future, the precision should get down to the sub eV level, and possibly even to the 0.1 eV level, depending on the distance to the SN, the features in the SN, and what detectors among SK, JUNO, HK, DUNE, and IceCube are online when the event is detected.

\section{Summary}
\label{sec:summary}
The standard three flavor oscillation model provides a compelling and consistent effective model that describes the relevant data.
It is also the only evidence for physics beyond the Standard Model that we have within the realm of particle physics.
First, it is of the utmost importance to determine the general nature of neutrino masses and mixings.
Second, the parameters should be determined precisely.
Finally, we should test to understand if there is more new physics associated with neutrino oscillations.

The first of these goals has been largely achieved and the field has made some progress on the other two, with more anticipated improvements on the horizon.
Most of the new parameters are approximately known and some are even known to good precision such as $\theta_{13}$.
The complex CP violating phase $\delta$ and the absolute neutrino mass scale remain unclear, but some preliminary hints exist and more data is en route.
Other parameters such as the octant of $\theta_{23}$ and the atmospheric mass ordering, i.e.~the sign of $\Delta m^2_{31}$, also remain undetermined and are likely to be measured in the future.
Significant improvements in the precision of many parameters are expected in the future as well, notably $\Delta m^2_{21}$, $|\Delta m^2_{31}|$, and $\theta_{12}$.

Determining each parameter and achieving high precision requires a detailed understanding of the complicated interplay among all the oscillation parameters along with the matter effect and each kind of experiment that measures them.
It is hoped that the precision measurements from upcoming neutrino oscillation experiments will shed light on the flavor puzzle.

\let\oldaddcontentsline\addcontentsline
\renewcommand{\addcontentsline}[3]{}
\section*{Acknowledgements}
We thank Stephen Parke for helpful comments.
This work is supported by the US Department of Energy under Grant Contract DE-SC0012704. 

\bibliographystyle{Numbered-Style}
\bibliography{Encyclopedia_Nu_Osc}

\begin{thebibliography*}{100}
\providecommand{\bibtype}[1]{}
\providecommand{\url}[1]{{\tt #1}}
\providecommand{\urlprefix}{URL }
\expandafter\ifx\csname urlstyle\endcsname\relax
  \providecommand{\doi}[1]{doi:\discretionary{}{}{}#1}\else
  \providecommand{\doi}{doi:\discretionary{}{}{}\begingroup
  \urlstyle{rm}\Url}\fi
\providecommand{\bibinfo}[2]{#2}
\providecommand{\eprint}[2][]{\url{#2}}
\makeatletter\def\@biblabel#1{\bibinfo{label}{[#1]}}\makeatother

\bibtype{Article}%
\bibitem{Pontecorvo:1957cp}
\bibinfo{author}{B. Pontecorvo}, \bibinfo{title}{{Mesonium and anti-mesonium}},
  \bibinfo{journal}{Sov. Phys. JETP} \bibinfo{volume}{6} (\bibinfo{year}{1957})
  \bibinfo{pages}{429}.

\bibtype{Article}%
\bibitem{Maki:1962mu}
\bibinfo{author}{Ziro Maki}, \bibinfo{author}{Masami Nakagawa},
  \bibinfo{author}{Shoichi Sakata}, \bibinfo{title}{{Remarks on the unified
  model of elementary particles}}, \bibinfo{journal}{Prog. Theor. Phys.}
  \bibinfo{volume}{28} (\bibinfo{year}{1962}) \bibinfo{pages}{870--880},
  \bibinfo{doi}{\doi{10.1143/PTP.28.870}}.

\bibtype{Article}%
\bibitem{Davis:1968cp}
\bibinfo{author}{Raymond Davis, Jr.}, \bibinfo{author}{Don~S. Harmer},
  \bibinfo{author}{Kenneth~C. Hoffman}, \bibinfo{title}{{Search for neutrinos
  from the sun}}, \bibinfo{journal}{Phys. Rev. Lett.} \bibinfo{volume}{20}
  (\bibinfo{year}{1968}) \bibinfo{pages}{1205--1209},
  \bibinfo{doi}{\doi{10.1103/PhysRevLett.20.1205}}.

\bibtype{Article}%
\bibitem{Cleveland:1998nv}
\bibinfo{author}{B.~T. Cleveland}, \bibinfo{author}{Timothy Daily},
  \bibinfo{author}{Raymond Davis, Jr.}, \bibinfo{author}{James~R. Distel},
  \bibinfo{author}{Kenneth Lande}, \bibinfo{author}{C.~K. Lee},
  \bibinfo{author}{Paul~S. Wildenhain}, \bibinfo{author}{Jack Ullman},
  \bibinfo{title}{{Measurement of the solar electron neutrino flux with the
  Homestake chlorine detector}}, \bibinfo{journal}{Astrophys. J.}
  \bibinfo{volume}{496} (\bibinfo{year}{1998}) \bibinfo{pages}{505--526},
  \bibinfo{doi}{\doi{10.1086/305343}}.

\bibtype{Article}%
\bibitem{Bahcall:1968hc}
\bibinfo{author}{John~N. Bahcall}, \bibinfo{author}{Neta~A. Bahcall},
  \bibinfo{author}{G. Shaviv}, \bibinfo{title}{{Present status of the
  theoretical predictions for the Cl-36 solar neutrino experiment}},
  \bibinfo{journal}{Phys. Rev. Lett.} \bibinfo{volume}{20}
  (\bibinfo{year}{1968}) \bibinfo{pages}{1209--1212},
  \bibinfo{doi}{\doi{10.1103/PhysRevLett.20.1209}}.

\bibtype{Article}%
\bibitem{Bahcall:2000nu}
\bibinfo{author}{John~N. Bahcall}, \bibinfo{author}{M.~H. Pinsonneault},
  \bibinfo{author}{Sarbani Basu}, \bibinfo{title}{{Solar models: Current epoch
  and time dependences, neutrinos, and helioseismological properties}},
  \bibinfo{journal}{Astrophys. J.} \bibinfo{volume}{555} (\bibinfo{year}{2001})
  \bibinfo{pages}{990--1012}, \bibinfo{doi}{\doi{10.1086/321493}},
  \eprint{astro-ph/0010346}.

\bibtype{Article}%
\bibitem{Bahcall:1968xb}
\bibinfo{author}{John~N. Bahcall}, \bibinfo{author}{Robert~M. May},
  \bibinfo{title}{{THE RATE OF THE PROTON-PROTON REACTION}},
  \bibinfo{journal}{Astrophys. J. Lett.} \bibinfo{volume}{152}
  (\bibinfo{year}{1968}) \bibinfo{pages}{L17--L20},
  \bibinfo{doi}{\doi{10.1086/180169}}.

\bibtype{Article}%
\bibitem{GALLEX:1998kcz}
\bibinfo{author}{W. Hampel}, et al. (\bibinfo{collaboration}{GALLEX}),
  \bibinfo{title}{{GALLEX solar neutrino observations: Results for GALLEX IV}},
  \bibinfo{journal}{Phys. Lett. B} \bibinfo{volume}{447} (\bibinfo{year}{1999})
  \bibinfo{pages}{127--133},
  \bibinfo{doi}{\doi{10.1016/S0370-2693(98)01579-2}}.

\bibtype{Article}%
\bibitem{SAGE:1999nng}
\bibinfo{author}{J.~N. Abdurashitov}, et al. (\bibinfo{collaboration}{SAGE}),
  \bibinfo{title}{{Measurement of the solar neutrino capture rate with gallium
  metal}}, \bibinfo{journal}{Phys. Rev. C} \bibinfo{volume}{60}
  (\bibinfo{year}{1999}) \bibinfo{pages}{055801},
  \bibinfo{doi}{\doi{10.1103/PhysRevC.60.055801}}, \eprint{astro-ph/9907113}.

\bibtype{Article}%
\bibitem{Gavrin:2001sz}
\bibinfo{author}{V.~N. Gavrin} (\bibinfo{collaboration}{SAGE}),
  \bibinfo{title}{{Solar neutrino results from SAGE}}, \bibinfo{journal}{Nucl.
  Phys. B Proc. Suppl.} \bibinfo{volume}{91} (\bibinfo{year}{2001})
  \bibinfo{pages}{36--43}, \bibinfo{doi}{\doi{10.1016/S0920-5632(00)00920-8}}.

\bibtype{Article}%
\bibitem{GNO:2000avz}
\bibinfo{author}{M. Altmann}, et al. (\bibinfo{collaboration}{GNO}),
  \bibinfo{title}{{GNO solar neutrino observations: Results for GNO I}},
  \bibinfo{journal}{Phys. Lett. B} \bibinfo{volume}{490} (\bibinfo{year}{2000})
  \bibinfo{pages}{16--26}, \bibinfo{doi}{\doi{10.1016/S0370-2693(00)00915-1}},
  \eprint{hep-ex/0006034}.

\bibtype{Article}%
\bibitem{Super-Kamiokande:1998kpq}
\bibinfo{author}{Y. Fukuda}, et al.
  (\bibinfo{collaboration}{Super-Kamiokande}), \bibinfo{title}{{Evidence for
  oscillation of atmospheric neutrinos}}, \bibinfo{journal}{Phys. Rev. Lett.}
  \bibinfo{volume}{81} (\bibinfo{year}{1998}) \bibinfo{pages}{1562--1567},
  \bibinfo{doi}{\doi{10.1103/PhysRevLett.81.1562}}, \eprint{hep-ex/9807003}.

\bibtype{Article}%
\bibitem{SNO:2001kpb}
\bibinfo{author}{Q.~R. Ahmad}, et al. (\bibinfo{collaboration}{SNO}),
  \bibinfo{title}{{Measurement of the rate of $\nu_e+d \to p+p+e^-$
  interactions produced by $^8$B solar neutrinos at the Sudbury Neutrino
  Observatory}}, \bibinfo{journal}{Phys. Rev. Lett.} \bibinfo{volume}{87}
  (\bibinfo{year}{2001}) \bibinfo{pages}{071301},
  \bibinfo{doi}{\doi{10.1103/PhysRevLett.87.071301}}, \eprint{nucl-ex/0106015}.

\bibtype{Article}%
\bibitem{SNO:2002tuh}
\bibinfo{author}{Q.~R. Ahmad}, et al. (\bibinfo{collaboration}{SNO}),
  \bibinfo{title}{{Direct evidence for neutrino flavor transformation from
  neutral current interactions in the Sudbury Neutrino Observatory}},
  \bibinfo{journal}{Phys. Rev. Lett.} \bibinfo{volume}{89}
  (\bibinfo{year}{2002}) \bibinfo{pages}{011301},
  \bibinfo{doi}{\doi{10.1103/PhysRevLett.89.011301}}, \eprint{nucl-ex/0204008}.

\bibtype{Article}%
\bibitem{Wolfenstein:1977ue}
\bibinfo{author}{L. Wolfenstein}, \bibinfo{title}{{Neutrino Oscillations in
  Matter}}, \bibinfo{journal}{Phys. Rev. D} \bibinfo{volume}{17}
  (\bibinfo{year}{1978}) \bibinfo{pages}{2369--2374},
  \bibinfo{doi}{\doi{10.1103/PhysRevD.17.2369}}.

\bibtype{Article}%
\bibitem{Mikheyev:1985zog}
\bibinfo{author}{S.~P. Mikheyev}, \bibinfo{author}{A.~Yu. Smirnov},
  \bibinfo{title}{{Resonance Amplification of Oscillations in Matter and
  Spectroscopy of Solar Neutrinos}}, \bibinfo{journal}{Sov. J. Nucl. Phys.}
  \bibinfo{volume}{42} (\bibinfo{year}{1985}) \bibinfo{pages}{913--917}.

\bibtype{Article}%
\bibitem{Higgs:1964pj}
\bibinfo{author}{Peter~W. Higgs}, \bibinfo{title}{{Broken Symmetries and the
  Masses of Gauge Bosons}}, \bibinfo{journal}{Phys. Rev. Lett.}
  \bibinfo{volume}{13} (\bibinfo{year}{1964}) \bibinfo{pages}{508--509},
  \bibinfo{doi}{\doi{10.1103/PhysRevLett.13.508}}.

\bibtype{Article}%
\bibitem{Englert:1964et}
\bibinfo{author}{F. Englert}, \bibinfo{author}{R. Brout},
  \bibinfo{title}{{Broken Symmetry and the Mass of Gauge Vector Mesons}},
  \bibinfo{journal}{Phys. Rev. Lett.} \bibinfo{volume}{13}
  (\bibinfo{year}{1964}) \bibinfo{pages}{321--323},
  \bibinfo{doi}{\doi{10.1103/PhysRevLett.13.321}}.

\bibtype{Article}%
\bibitem{ATLAS:2012yve}
\bibinfo{author}{Georges Aad}, et al. (\bibinfo{collaboration}{ATLAS}),
  \bibinfo{title}{{Observation of a new particle in the search for the Standard
  Model Higgs boson with the ATLAS detector at the LHC}},
  \bibinfo{journal}{Phys. Lett. B} \bibinfo{volume}{716} (\bibinfo{year}{2012})
  \bibinfo{pages}{1--29}, \bibinfo{doi}{\doi{10.1016/j.physletb.2012.08.020}},
  \eprint{1207.7214}.

\bibtype{Article}%
\bibitem{CMS:2012qbp}
\bibinfo{author}{Serguei Chatrchyan}, et al. (\bibinfo{collaboration}{CMS}),
  \bibinfo{title}{{Observation of a New Boson at a Mass of 125 GeV with the CMS
  Experiment at the LHC}}, \bibinfo{journal}{Phys. Lett. B}
  \bibinfo{volume}{716} (\bibinfo{year}{2012}) \bibinfo{pages}{30--61},
  \bibinfo{doi}{\doi{10.1016/j.physletb.2012.08.021}}, \eprint{1207.7235}.

\bibtype{Article}%
\bibitem{Denton:2022een}
\bibinfo{author}{Peter~B. Denton}, \bibinfo{author}{Megan Friend},
  \bibinfo{author}{Mark~D. Messier}, \bibinfo{author}{Hirohisa~A. Tanaka},
  \bibinfo{author}{Sebastian B\"oser}, \bibinfo{author}{Jo\~ao A.~B. Coelho},
  \bibinfo{author}{Mathieu Perrin-Terrin}, \bibinfo{author}{Tom Stuttard},
  \bibinfo{title}{{Snowmass Neutrino Frontier: NF01 Topical Group Report on
  Three-Flavor Neutrino Oscillations}}  (\bibinfo{year}{2022}),
  \eprint{2212.00809}.

\bibtype{Article}%
\bibitem{Weinberg:1967tq}
\bibinfo{author}{Steven Weinberg}, \bibinfo{title}{{A Model of Leptons}},
  \bibinfo{journal}{Phys. Rev. Lett.} \bibinfo{volume}{19}
  (\bibinfo{year}{1967}) \bibinfo{pages}{1264--1266},
  \bibinfo{doi}{\doi{10.1103/PhysRevLett.19.1264}}.

\bibtype{Misc}%
\bibitem{NYT_Weinberg}
\bibinfo{author}{Dylan~Loeb McClain}, \bibinfo{title}{{Steven Weinberg,
  Groundbreaking Nobelist in Physics, Dies at 88}} \bibinfo{year}{2021},
  \bibinfo{url}{\urlprefix\url{https://www.nytimes.com/2021/07/25/science/steven-weinberg-groundbreaking-nobelist-in-physics-dies-at-88.html}}.

\bibtype{Article}%
\bibitem{Majorana:1932ga}
\bibinfo{author}{Ettore Majorana}, \bibinfo{title}{{Oriented atoms in a
  variable magnetic field}}, \bibinfo{journal}{Nuovo Cim.} \bibinfo{volume}{9}
  (\bibinfo{year}{1932}) \bibinfo{pages}{43--50},
  \bibinfo{doi}{\doi{10.1007/BF02960953}}.

\bibtype{Article}%
\bibitem{Kayser:1982br}
\bibinfo{author}{Boris Kayser}, \bibinfo{title}{{Majorana Neutrinos and their
  Electromagnetic Properties}}, \bibinfo{journal}{Phys. Rev. D}
  \bibinfo{volume}{26} (\bibinfo{year}{1982}) \bibinfo{pages}{1662},
  \bibinfo{doi}{\doi{10.1103/PhysRevD.26.1662}}.

\bibtype{Article}%
\bibitem{Shrock:1982sc}
\bibinfo{author}{Robert~E. Shrock}, \bibinfo{title}{{Electromagnetic Properties
  and Decays of Dirac and Majorana Neutrinos in a General Class of Gauge
  Theories}}, \bibinfo{journal}{Nucl. Phys. B} \bibinfo{volume}{206}
  (\bibinfo{year}{1982}) \bibinfo{pages}{359--379},
  \bibinfo{doi}{\doi{10.1016/0550-3213(82)90273-5}}.

\bibtype{Article}%
\bibitem{Minkowski:1977sc}
\bibinfo{author}{Peter Minkowski}, \bibinfo{title}{{$\mu \to e\gamma$ at a Rate
  of One Out of $10^{9}$ Muon Decays?}}, \bibinfo{journal}{Phys. Lett. B}
  \bibinfo{volume}{67} (\bibinfo{year}{1977}) \bibinfo{pages}{421--428},
  \bibinfo{doi}{\doi{10.1016/0370-2693(77)90435-X}}.

\bibtype{Article}%
\bibitem{Yanagida:1979as}
\bibinfo{author}{Tsutomu Yanagida}, \bibinfo{title}{{Horizontal gauge symmetry
  and masses of neutrinos}}, \bibinfo{journal}{Conf. Proc. C}
  \bibinfo{volume}{7902131} (\bibinfo{year}{1979}) \bibinfo{pages}{95--99}.

\bibtype{Article}%
\bibitem{Gell-Mann:1979vob}
\bibinfo{author}{Murray Gell-Mann}, \bibinfo{author}{Pierre Ramond},
  \bibinfo{author}{Richard Slansky}, \bibinfo{title}{{Complex Spinors and
  Unified Theories}}, \bibinfo{journal}{Conf. Proc. C} \bibinfo{volume}{790927}
  (\bibinfo{year}{1979}) \bibinfo{pages}{315--321}, \eprint{1306.4669}.

\bibtype{Article}%
\bibitem{Yanagida:1980xy}
\bibinfo{author}{Tsutomu Yanagida}, \bibinfo{title}{{Horizontal Symmetry and
  Masses of Neutrinos}}, \bibinfo{journal}{Prog. Theor. Phys.}
  \bibinfo{volume}{64} (\bibinfo{year}{1980}) \bibinfo{pages}{1103},
  \bibinfo{doi}{\doi{10.1143/PTP.64.1103}}.

\bibtype{Article}%
\bibitem{Mohapatra:1979ia}
\bibinfo{author}{Rabindra~N. Mohapatra}, \bibinfo{author}{Goran Senjanovic},
  \bibinfo{title}{{Neutrino Mass and Spontaneous Parity Nonconservation}},
  \bibinfo{journal}{Phys. Rev. Lett.} \bibinfo{volume}{44}
  (\bibinfo{year}{1980}) \bibinfo{pages}{912},
  \bibinfo{doi}{\doi{10.1103/PhysRevLett.44.912}}.

\bibtype{Article}%
\bibitem{Arguelles:2019xgp}
\bibinfo{author}{C.~A. Arg\"uelles}, et al., \bibinfo{title}{{New opportunities
  at the next-generation neutrino experiments I: BSM neutrino physics and dark
  matter}}, \bibinfo{journal}{Rept. Prog. Phys.} \bibinfo{volume}{83}
  (\bibinfo{number}{12}) (\bibinfo{year}{2020}) \bibinfo{pages}{124201},
  \bibinfo{doi}{\doi{10.1088/1361-6633/ab9d12}}, \eprint{1907.08311}.

\bibtype{Article}%
\bibitem{Denton:2020igp}
\bibinfo{author}{Peter~B. Denton}, \bibinfo{author}{Rebekah Pestes},
  \bibinfo{title}{{The impact of different parameterizations on the
  interpretation of CP violation in neutrino oscillations}},
  \bibinfo{journal}{JHEP} \bibinfo{volume}{05} (\bibinfo{year}{2021})
  \bibinfo{pages}{139}, \bibinfo{doi}{\doi{10.1007/JHEP05(2021)139}},
  \eprint{2006.09384}.

\bibtype{Article}%
\bibitem{Jarlskog:1985ht}
\bibinfo{author}{C. Jarlskog}, \bibinfo{title}{{Commutator of the Quark Mass
  Matrices in the Standard Electroweak Model and a Measure of Maximal CP
  Nonconservation}}, \bibinfo{journal}{Phys. Rev. Lett.} \bibinfo{volume}{55}
  (\bibinfo{year}{1985}) \bibinfo{pages}{1039},
  \bibinfo{doi}{\doi{10.1103/PhysRevLett.55.1039}}.

\bibtype{Article}%
\bibitem{Krastev:1988yu}
\bibinfo{author}{P.~I. Krastev}, \bibinfo{author}{S.~T. Petcov},
  \bibinfo{title}{{Resonance Amplification and t Violation Effects in Three
  Neutrino Oscillations in the Earth}}, \bibinfo{journal}{Phys. Lett. B}
  \bibinfo{volume}{205} (\bibinfo{year}{1988}) \bibinfo{pages}{84--92},
  \bibinfo{doi}{\doi{10.1016/0370-2693(88)90404-2}}.

\bibtype{Article}%
\bibitem{Super-Kamiokande:2004orf}
\bibinfo{author}{Y. Ashie}, et al. (\bibinfo{collaboration}{Super-Kamiokande}),
  \bibinfo{title}{{Evidence for an oscillatory signature in atmospheric
  neutrino oscillation}}, \bibinfo{journal}{Phys. Rev. Lett.}
  \bibinfo{volume}{93} (\bibinfo{year}{2004}) \bibinfo{pages}{101801},
  \bibinfo{doi}{\doi{10.1103/PhysRevLett.93.101801}}, \eprint{hep-ex/0404034}.

\bibtype{Article}%
\bibitem{Super-Kamiokande:2005mbp}
\bibinfo{author}{Y. Ashie}, et al. (\bibinfo{collaboration}{Super-Kamiokande}),
  \bibinfo{title}{{A Measurement of atmospheric neutrino oscillation parameters
  by SUPER-KAMIOKANDE I}}, \bibinfo{journal}{Phys. Rev. D} \bibinfo{volume}{71}
  (\bibinfo{year}{2005}) \bibinfo{pages}{112005},
  \bibinfo{doi}{\doi{10.1103/PhysRevD.71.112005}}, \eprint{hep-ex/0501064}.

\bibtype{Article}%
\bibitem{Super-Kamiokande:2006jvq}
\bibinfo{author}{J. Hosaka}, et al.
  (\bibinfo{collaboration}{Super-Kamiokande}), \bibinfo{title}{{Three flavor
  neutrino oscillation analysis of atmospheric neutrinos in Super-Kamiokande}},
  \bibinfo{journal}{Phys. Rev. D} \bibinfo{volume}{74} (\bibinfo{year}{2006})
  \bibinfo{pages}{032002}, \bibinfo{doi}{\doi{10.1103/PhysRevD.74.032002}},
  \eprint{hep-ex/0604011}.

\bibtype{Article}%
\bibitem{Super-Kamiokande:2010orq}
\bibinfo{author}{R. Wendell}, et al.
  (\bibinfo{collaboration}{Super-Kamiokande}), \bibinfo{title}{{Atmospheric
  neutrino oscillation analysis with sub-leading effects in Super-Kamiokande I,
  II, and III}}, \bibinfo{journal}{Phys. Rev. D} \bibinfo{volume}{81}
  (\bibinfo{year}{2010}) \bibinfo{pages}{092004},
  \bibinfo{doi}{\doi{10.1103/PhysRevD.81.092004}}, \eprint{1002.3471}.

\bibtype{Article}%
\bibitem{Super-Kamiokande:2017yvm}
\bibinfo{author}{K. Abe}, et al. (\bibinfo{collaboration}{Super-Kamiokande}),
  \bibinfo{title}{{Atmospheric neutrino oscillation analysis with external
  constraints in Super-Kamiokande I-IV}}, \bibinfo{journal}{Phys. Rev. D}
  \bibinfo{volume}{97} (\bibinfo{number}{7}) (\bibinfo{year}{2018})
  \bibinfo{pages}{072001}, \bibinfo{doi}{\doi{10.1103/PhysRevD.97.072001}},
  \eprint{1710.09126}.

\bibtype{Article}%
\bibitem{Super-Kamiokande:2023ahc}
\bibinfo{author}{T. Wester}, et al.
  (\bibinfo{collaboration}{Super-Kamiokande}), \bibinfo{title}{{Atmospheric
  neutrino oscillation analysis with neutron tagging and an expanded fiducial
  volume in Super-Kamiokande I\textendash{}V}}, \bibinfo{journal}{Phys. Rev. D}
  \bibinfo{volume}{109} (\bibinfo{number}{7}) (\bibinfo{year}{2024})
  \bibinfo{pages}{072014}, \bibinfo{doi}{\doi{10.1103/PhysRevD.109.072014}},
  \eprint{2311.05105}.

\bibtype{Article}%
\bibitem{IceCube:2013pav}
\bibinfo{author}{M.~G. Aartsen}, et al. (\bibinfo{collaboration}{IceCube}),
  \bibinfo{title}{{Measurement of Atmospheric Neutrino Oscillations with
  IceCube}}, \bibinfo{journal}{Phys. Rev. Lett.} \bibinfo{volume}{111}
  (\bibinfo{number}{8}) (\bibinfo{year}{2013}) \bibinfo{pages}{081801},
  \bibinfo{doi}{\doi{10.1103/PhysRevLett.111.081801}}, \eprint{1305.3909}.

\bibtype{Article}%
\bibitem{IceCube:2014flw}
\bibinfo{author}{M.~G. Aartsen}, et al. (\bibinfo{collaboration}{IceCube}),
  \bibinfo{title}{{Determining neutrino oscillation parameters from atmospheric
  muon neutrino disappearance with three years of IceCube DeepCore data}},
  \bibinfo{journal}{Phys. Rev. D} \bibinfo{volume}{91} (\bibinfo{number}{7})
  (\bibinfo{year}{2015}) \bibinfo{pages}{072004},
  \bibinfo{doi}{\doi{10.1103/PhysRevD.91.072004}}, \eprint{1410.7227}.

\bibtype{Article}%
\bibitem{IceCube:2017lak}
\bibinfo{author}{M.~G. Aartsen}, et al. (\bibinfo{collaboration}{IceCube}),
  \bibinfo{title}{{Measurement of Atmospheric Neutrino Oscillations at
  6\textendash{}56 GeV with IceCube DeepCore}}, \bibinfo{journal}{Phys. Rev.
  Lett.} \bibinfo{volume}{120} (\bibinfo{number}{7}) (\bibinfo{year}{2018})
  \bibinfo{pages}{071801}, \bibinfo{doi}{\doi{10.1103/PhysRevLett.120.071801}},
  \eprint{1707.07081}.

\bibtype{Article}%
\bibitem{IceCubeCollaboration:2023wtb}
\bibinfo{author}{R. Abbasi}, et al. (\bibinfo{collaboration}{(IceCube
  Collaboration)*, IceCube}), \bibinfo{title}{{Measurement of atmospheric
  neutrino mixing with improved IceCube DeepCore calibration and data
  processing}}, \bibinfo{journal}{Phys. Rev. D} \bibinfo{volume}{108}
  (\bibinfo{number}{1}) (\bibinfo{year}{2023}) \bibinfo{pages}{012014},
  \bibinfo{doi}{\doi{10.1103/PhysRevD.108.012014}}, \eprint{2304.12236}.

\bibtype{Article}%
\bibitem{IceCubeCollaboration:2024ssx}
\bibinfo{author}{R. Abbasi}, et al. (\bibinfo{collaboration}{(IceCube
  Collaboration){\ensuremath{\parallel}}, IceCube}),
  \bibinfo{title}{{Measurement of Atmospheric Neutrino Oscillation Parameters
  Using Convolutional Neural Networks with 9.3 Years of Data in IceCube
  DeepCore}}, \bibinfo{journal}{Phys. Rev. Lett.} \bibinfo{volume}{134}
  (\bibinfo{number}{9}) (\bibinfo{year}{2025}) \bibinfo{pages}{091801},
  \bibinfo{doi}{\doi{10.1103/PhysRevLett.134.091801}}, \eprint{2405.02163}.

\bibtype{Article}%
\bibitem{Nunokawa:2005nx}
\bibinfo{author}{Hiroshi Nunokawa}, \bibinfo{author}{Stephen~J. Parke},
  \bibinfo{author}{Renata Zukanovich~Funchal}, \bibinfo{title}{{Another
  possible way to determine the neutrino mass hierarchy}},
  \bibinfo{journal}{Phys. Rev. D} \bibinfo{volume}{72} (\bibinfo{year}{2005})
  \bibinfo{pages}{013009}, \bibinfo{doi}{\doi{10.1103/PhysRevD.72.013009}},
  \eprint{hep-ph/0503283}.

\bibtype{Article}%
\bibitem{Parke:2024xre}
\bibinfo{author}{Stephen~J. Parke}, \bibinfo{author}{Renata
  Zukanovich-Funchal}, \bibinfo{title}{{Mass ordering sum rule for the neutrino
  disappearance channels in T2K, NOvA, and JUNO}}, \bibinfo{journal}{Phys. Rev.
  D} \bibinfo{volume}{111} (\bibinfo{number}{1}) (\bibinfo{year}{2025})
  \bibinfo{pages}{013008}, \bibinfo{doi}{\doi{10.1103/PhysRevD.111.013008}},
  \eprint{2404.08733}.

\bibtype{Article}%
\bibitem{Denton:2024thm}
\bibinfo{author}{Peter~B. Denton}, \bibinfo{author}{Stephen~J. Parke},
  \bibinfo{title}{{Smallness of matter effects in long-baseline muon neutrino
  disappearance}}, \bibinfo{journal}{Phys. Rev. D} \bibinfo{volume}{109}
  (\bibinfo{number}{5}) (\bibinfo{year}{2024}) \bibinfo{pages}{053002},
  \bibinfo{doi}{\doi{10.1103/PhysRevD.109.053002}}, \eprint{2401.10326}.

\bibtype{Article}%
\bibitem{Parke:2016joa}
\bibinfo{author}{Stephen Parke}, \bibinfo{title}{{What is $\Delta m^2_{ee}$
  ?}}, \bibinfo{journal}{Phys. Rev. D} \bibinfo{volume}{93}
  (\bibinfo{number}{5}) (\bibinfo{year}{2016}) \bibinfo{pages}{053008},
  \bibinfo{doi}{\doi{10.1103/PhysRevD.93.053008}}, \eprint{1601.07464}.

\bibtype{Article}%
\bibitem{DayaBay:2013yxg}
\bibinfo{author}{F.~P. An}, et al. (\bibinfo{collaboration}{Daya Bay}),
  \bibinfo{title}{{Spectral measurement of electron antineutrino oscillation
  amplitude and frequency at Daya Bay}}, \bibinfo{journal}{Phys. Rev. Lett.}
  \bibinfo{volume}{112} (\bibinfo{year}{2014}) \bibinfo{pages}{061801},
  \bibinfo{doi}{\doi{10.1103/PhysRevLett.112.061801}}, \eprint{1310.6732}.

\bibtype{Article}%
\bibitem{DayaBay:2015ayh}
\bibinfo{author}{F.~P. An}, et al. (\bibinfo{collaboration}{Daya Bay}),
  \bibinfo{title}{{New Measurement of Antineutrino Oscillation with the Full
  Detector Configuration at Daya Bay}}, \bibinfo{journal}{Phys. Rev. Lett.}
  \bibinfo{volume}{115} (\bibinfo{number}{11}) (\bibinfo{year}{2015})
  \bibinfo{pages}{111802}, \bibinfo{doi}{\doi{10.1103/PhysRevLett.115.111802}},
  \eprint{1505.03456}.

\bibtype{Article}%
\bibitem{Li:2016txk}
\bibinfo{author}{Yu-Feng Li}, \bibinfo{author}{Yifang Wang},
  \bibinfo{author}{Zhi-zhong Xing}, \bibinfo{title}{{Terrestrial matter effects
  on reactor antineutrino oscillations at JUNO or RENO-50: how small is
  small?}}, \bibinfo{journal}{Chin. Phys. C} \bibinfo{volume}{40}
  (\bibinfo{number}{9}) (\bibinfo{year}{2016}) \bibinfo{pages}{091001},
  \bibinfo{doi}{\doi{10.1088/1674-1137/40/9/091001}}, \eprint{1605.00900}.

\bibtype{Article}%
\bibitem{Khan:2019doq}
\bibinfo{author}{Amir~N. Khan}, \bibinfo{author}{Hiroshi Nunokawa},
  \bibinfo{author}{Stephen~J Parke}, \bibinfo{title}{{Why matter effects matter
  for JUNO}}, \bibinfo{journal}{Phys. Lett. B} \bibinfo{volume}{803}
  (\bibinfo{year}{2020}) \bibinfo{pages}{135354},
  \bibinfo{doi}{\doi{10.1016/j.physletb.2020.135354}}, \eprint{1910.12900}.

\bibtype{Article}%
\bibitem{K2K:2004iot}
\bibinfo{author}{E. Aliu}, et al. (\bibinfo{collaboration}{K2K}),
  \bibinfo{title}{{Evidence for muon neutrino oscillation in an
  accelerator-based experiment}}, \bibinfo{journal}{Phys. Rev. Lett.}
  \bibinfo{volume}{94} (\bibinfo{year}{2005}) \bibinfo{pages}{081802},
  \bibinfo{doi}{\doi{10.1103/PhysRevLett.94.081802}}, \eprint{hep-ex/0411038}.

\bibtype{Article}%
\bibitem{K2K:2006yov}
\bibinfo{author}{M.~H. Ahn}, et al. (\bibinfo{collaboration}{K2K}),
  \bibinfo{title}{{Measurement of Neutrino Oscillation by the K2K Experiment}},
  \bibinfo{journal}{Phys. Rev. D} \bibinfo{volume}{74} (\bibinfo{year}{2006})
  \bibinfo{pages}{072003}, \bibinfo{doi}{\doi{10.1103/PhysRevD.74.072003}},
  \eprint{hep-ex/0606032}.

\bibtype{Article}%
\bibitem{MINOS:2007ixr}
\bibinfo{author}{P. Adamson}, et al. (\bibinfo{collaboration}{MINOS}),
  \bibinfo{title}{{A Study of Muon Neutrino Disappearance Using the Fermilab
  Main Injector Neutrino Beam}}, \bibinfo{journal}{Phys. Rev. D}
  \bibinfo{volume}{77} (\bibinfo{year}{2008}) \bibinfo{pages}{072002},
  \bibinfo{doi}{\doi{10.1103/PhysRevD.77.072002}}, \eprint{0711.0769}.

\bibtype{Article}%
\bibitem{MINOS:2008kxu}
\bibinfo{author}{P. Adamson}, et al. (\bibinfo{collaboration}{MINOS}),
  \bibinfo{title}{{Measurement of Neutrino Oscillations with the MINOS
  Detectors in the NuMI Beam}}, \bibinfo{journal}{Phys. Rev. Lett.}
  \bibinfo{volume}{101} (\bibinfo{year}{2008}) \bibinfo{pages}{131802},
  \bibinfo{doi}{\doi{10.1103/PhysRevLett.101.131802}}, \eprint{0806.2237}.

\bibtype{Article}%
\bibitem{MINOS:2011neo}
\bibinfo{author}{P. Adamson}, et al. (\bibinfo{collaboration}{MINOS}),
  \bibinfo{title}{{Measurement of the Neutrino Mass Splitting and Flavor Mixing
  by MINOS}}, \bibinfo{journal}{Phys. Rev. Lett.} \bibinfo{volume}{106}
  (\bibinfo{year}{2011}) \bibinfo{pages}{181801},
  \bibinfo{doi}{\doi{10.1103/PhysRevLett.106.181801}}, \eprint{1103.0340}.

\bibtype{Article}%
\bibitem{MINOS:2011qho}
\bibinfo{author}{P. Adamson}, et al. (\bibinfo{collaboration}{MINOS}),
  \bibinfo{title}{{First Direct Observation of Muon Antineutrino
  Disappearance}}, \bibinfo{journal}{Phys. Rev. Lett.} \bibinfo{volume}{107}
  (\bibinfo{year}{2011}) \bibinfo{pages}{021801},
  \bibinfo{doi}{\doi{10.1103/PhysRevLett.107.021801}}, \eprint{1104.0344}.

\bibtype{Article}%
\bibitem{MINOS:2012dbe}
\bibinfo{author}{P. Adamson}, et al. (\bibinfo{collaboration}{MINOS}),
  \bibinfo{title}{{An improved measurement of muon antineutrino disappearance
  in MINOS}}, \bibinfo{journal}{Phys. Rev. Lett.} \bibinfo{volume}{108}
  (\bibinfo{year}{2012}) \bibinfo{pages}{191801},
  \bibinfo{doi}{\doi{10.1103/PhysRevLett.108.191801}}, \eprint{1202.2772}.

\bibtype{Article}%
\bibitem{MINOS:2013utc}
\bibinfo{author}{P. Adamson}, et al. (\bibinfo{collaboration}{MINOS}),
  \bibinfo{title}{{Measurement of Neutrino and Antineutrino Oscillations Using
  Beam and Atmospheric Data in MINOS}}, \bibinfo{journal}{Phys. Rev. Lett.}
  \bibinfo{volume}{110} (\bibinfo{number}{25}) (\bibinfo{year}{2013})
  \bibinfo{pages}{251801}, \bibinfo{doi}{\doi{10.1103/PhysRevLett.110.251801}},
  \eprint{1304.6335}.

\bibtype{Article}%
\bibitem{MINOS:2014rjg}
\bibinfo{author}{P. Adamson}, et al. (\bibinfo{collaboration}{MINOS}),
  \bibinfo{title}{{Combined analysis of $\nu_{\mu}$ disappearance and
  $\nu_{\mu} \rightarrow \nu_{e}$ appearance in MINOS using accelerator and
  atmospheric neutrinos}}, \bibinfo{journal}{Phys. Rev. Lett.}
  \bibinfo{volume}{112} (\bibinfo{year}{2014}) \bibinfo{pages}{191801},
  \bibinfo{doi}{\doi{10.1103/PhysRevLett.112.191801}}, \eprint{1403.0867}.

\bibtype{Article}%
\bibitem{T2K:2012qoq}
\bibinfo{author}{K. Abe}, et al. (\bibinfo{collaboration}{T2K}),
  \bibinfo{title}{{First Muon-Neutrino Disappearance Study with an Off-Axis
  Beam}}, \bibinfo{journal}{Phys. Rev. D} \bibinfo{volume}{85}
  (\bibinfo{year}{2012}) \bibinfo{pages}{031103},
  \bibinfo{doi}{\doi{10.1103/PhysRevD.85.031103}}, \eprint{1201.1386}.

\bibtype{Article}%
\bibitem{T2K:2013bzi}
\bibinfo{author}{K. Abe}, et al. (\bibinfo{collaboration}{T2K}),
  \bibinfo{title}{{Measurement of Neutrino Oscillation Parameters from Muon
  Neutrino Disappearance with an Off-axis Beam}}, \bibinfo{journal}{Phys. Rev.
  Lett.} \bibinfo{volume}{111} (\bibinfo{number}{21}) (\bibinfo{year}{2013})
  \bibinfo{pages}{211803}, \bibinfo{doi}{\doi{10.1103/PhysRevLett.111.211803}},
  \eprint{1308.0465}.

\bibtype{Article}%
\bibitem{T2K:2014ghj}
\bibinfo{author}{K. Abe}, et al. (\bibinfo{collaboration}{T2K}),
  \bibinfo{title}{{Precise Measurement of the Neutrino Mixing Parameter
  $\theta_{23}$ from Muon Neutrino Disappearance in an Off-Axis Beam}},
  \bibinfo{journal}{Phys. Rev. Lett.} \bibinfo{volume}{112}
  (\bibinfo{number}{18}) (\bibinfo{year}{2014}) \bibinfo{pages}{181801},
  \bibinfo{doi}{\doi{10.1103/PhysRevLett.112.181801}}, \eprint{1403.1532}.

\bibtype{Article}%
\bibitem{T2K:2015sqm}
\bibinfo{author}{K. Abe}, et al. (\bibinfo{collaboration}{T2K}),
  \bibinfo{title}{{Measurements of neutrino oscillation in appearance and
  disappearance channels by the T2K experiment with 6.6\texttimes{}10$^{20}$
  protons on target}}, \bibinfo{journal}{Phys. Rev. D} \bibinfo{volume}{91}
  (\bibinfo{number}{7}) (\bibinfo{year}{2015}) \bibinfo{pages}{072010},
  \bibinfo{doi}{\doi{10.1103/PhysRevD.91.072010}}, \eprint{1502.01550}.

\bibtype{Article}%
\bibitem{T2K:2017krm}
\bibinfo{author}{K. Abe}, et al. (\bibinfo{collaboration}{T2K}),
  \bibinfo{title}{{Updated T2K measurements of muon neutrino and antineutrino
  disappearance using 1.5$\times$10$^{21}$ protons on target}},
  \bibinfo{journal}{Phys. Rev. D} \bibinfo{volume}{96} (\bibinfo{number}{1})
  (\bibinfo{year}{2017}) \bibinfo{pages}{011102},
  \bibinfo{doi}{\doi{10.1103/PhysRevD.96.011102}}, \eprint{1704.06409}.

\bibtype{Article}%
\bibitem{T2K:2020nqo}
\bibinfo{author}{K. Abe}, et al. (\bibinfo{collaboration}{T2K}),
  \bibinfo{title}{{T2K measurements of muon neutrino and antineutrino
  disappearance using $3.13\times 10^{21}$ protons on target}},
  \bibinfo{journal}{Phys. Rev. D} \bibinfo{volume}{103} (\bibinfo{number}{1})
  (\bibinfo{year}{2021}) \bibinfo{pages}{L011101},
  \bibinfo{doi}{\doi{10.1103/PhysRevD.103.L011101}}, \eprint{2008.07921}.

\bibtype{Article}%
\bibitem{T2K:2023mcm}
\bibinfo{author}{K. Abe}, et al. (\bibinfo{collaboration}{T2K}),
  \bibinfo{title}{{Updated T2K measurements of muon neutrino and antineutrino
  disappearance using 3.6\texttimes{}1021 protons on target}},
  \bibinfo{journal}{Phys. Rev. D} \bibinfo{volume}{108} (\bibinfo{number}{7})
  (\bibinfo{year}{2023}) \bibinfo{pages}{072011},
  \bibinfo{doi}{\doi{10.1103/PhysRevD.108.072011}}, \eprint{2305.09916}.

\bibtype{Article}%
\bibitem{NOvA:2016vij}
\bibinfo{author}{P. Adamson}, et al. (\bibinfo{collaboration}{NOvA}),
  \bibinfo{title}{{First measurement of muon-neutrino disappearance in NOvA}},
  \bibinfo{journal}{Phys. Rev. D} \bibinfo{volume}{93} (\bibinfo{number}{5})
  (\bibinfo{year}{2016}) \bibinfo{pages}{051104},
  \bibinfo{doi}{\doi{10.1103/PhysRevD.93.051104}}, \eprint{1601.05037}.

\bibtype{Article}%
\bibitem{NOvA:2017abs}
\bibinfo{author}{P. Adamson}, et al. (\bibinfo{collaboration}{NOvA}),
  \bibinfo{title}{{Constraints on Oscillation Parameters from $\nu_e$
  Appearance and $\nu_\mu$ Disappearance in NOvA}}, \bibinfo{journal}{Phys.
  Rev. Lett.} \bibinfo{volume}{118} (\bibinfo{number}{23})
  (\bibinfo{year}{2017}) \bibinfo{pages}{231801},
  \bibinfo{doi}{\doi{10.1103/PhysRevLett.118.231801}}, \eprint{1703.03328}.

\bibtype{Article}%
\bibitem{NOvA:2018gge}
\bibinfo{author}{M.~A. Acero}, et al. (\bibinfo{collaboration}{NOvA}),
  \bibinfo{title}{{New constraints on oscillation parameters from $\nu_e$
  appearance and $\nu_\mu$ disappearance in the NOvA experiment}},
  \bibinfo{journal}{Phys. Rev. D} \bibinfo{volume}{98} (\bibinfo{year}{2018})
  \bibinfo{pages}{032012}, \bibinfo{doi}{\doi{10.1103/PhysRevD.98.032012}},
  \eprint{1806.00096}.

\bibtype{Article}%
\bibitem{NOvA:2019cyt}
\bibinfo{author}{M.~A. Acero}, et al. (\bibinfo{collaboration}{NOvA}),
  \bibinfo{title}{{First Measurement of Neutrino Oscillation Parameters using
  Neutrinos and Antineutrinos by NOvA}}, \bibinfo{journal}{Phys. Rev. Lett.}
  \bibinfo{volume}{123} (\bibinfo{number}{15}) (\bibinfo{year}{2019})
  \bibinfo{pages}{151803}, \bibinfo{doi}{\doi{10.1103/PhysRevLett.123.151803}},
  \eprint{1906.04907}.

\bibtype{Article}%
\bibitem{NOvA:2021nfi}
\bibinfo{author}{M.~A. Acero}, et al. (\bibinfo{collaboration}{NOvA}),
  \bibinfo{title}{{Improved measurement of neutrino oscillation parameters by
  the NOvA experiment}}, \bibinfo{journal}{Phys. Rev. D} \bibinfo{volume}{106}
  (\bibinfo{number}{3}) (\bibinfo{year}{2022}) \bibinfo{pages}{032004},
  \bibinfo{doi}{\doi{10.1103/PhysRevD.106.032004}}, \eprint{2108.08219}.

\bibtype{Article}%
\bibitem{LSND:2001aii}
\bibinfo{author}{A. Aguilar}, et al. (\bibinfo{collaboration}{LSND}),
  \bibinfo{title}{{Evidence for neutrino oscillations from the observation of
  $\bar{\nu}_e$ appearance in a $\bar{\nu}_\mu$ beam}}, \bibinfo{journal}{Phys.
  Rev. D} \bibinfo{volume}{64} (\bibinfo{year}{2001}) \bibinfo{pages}{112007},
  \bibinfo{doi}{\doi{10.1103/PhysRevD.64.112007}}, \eprint{hep-ex/0104049}.

\bibtype{Article}%
\bibitem{MiniBooNE:2020pnu}
\bibinfo{author}{A.~A. Aguilar-Arevalo}, et al.
  (\bibinfo{collaboration}{MiniBooNE}), \bibinfo{title}{{Updated MiniBooNE
  neutrino oscillation results with increased data and new background
  studies}}, \bibinfo{journal}{Phys. Rev. D} \bibinfo{volume}{103}
  (\bibinfo{number}{5}) (\bibinfo{year}{2021}) \bibinfo{pages}{052002},
  \bibinfo{doi}{\doi{10.1103/PhysRevD.103.052002}}, \eprint{2006.16883}.

\bibtype{Article}%
\bibitem{IceCube:2020phf}
\bibinfo{author}{M.~G. Aartsen}, et al. (\bibinfo{collaboration}{IceCube}),
  \bibinfo{title}{{eV-Scale Sterile Neutrino Search Using Eight Years of
  Atmospheric Muon Neutrino Data from the IceCube Neutrino Observatory}},
  \bibinfo{journal}{Phys. Rev. Lett.} \bibinfo{volume}{125}
  (\bibinfo{number}{14}) (\bibinfo{year}{2020}) \bibinfo{pages}{141801},
  \bibinfo{doi}{\doi{10.1103/PhysRevLett.125.141801}}, \eprint{2005.12942}.

\bibtype{Article}%
\bibitem{MINOS:2017cae}
\bibinfo{author}{P. Adamson}, et al. (\bibinfo{collaboration}{MINOS+}),
  \bibinfo{title}{{Search for sterile neutrinos in MINOS and MINOS+ using a
  two-detector fit}}, \bibinfo{journal}{Phys. Rev. Lett.} \bibinfo{volume}{122}
  (\bibinfo{number}{9}) (\bibinfo{year}{2019}) \bibinfo{pages}{091803},
  \bibinfo{doi}{\doi{10.1103/PhysRevLett.122.091803}}, \eprint{1710.06488}.

\bibtype{Article}%
\bibitem{MicroBooNE:2022sdp}
\bibinfo{author}{P. Abratenko}, et al. (\bibinfo{collaboration}{MicroBooNE}),
  \bibinfo{title}{{First Constraints on Light Sterile Neutrino Oscillations
  from Combined Appearance and Disappearance Searches with the MicroBooNE
  Detector}}, \bibinfo{journal}{Phys. Rev. Lett.} \bibinfo{volume}{130}
  (\bibinfo{number}{1}) (\bibinfo{year}{2023}) \bibinfo{pages}{011801},
  \bibinfo{doi}{\doi{10.1103/PhysRevLett.130.011801}}, \eprint{2210.10216}.

\bibtype{Article}%
\bibitem{Dentler:2018sju}
\bibinfo{author}{Mona Dentler}, \bibinfo{author}{\'Alvaro
  Hern\'andez-Cabezudo}, \bibinfo{author}{Joachim Kopp}, \bibinfo{author}{Pedro
  A.~N. Machado}, \bibinfo{author}{Michele Maltoni}, \bibinfo{author}{Ivan
  Martinez-Soler}, \bibinfo{author}{Thomas Schwetz}, \bibinfo{title}{{Updated
  Global Analysis of Neutrino Oscillations in the Presence of eV-Scale Sterile
  Neutrinos}}, \bibinfo{journal}{JHEP} \bibinfo{volume}{08}
  (\bibinfo{year}{2018}) \bibinfo{pages}{010},
  \bibinfo{doi}{\doi{10.1007/JHEP08(2018)010}}, \eprint{1803.10661}.

\bibtype{Article}%
\bibitem{Diaz:2019fwt}
\bibinfo{author}{A. Diaz}, \bibinfo{author}{C.~A. Arg{\"u}elles},
  \bibinfo{author}{G.~H. Collin}, \bibinfo{author}{J.~M. Conrad},
  \bibinfo{author}{M.~H. Shaevitz}, \bibinfo{title}{{Where Are We With Light
  Sterile Neutrinos?}}, \bibinfo{journal}{Phys. Rept.} \bibinfo{volume}{884}
  (\bibinfo{year}{2020}) \bibinfo{pages}{1--59},
  \bibinfo{doi}{\doi{10.1016/j.physrep.2020.08.005}}, \eprint{1906.00045}.

\bibtype{Article}%
\bibitem{Boser:2019rta}
\bibinfo{author}{Sebastian B{\"o}ser}, \bibinfo{author}{Christian Buck},
  \bibinfo{author}{Carlo Giunti}, \bibinfo{author}{Julien Lesgourgues},
  \bibinfo{author}{Livia Ludhova}, \bibinfo{author}{Susanne Mertens},
  \bibinfo{author}{Anne Schukraft}, \bibinfo{author}{Michael Wurm},
  \bibinfo{title}{{Status of Light Sterile Neutrino Searches}},
  \bibinfo{journal}{Prog. Part. Nucl. Phys.} \bibinfo{volume}{111}
  (\bibinfo{year}{2020}) \bibinfo{pages}{103736},
  \bibinfo{doi}{\doi{10.1016/j.ppnp.2019.103736}}, \eprint{1906.01739}.

\bibtype{Article}%
\bibitem{SAGE:2009eeu}
\bibinfo{author}{J.~N. Abdurashitov}, et al. (\bibinfo{collaboration}{SAGE}),
  \bibinfo{title}{{Measurement of the solar neutrino capture rate with gallium
  metal. III: Results for the 2002--2007 data-taking period}},
  \bibinfo{journal}{Phys. Rev. C} \bibinfo{volume}{80} (\bibinfo{year}{2009})
  \bibinfo{pages}{015807}, \bibinfo{doi}{\doi{10.1103/PhysRevC.80.015807}},
  \eprint{0901.2200}.

\bibtype{Article}%
\bibitem{Kaether:2010ag}
\bibinfo{author}{F. Kaether}, \bibinfo{author}{W. Hampel}, \bibinfo{author}{G.
  Heusser}, \bibinfo{author}{J. Kiko}, \bibinfo{author}{T. Kirsten},
  \bibinfo{title}{{Reanalysis of the GALLEX solar neutrino flux and source
  experiments}}, \bibinfo{journal}{Phys. Lett. B} \bibinfo{volume}{685}
  (\bibinfo{year}{2010}) \bibinfo{pages}{47--54},
  \bibinfo{doi}{\doi{10.1016/j.physletb.2010.01.030}}, \eprint{1001.2731}.

\bibtype{Article}%
\bibitem{Barinov:2021asz}
\bibinfo{author}{V.~V. Barinov}, et al., \bibinfo{title}{{Results from the
  Baksan Experiment on Sterile Transitions (BEST)}}, \bibinfo{journal}{Phys.
  Rev. Lett.} \bibinfo{volume}{128} (\bibinfo{number}{23})
  (\bibinfo{year}{2022}) \bibinfo{pages}{232501},
  \bibinfo{doi}{\doi{10.1103/PhysRevLett.128.232501}}, \eprint{2109.11482}.

\bibtype{Article}%
\bibitem{Hagstotz:2020ukm}
\bibinfo{author}{Steffen Hagstotz}, \bibinfo{author}{Pablo~F. de Salas},
  \bibinfo{author}{Stefano Gariazzo}, \bibinfo{author}{Martina Gerbino},
  \bibinfo{author}{Massimiliano Lattanzi}, \bibinfo{author}{Sunny Vagnozzi},
  \bibinfo{author}{Katherine Freese}, \bibinfo{author}{Sergio Pastor},
  \bibinfo{title}{{Bounds on light sterile neutrino mass and mixing from
  cosmology and laboratory searches}}, \bibinfo{journal}{Phys. Rev. D}
  \bibinfo{volume}{104} (\bibinfo{number}{12}) (\bibinfo{year}{2021})
  \bibinfo{pages}{123524}, \bibinfo{doi}{\doi{10.1103/PhysRevD.104.123524}},
  \eprint{2003.02289}.

\bibtype{Article}%
\bibitem{ALEPH:2005ab}
\bibinfo{author}{S. Schael}, et al. (\bibinfo{collaboration}{ALEPH, DELPHI, L3,
  OPAL, SLD, LEP Electroweak Working Group, SLD Electroweak Group, SLD Heavy
  Flavour Group}), \bibinfo{title}{{Precision electroweak measurements on the
  $Z$ resonance}}, \bibinfo{journal}{Phys. Rept.} \bibinfo{volume}{427}
  (\bibinfo{year}{2006}) \bibinfo{pages}{257--454},
  \bibinfo{doi}{\doi{10.1016/j.physrep.2005.12.006}}, \eprint{hep-ex/0509008}.

\bibtype{Article}%
\bibitem{JUNO:2015zny}
\bibinfo{author}{Fengpeng An}, et al. (\bibinfo{collaboration}{JUNO}),
  \bibinfo{title}{{Neutrino Physics with JUNO}}, \bibinfo{journal}{J. Phys. G}
  \bibinfo{volume}{43} (\bibinfo{number}{3}) (\bibinfo{year}{2016})
  \bibinfo{pages}{030401}, \bibinfo{doi}{\doi{10.1088/0954-3899/43/3/030401}},
  \eprint{1507.05613}.

\bibtype{Article}%
\bibitem{JUNO:2025gmd}
\bibinfo{author}{Angel Abusleme}, et al. (\bibinfo{collaboration}{JUNO}),
  \bibinfo{title}{{First measurement of reactor neutrino oscillations at JUNO}}
   (\bibinfo{year}{2025}), \eprint{2511.14593}.

\bibtype{Article}%
\bibitem{Abe:2011ts}
\bibinfo{author}{K. Abe}, et al., \bibinfo{title}{{Letter of Intent: The
  Hyper-Kamiokande Experiment --- Detector Design and Physics Potential ---}}
  (\bibinfo{year}{2011}), \eprint{1109.3262}.

\bibtype{Article}%
\bibitem{DUNE:2020ypp}
\bibinfo{author}{Babak Abi}, et al. (\bibinfo{collaboration}{DUNE}),
  \bibinfo{title}{{Deep Underground Neutrino Experiment (DUNE), Far Detector
  Technical Design Report, Volume II: DUNE Physics}}  (\bibinfo{year}{2020}),
  \eprint{2002.03005}.

\bibtype{Article}%
\bibitem{T2K:2011ypd}
\bibinfo{author}{K. Abe}, et al. (\bibinfo{collaboration}{T2K}),
  \bibinfo{title}{{Indication of Electron Neutrino Appearance from an
  Accelerator-produced Off-axis Muon Neutrino Beam}}, \bibinfo{journal}{Phys.
  Rev. Lett.} \bibinfo{volume}{107} (\bibinfo{year}{2011})
  \bibinfo{pages}{041801}, \bibinfo{doi}{\doi{10.1103/PhysRevLett.107.041801}},
  \eprint{1106.2822}.

\bibtype{Article}%
\bibitem{T2K:2013bqz}
\bibinfo{author}{K. Abe}, et al. (\bibinfo{collaboration}{T2K}),
  \bibinfo{title}{{Evidence of Electron Neutrino Appearance in a Muon Neutrino
  Beam}}, \bibinfo{journal}{Phys. Rev. D} \bibinfo{volume}{88}
  (\bibinfo{number}{3}) (\bibinfo{year}{2013}) \bibinfo{pages}{032002},
  \bibinfo{doi}{\doi{10.1103/PhysRevD.88.032002}}, \eprint{1304.0841}.

\bibtype{Article}%
\bibitem{T2K:2013ppw}
\bibinfo{author}{K. Abe}, et al. (\bibinfo{collaboration}{T2K}),
  \bibinfo{title}{{Observation of Electron Neutrino Appearance in a Muon
  Neutrino Beam}}, \bibinfo{journal}{Phys. Rev. Lett.} \bibinfo{volume}{112}
  (\bibinfo{year}{2014}) \bibinfo{pages}{061802},
  \bibinfo{doi}{\doi{10.1103/PhysRevLett.112.061802}}, \eprint{1311.4750}.

\bibtype{Article}%
\bibitem{T2K:2017hed}
\bibinfo{author}{K. Abe}, et al. (\bibinfo{collaboration}{T2K}),
  \bibinfo{title}{{Combined Analysis of Neutrino and Antineutrino Oscillations
  at T2K}}, \bibinfo{journal}{Phys. Rev. Lett.} \bibinfo{volume}{118}
  (\bibinfo{number}{15}) (\bibinfo{year}{2017}) \bibinfo{pages}{151801},
  \bibinfo{doi}{\doi{10.1103/PhysRevLett.118.151801}}, \eprint{1701.00432}.

\bibtype{Article}%
\bibitem{T2K:2017rgv}
\bibinfo{author}{K. Abe}, et al. (\bibinfo{collaboration}{T2K}),
  \bibinfo{title}{{Measurement of neutrino and antineutrino oscillations by the
  T2K experiment including a new additional sample of $\nu_e$ interactions at
  the far detector}}, \bibinfo{journal}{Phys. Rev. D} \bibinfo{volume}{96}
  (\bibinfo{number}{9}) (\bibinfo{year}{2017}) \bibinfo{pages}{092006},
  \bibinfo{doi}{\doi{10.1103/PhysRevD.96.092006}}, \bibinfo{note}{[Erratum:
  Phys.Rev.D 98, 019902 (2018)]}, \eprint{1707.01048}.

\bibtype{Article}%
\bibitem{T2K:2018rhz}
\bibinfo{author}{K. Abe}, et al. (\bibinfo{collaboration}{T2K}),
  \bibinfo{title}{{Search for CP Violation in Neutrino and Antineutrino
  Oscillations by the T2K Experiment with $2.2\times10^{21}$ Protons on
  Target}}, \bibinfo{journal}{Phys. Rev. Lett.} \bibinfo{volume}{121}
  (\bibinfo{number}{17}) (\bibinfo{year}{2018}) \bibinfo{pages}{171802},
  \bibinfo{doi}{\doi{10.1103/PhysRevLett.121.171802}}, \eprint{1807.07891}.

\bibtype{Article}%
\bibitem{T2K:2019ird}
\bibinfo{author}{K. Abe}, et al. (\bibinfo{collaboration}{T2K}),
  \bibinfo{title}{{Search for Electron Antineutrino Appearance in a
  Long-baseline Muon Antineutrino Beam}}, \bibinfo{journal}{Phys. Rev. Lett.}
  \bibinfo{volume}{124} (\bibinfo{number}{16}) (\bibinfo{year}{2020})
  \bibinfo{pages}{161802}, \bibinfo{doi}{\doi{10.1103/PhysRevLett.124.161802}},
  \eprint{1911.07283}.

\bibtype{Article}%
\bibitem{T2K:2019bcf}
\bibinfo{author}{K. Abe}, et al. (\bibinfo{collaboration}{T2K}),
  \bibinfo{title}{{Constraint on the matter\textendash{}antimatter
  symmetry-violating phase in neutrino oscillations}},
  \bibinfo{journal}{Nature} \bibinfo{volume}{580} (\bibinfo{number}{7803})
  (\bibinfo{year}{2020}) \bibinfo{pages}{339--344},
  \bibinfo{doi}{\doi{10.1038/s41586-020-2177-0}}, \bibinfo{note}{[Erratum:
  Nature 583, E16 (2020)]}, \eprint{1910.03887}.

\bibtype{Article}%
\bibitem{T2K:2023smv}
\bibinfo{author}{K. Abe}, et al. (\bibinfo{collaboration}{T2K}),
  \bibinfo{title}{{Measurements of neutrino oscillation parameters from the T2K
  experiment using $3.6\times 10^{21}$ protons on target}},
  \bibinfo{journal}{Eur. Phys. J. C} \bibinfo{volume}{83} (\bibinfo{number}{9})
  (\bibinfo{year}{2023}) \bibinfo{pages}{782},
  \bibinfo{doi}{\doi{10.1140/epjc/s10052-023-11819-x}}, \eprint{2303.03222}.

\bibtype{Article}%
\bibitem{NOvA:2016kwd}
\bibinfo{author}{P. Adamson}, et al. (\bibinfo{collaboration}{NOvA}),
  \bibinfo{title}{{First measurement of electron neutrino appearance in NOvA}},
  \bibinfo{journal}{Phys. Rev. Lett.} \bibinfo{volume}{116}
  (\bibinfo{number}{15}) (\bibinfo{year}{2016}) \bibinfo{pages}{151806},
  \bibinfo{doi}{\doi{10.1103/PhysRevLett.116.151806}}, \eprint{1601.05022}.

\bibtype{Article}%
\bibitem{MammenAbraham:2022xoc}
\bibinfo{author}{Roshan Mammen~Abraham}, et al., \bibinfo{title}{{Tau neutrinos
  in the next decade: from GeV to EeV}}, \bibinfo{journal}{J. Phys. G}
  \bibinfo{volume}{49} (\bibinfo{number}{11}) (\bibinfo{year}{2022})
  \bibinfo{pages}{110501}, \bibinfo{doi}{\doi{10.1088/1361-6471/ac89d2}},
  \eprint{2203.05591}.

\bibtype{Article}%
\bibitem{OPERA:2010pne}
\bibinfo{author}{N. Agafonova}, et al. (\bibinfo{collaboration}{OPERA}),
  \bibinfo{title}{{Observation of a first $\nu_\tau$ candidate in the OPERA
  experiment in the CNGS beam}}, \bibinfo{journal}{Phys. Lett. B}
  \bibinfo{volume}{691} (\bibinfo{year}{2010}) \bibinfo{pages}{138--145},
  \bibinfo{doi}{\doi{10.1016/j.physletb.2010.06.022}}, \eprint{1006.1623}.

\bibtype{Article}%
\bibitem{OPERA:2013tlg}
\bibinfo{author}{N. Agafonova}, et al. (\bibinfo{collaboration}{OPERA}),
  \bibinfo{title}{{New results on $\nu_\mu \to \nu_\tau$ appearance with the
  OPERA experiment in the CNGS beam}}, \bibinfo{journal}{JHEP}
  \bibinfo{volume}{11} (\bibinfo{year}{2013}) \bibinfo{pages}{036},
  \bibinfo{doi}{\doi{10.1007/JHEP11(2013)036}}, \bibinfo{note}{[Erratum: JHEP
  04, 014 (2014)]}, \eprint{1308.2553}.

\bibtype{Article}%
\bibitem{OPERA:2014fax}
\bibinfo{author}{N. Agafonova}, et al. (\bibinfo{collaboration}{OPERA}),
  \bibinfo{title}{{Evidence for $\nu_\mu \to \nu_\tau$ appearance in the CNGS
  neutrino beam with the OPERA experiment}}, \bibinfo{journal}{Phys. Rev. D}
  \bibinfo{volume}{89} (\bibinfo{number}{5}) (\bibinfo{year}{2014})
  \bibinfo{pages}{051102}, \bibinfo{doi}{\doi{10.1103/PhysRevD.89.051102}},
  \eprint{1401.2079}.

\bibtype{Article}%
\bibitem{OPERA:2015wbl}
\bibinfo{author}{N. Agafonova}, et al. (\bibinfo{collaboration}{OPERA}),
  \bibinfo{title}{{Discovery of $\tau$ Neutrino Appearance in the CNGS Neutrino
  Beam with the OPERA Experiment}}, \bibinfo{journal}{Phys. Rev. Lett.}
  \bibinfo{volume}{115} (\bibinfo{number}{12}) (\bibinfo{year}{2015})
  \bibinfo{pages}{121802}, \bibinfo{doi}{\doi{10.1103/PhysRevLett.115.121802}},
  \eprint{1507.01417}.

\bibtype{Article}%
\bibitem{OPERA:2018nar}
\bibinfo{author}{N. Agafonova}, et al. (\bibinfo{collaboration}{OPERA}),
  \bibinfo{title}{{Final Results of the OPERA Experiment on $\nu_\tau$
  Appearance in the CNGS Neutrino Beam}}, \bibinfo{journal}{Phys. Rev. Lett.}
  \bibinfo{volume}{120} (\bibinfo{number}{21}) (\bibinfo{year}{2018})
  \bibinfo{pages}{211801}, \bibinfo{doi}{\doi{10.1103/PhysRevLett.120.211801}},
  \bibinfo{note}{[Erratum: Phys.Rev.Lett. 121, 139901 (2018)]},
  \eprint{1804.04912}.

\bibtype{Article}%
\bibitem{Super-Kamiokande:2017edb}
\bibinfo{author}{Z. Li}, et al. (\bibinfo{collaboration}{Super-Kamiokande}),
  \bibinfo{title}{{Measurement of the tau neutrino cross section in atmospheric
  neutrino oscillations with Super-Kamiokande}}, \bibinfo{journal}{Phys. Rev.
  D} \bibinfo{volume}{98} (\bibinfo{number}{5}) (\bibinfo{year}{2018})
  \bibinfo{pages}{052006}, \bibinfo{doi}{\doi{10.1103/PhysRevD.98.052006}},
  \eprint{1711.09436}.

\bibtype{Article}%
\bibitem{IceCube:2019dqi}
\bibinfo{author}{M.~G. Aartsen}, et al. (\bibinfo{collaboration}{IceCube}),
  \bibinfo{title}{{Measurement of Atmospheric Tau Neutrino Appearance with
  IceCube DeepCore}}, \bibinfo{journal}{Phys. Rev. D} \bibinfo{volume}{99}
  (\bibinfo{number}{3}) (\bibinfo{year}{2019}) \bibinfo{pages}{032007},
  \bibinfo{doi}{\doi{10.1103/PhysRevD.99.032007}}, \eprint{1901.05366}.

\bibtype{Article}%
\bibitem{IceCube:2020fpi}
\bibinfo{author}{R. Abbasi}, et al. (\bibinfo{collaboration}{IceCube}),
  \bibinfo{title}{{Detection of astrophysical tau neutrino candidates in
  IceCube}}, \bibinfo{journal}{Eur. Phys. J. C} \bibinfo{volume}{82}
  (\bibinfo{number}{11}) (\bibinfo{year}{2022}) \bibinfo{pages}{1031},
  \bibinfo{doi}{\doi{10.1140/epjc/s10052-022-10795-y}}, \eprint{2011.03561}.

\bibtype{Article}%
\bibitem{IceCube:2024nhk}
\bibinfo{author}{R. Abbasi}, et al. (\bibinfo{collaboration}{IceCube}),
  \bibinfo{title}{{Observation of Seven Astrophysical Tau Neutrino Candidates
  with IceCube}}, \bibinfo{journal}{Phys. Rev. Lett.} \bibinfo{volume}{132}
  (\bibinfo{number}{15}) (\bibinfo{year}{2024}) \bibinfo{pages}{151001},
  \bibinfo{doi}{\doi{10.1103/PhysRevLett.132.151001}}, \eprint{2403.02516}.

\bibtype{Article}%
\bibitem{Sarantakos:1982bp}
\bibinfo{author}{S. Sarantakos}, \bibinfo{author}{A. Sirlin},
  \bibinfo{author}{W.~J. Marciano}, \bibinfo{title}{{Radiative Corrections to
  Neutrino-Lepton Scattering in the SU(2)-L x U(1) Theory}},
  \bibinfo{journal}{Nucl. Phys. B} \bibinfo{volume}{217} (\bibinfo{year}{1983})
  \bibinfo{pages}{84--116}, \bibinfo{doi}{\doi{10.1016/0550-3213(83)90079-2}}.

\bibtype{Article}%
\bibitem{Brdar:2023ttb}
\bibinfo{author}{Vedran Brdar}, \bibinfo{author}{Xun-Jie Xu},
  \bibinfo{title}{{Beyond tree level with solar neutrinos: Towards measuring
  the flavor composition and CP violation}}, \bibinfo{journal}{Phys. Lett. B}
  \bibinfo{volume}{846} (\bibinfo{year}{2023}) \bibinfo{pages}{138255},
  \bibinfo{doi}{\doi{10.1016/j.physletb.2023.138255}}, \eprint{2306.03160}.

\bibtype{Article}%
\bibitem{Kelly:2024tvh}
\bibinfo{author}{Kevin~J. Kelly}, \bibinfo{author}{Nityasa Mishra},
  \bibinfo{author}{Mudit Rai}, \bibinfo{author}{Louis~E. Strigari},
  \bibinfo{title}{{{\ensuremath{\nu}}{\ensuremath{\mu}} and
  {\ensuremath{\nu}}{\ensuremath{\tau}} elastic scattering in Borexino}},
  \bibinfo{journal}{Phys. Rev. D} \bibinfo{volume}{110} (\bibinfo{number}{11})
  (\bibinfo{year}{2024}) \bibinfo{pages}{113004},
  \bibinfo{doi}{\doi{10.1103/PhysRevD.110.113004}}, \eprint{2407.03174}.

\bibtype{Article}%
\bibitem{Honda:2015fha}
\bibinfo{author}{M. Honda}, \bibinfo{author}{M. Sajjad~Athar},
  \bibinfo{author}{T. Kajita}, \bibinfo{author}{K. Kasahara},
  \bibinfo{author}{S. Midorikawa}, \bibinfo{title}{{Atmospheric neutrino flux
  calculation using the NRLMSISE-00 atmospheric model}},
  \bibinfo{journal}{Phys. Rev. D} \bibinfo{volume}{92} (\bibinfo{number}{2})
  (\bibinfo{year}{2015}) \bibinfo{pages}{023004},
  \bibinfo{doi}{\doi{10.1103/PhysRevD.92.023004}}, \eprint{1502.03916}.

\bibtype{Article}%
\bibitem{Lipkin:2005kg}
\bibinfo{author}{Harry~J. Lipkin}, \bibinfo{title}{{Quantum theory of neutrino
  oscillations for pedestrians: Simple answers to confusing questions}},
  \bibinfo{journal}{Phys. Lett. B} \bibinfo{volume}{642} (\bibinfo{year}{2006})
  \bibinfo{pages}{366--371},
  \bibinfo{doi}{\doi{10.1016/j.physletb.2006.09.054}}, \eprint{hep-ph/0505141}.

\bibtype{Article}%
\bibitem{Akhmedov:2009rb}
\bibinfo{author}{Evgeny~Kh. Akhmedov}, \bibinfo{author}{Alexei~Yu. Smirnov},
  \bibinfo{title}{{Paradoxes of neutrino oscillations}},
  \bibinfo{journal}{Phys. Atom. Nucl.} \bibinfo{volume}{72}
  (\bibinfo{year}{2009}) \bibinfo{pages}{1363--1381},
  \bibinfo{doi}{\doi{10.1134/S1063778809080122}}, \eprint{0905.1903}.

\bibtype{Article}%
\bibitem{Akhmedov:2017mcc}
\bibinfo{author}{Evgeny Akhmedov}, \bibinfo{author}{Joachim Kopp},
  \bibinfo{author}{Manfred Lindner}, \bibinfo{title}{{Collective neutrino
  oscillations and neutrino wave packets}}, \bibinfo{journal}{JCAP}
  \bibinfo{volume}{09} (\bibinfo{year}{2017}) \bibinfo{pages}{017},
  \bibinfo{doi}{\doi{10.1088/1475-7516/2017/09/017}}, \eprint{1702.08338}.

\bibtype{Article}%
\bibitem{Akhmedov:2022bjs}
\bibinfo{author}{Evgeny Akhmedov}, \bibinfo{author}{Alexei~Y. Smirnov},
  \bibinfo{title}{{Damping of neutrino oscillations, decoherence and the
  lengths of neutrino wave packets}}, \bibinfo{journal}{JHEP}
  \bibinfo{volume}{11} (\bibinfo{year}{2022}) \bibinfo{pages}{082},
  \bibinfo{doi}{\doi{10.1007/JHEP11(2022)082}}, \eprint{2208.03736}.

\bibtype{Article}%
\bibitem{DayaBay:2016ouy}
\bibinfo{author}{Feng~Peng An}, et al. (\bibinfo{collaboration}{Daya Bay}),
  \bibinfo{title}{{Study of the wave packet treatment of neutrino oscillation
  at Daya Bay}}, \bibinfo{journal}{Eur. Phys. J. C} \bibinfo{volume}{77}
  (\bibinfo{number}{9}) (\bibinfo{year}{2017}) \bibinfo{pages}{606},
  \bibinfo{doi}{\doi{10.1140/epjc/s10052-017-4970-y}}, \eprint{1608.01661}.

\bibtype{Article}%
\bibitem{Stuttard:2020qfv}
\bibinfo{author}{Thomas Stuttard}, \bibinfo{author}{Mikkel Jensen},
  \bibinfo{title}{{Neutrino decoherence from quantum gravitational stochastic
  perturbations}}, \bibinfo{journal}{Phys. Rev. D} \bibinfo{volume}{102}
  (\bibinfo{number}{11}) (\bibinfo{year}{2020}) \bibinfo{pages}{115003},
  \bibinfo{doi}{\doi{10.1103/PhysRevD.102.115003}}, \eprint{2007.00068}.

\bibtype{Article}%
\bibitem{deGouvea:2020hfl}
\bibinfo{author}{Andre de Gouvea}, \bibinfo{author}{Valentina de Romeri},
  \bibinfo{author}{Christoph~Andreas Ternes}, \bibinfo{title}{{Probing neutrino
  quantum decoherence at reactor experiments}}, \bibinfo{journal}{JHEP}
  \bibinfo{volume}{08} (\bibinfo{year}{2020}) \bibinfo{pages}{018},
  \bibinfo{doi}{\doi{10.1007/JHEP08(2020)049}}, \eprint{2005.03022}.

\bibtype{Article}%
\bibitem{deGouvea:2021uvg}
\bibinfo{author}{Andr\'e de Gouv\^ea}, \bibinfo{author}{Valentina De~Romeri},
  \bibinfo{author}{Christoph~A. Ternes}, \bibinfo{title}{{Combined analysis of
  neutrino decoherence at reactor experiments}}, \bibinfo{journal}{JHEP}
  \bibinfo{volume}{06} (\bibinfo{year}{2021}) \bibinfo{pages}{042},
  \bibinfo{doi}{\doi{10.1007/JHEP06(2021)042}}, \eprint{2104.05806}.

\bibtype{Article}%
\bibitem{DeRomeri:2023dht}
\bibinfo{author}{Valentina De~Romeri}, \bibinfo{author}{Carlo Giunti},
  \bibinfo{author}{Thomas Stuttard}, \bibinfo{author}{Christoph~A. Ternes},
  \bibinfo{title}{{Neutrino oscillation bounds on quantum decoherence}},
  \bibinfo{journal}{JHEP} \bibinfo{volume}{09} (\bibinfo{year}{2023})
  \bibinfo{pages}{097}, \bibinfo{doi}{\doi{10.1007/JHEP09(2023)097}},
  \eprint{2306.14699}.

\bibtype{Article}%
\bibitem{deGouvea:2024syg}
\bibinfo{author}{Andr\'e de Gouv\^ea}, \bibinfo{author}{Valentina De~Romeri},
  \bibinfo{author}{Christoph~A. Ternes}, \bibinfo{title}{{Addendum to: Combined
  analysis of neutrino decoherence at reactor experiments}},
  \bibinfo{journal}{JHEP} \bibinfo{volume}{11} (\bibinfo{year}{2024})
  \bibinfo{pages}{095}, \bibinfo{doi}{\doi{10.1007/JHEP11(2024)095}},
  \eprint{2410.01357}.

\bibtype{Article}%
\bibitem{Minakata:2015gra}
\bibinfo{author}{Hisakazu Minakata}, \bibinfo{author}{Stephen~J Parke},
  \bibinfo{title}{{Simple and Compact Expressions for Neutrino Oscillation
  Probabilities in Matter}}, \bibinfo{journal}{JHEP} \bibinfo{volume}{01}
  (\bibinfo{year}{2016}) \bibinfo{pages}{180},
  \bibinfo{doi}{\doi{10.1007/JHEP01(2016)180}}, \eprint{1505.01826}.

\bibtype{Article}%
\bibitem{Denton:2016wmg}
\bibinfo{author}{Peter~B. Denton}, \bibinfo{author}{Hisakazu Minakata},
  \bibinfo{author}{Stephen~J. Parke}, \bibinfo{title}{{Compact Perturbative
  Expressions For Neutrino Oscillations in Matter}}, \bibinfo{journal}{JHEP}
  \bibinfo{volume}{06} (\bibinfo{year}{2016}) \bibinfo{pages}{051},
  \bibinfo{doi}{\doi{10.1007/JHEP06(2016)051}}, \eprint{1604.08167}.

\bibtype{Article}%
\bibitem{Denton:2018hal}
\bibinfo{author}{Peter~B. Denton}, \bibinfo{author}{Stephen~J. Parke},
  \bibinfo{title}{{Addendum to ``Compact perturbative expressions for neutrino
  oscillations in matter''}}, \bibinfo{journal}{JHEP}  (\bibinfo{year}{2018}),
  \bibinfo{doi}{\doi{10.1007/JHEP06(2018)109}}, \bibinfo{note}{[Addendum: JHEP
  06, 109 (2018)]}, \eprint{1801.06514}.

\bibtype{Article}%
\bibitem{Denton:2018fex}
\bibinfo{author}{Peter~B. Denton}, \bibinfo{author}{Stephen~J. Parke},
  \bibinfo{author}{Xining Zhang}, \bibinfo{title}{{Rotations Versus
  Perturbative Expansions for Calculating Neutrino Oscillation Probabilities in
  Matter}}, \bibinfo{journal}{Phys. Rev. D} \bibinfo{volume}{98}
  (\bibinfo{number}{3}) (\bibinfo{year}{2018}) \bibinfo{pages}{033001},
  \bibinfo{doi}{\doi{10.1103/PhysRevD.98.033001}}, \eprint{1806.01277}.

\bibtype{Article}%
\bibitem{Denton:2018cpu}
\bibinfo{author}{Peter~B. Denton}, \bibinfo{author}{Stephen~J. Parke},
  \bibinfo{title}{{The effective $\Delta m^2_{ee}$ in matter}},
  \bibinfo{journal}{Phys. Rev. D} \bibinfo{volume}{98} (\bibinfo{number}{9})
  (\bibinfo{year}{2018}) \bibinfo{pages}{093001},
  \bibinfo{doi}{\doi{10.1103/PhysRevD.98.093001}}, \eprint{1808.09453}.

\bibtype{Article}%
\bibitem{Denton:2019yiw}
\bibinfo{author}{Peter~B Denton}, \bibinfo{author}{Stephen~J Parke},
  \bibinfo{title}{{Simple and Precise Factorization of the Jarlskog Invariant
  for Neutrino Oscillations in Matter}}, \bibinfo{journal}{Phys. Rev. D}
  \bibinfo{volume}{100} (\bibinfo{number}{5}) (\bibinfo{year}{2019})
  \bibinfo{pages}{053004}, \bibinfo{doi}{\doi{10.1103/PhysRevD.100.053004}},
  \eprint{1902.07185}.

\bibtype{Article}%
\bibitem{Denton:2019ovn}
\bibinfo{author}{Peter~B Denton}, \bibinfo{author}{Stephen~J Parke},
  \bibinfo{author}{Xining Zhang}, \bibinfo{title}{{Neutrino oscillations in
  matter via eigenvalues}}, \bibinfo{journal}{Phys. Rev. D}
  \bibinfo{volume}{101} (\bibinfo{number}{9}) (\bibinfo{year}{2020})
  \bibinfo{pages}{093001}, \bibinfo{doi}{\doi{10.1103/PhysRevD.101.093001}},
  \eprint{1907.02534}.

\bibtype{Article}%
\bibitem{Denton:2019qzn}
\bibinfo{author}{Peter~B. Denton}, \bibinfo{author}{Stephen~J. Parke},
  \bibinfo{author}{Xining Zhang}, \bibinfo{title}{{Fibonacci fast convergence
  for neutrino oscillations in matter}}, \bibinfo{journal}{Phys. Lett. B}
  \bibinfo{volume}{807} (\bibinfo{year}{2020}) \bibinfo{pages}{135592},
  \bibinfo{doi}{\doi{10.1016/j.physletb.2020.135592}}, \eprint{1909.02009}.

\bibtype{Article}%
\bibitem{Agarwalla:2013tza}
\bibinfo{author}{Sanjib~Kumar Agarwalla}, \bibinfo{author}{Yee Kao},
  \bibinfo{author}{Tatsu Takeuchi}, \bibinfo{title}{{Analytical approximation
  of the neutrino oscillation matter effects at large $\theta_{13}$}},
  \bibinfo{journal}{JHEP} \bibinfo{volume}{04} (\bibinfo{year}{2014})
  \bibinfo{pages}{047}, \bibinfo{doi}{\doi{10.1007/JHEP04(2014)047}},
  \eprint{1302.6773}.

\bibtype{Article}%
\bibitem{Barenboim:2019pfp}
\bibinfo{author}{Gabriela Barenboim}, \bibinfo{author}{Peter~B Denton},
  \bibinfo{author}{Stephen~J Parke}, \bibinfo{author}{Christoph~Andreas
  Ternes}, \bibinfo{title}{{Neutrino Oscillation Probabilities through the
  Looking Glass}}, \bibinfo{journal}{Phys. Lett. B} \bibinfo{volume}{791}
  (\bibinfo{year}{2019}) \bibinfo{pages}{351--360},
  \bibinfo{doi}{\doi{10.1016/j.physletb.2019.03.002}}, \eprint{1902.00517}.

\bibtype{Article}%
\bibitem{Denton:2024pzc}
\bibinfo{author}{Peter~B. Denton}, \bibinfo{author}{Stephen~J. Parke},
  \bibinfo{title}{{Fast and accurate algorithm for calculating long-baseline
  neutrino oscillation probabilities with matter effects}},
  \bibinfo{journal}{Phys. Rev. D} \bibinfo{volume}{110} (\bibinfo{number}{7})
  (\bibinfo{year}{2024}) \bibinfo{pages}{073005},
  \bibinfo{doi}{\doi{10.1103/PhysRevD.110.073005}}, \eprint{2405.02400}.

\bibtype{Article}%
\bibitem{Parke:1986jy}
\bibinfo{author}{Stephen~J. Parke}, \bibinfo{title}{{Nonadiabatic Level
  Crossing in Resonant Neutrino Oscillations}}, \bibinfo{journal}{Phys. Rev.
  Lett.} \bibinfo{volume}{57} (\bibinfo{year}{1986})
  \bibinfo{pages}{1275--1278},
  \bibinfo{doi}{\doi{10.1103/PhysRevLett.57.1275}}, \eprint{2212.06978}.

\bibtype{Article}%
\bibitem{Petcov:1987zj}
\bibinfo{author}{S.~T. Petcov}, \bibinfo{title}{{Exact analytic description of
  two neutrino oscillations in matter with exponentially varying density}},
  \bibinfo{journal}{Phys. Lett. B} \bibinfo{volume}{200} (\bibinfo{year}{1988})
  \bibinfo{pages}{373--379}, \bibinfo{doi}{\doi{10.1016/0370-2693(88)90791-5}}.

\bibtype{Article}%
\bibitem{Kuo:1988pn}
\bibinfo{author}{Tzee-Ke Kuo}, \bibinfo{author}{James~T. Pantaleone},
  \bibinfo{title}{{Nonadiabatic Neutrino Oscillations in Matter}},
  \bibinfo{journal}{Phys. Rev. D} \bibinfo{volume}{39} (\bibinfo{year}{1989})
  \bibinfo{pages}{1930}, \bibinfo{doi}{\doi{10.1103/PhysRevD.39.1930}}.

\bibtype{Article}%
\bibitem{Lehnert:2007fv}
\bibinfo{author}{Ralf Lehnert}, \bibinfo{author}{Thomas~J. Weiler},
  \bibinfo{title}{{Neutrino flavor ratios as diagnostic of solar WIMP
  annihilation}}, \bibinfo{journal}{Phys. Rev. D} \bibinfo{volume}{77}
  (\bibinfo{year}{2008}) \bibinfo{pages}{125004},
  \bibinfo{doi}{\doi{10.1103/PhysRevD.77.125004}}, \eprint{0708.1035}.

\bibtype{Article}%
\bibitem{Denton:2025cbo}
\bibinfo{author}{Peter~B. Denton}, \bibinfo{author}{Charles Gourley},
  \bibinfo{title}{{Determining the density of the sun with neutrinos}},
  \bibinfo{journal}{Phys. Lett. B} \bibinfo{volume}{866} (\bibinfo{year}{2025})
  \bibinfo{pages}{139560}, \bibinfo{doi}{\doi{10.1016/j.physletb.2025.139560}},
  \eprint{2502.17546}.

\bibtype{Article}%
\bibitem{Zaidel:2025kdk}
\bibinfo{author}{Melanie~A. Zaidel}, \bibinfo{author}{John~F. Beacom},
  \bibinfo{title}{{Probing the conditions of the solar core using B8
  neutrinos}}, \bibinfo{journal}{Phys. Rev. C} \bibinfo{volume}{112}
  (\bibinfo{number}{5}) (\bibinfo{year}{2025}) \bibinfo{pages}{055801},
  \bibinfo{doi}{\doi{10.1103/r3h6-tw7p}}, \eprint{2504.10583}.

\bibtype{Article}%
\bibitem{BOREXINO:2014pcl}
\bibinfo{author}{G. Bellini}, et al. (\bibinfo{collaboration}{BOREXINO}),
  \bibinfo{title}{{Neutrinos from the primary proton\textendash{}proton fusion
  process in the Sun}}, \bibinfo{journal}{Nature} \bibinfo{volume}{512}
  (\bibinfo{number}{7515}) (\bibinfo{year}{2014}) \bibinfo{pages}{383--386},
  \bibinfo{doi}{\doi{10.1038/nature13702}}.

\bibtype{Article}%
\bibitem{Bouchez:1986kb}
\bibinfo{author}{J. Bouchez}, \bibinfo{author}{M. Cribier}, \bibinfo{author}{J.
  Rich}, \bibinfo{author}{M. Spiro}, \bibinfo{author}{D. Vignaud},
  \bibinfo{author}{W. Hampel}, \bibinfo{title}{{Matter Effects for Solar
  Neutrino Oscillations}}, \bibinfo{journal}{Z. Phys. C} \bibinfo{volume}{32}
  (\bibinfo{year}{1986}) \bibinfo{pages}{499},
  \bibinfo{doi}{\doi{10.1007/BF01550771}}.

\bibtype{Article}%
\bibitem{Baltz:1986hn}
\bibinfo{author}{A.~J. Baltz}, \bibinfo{author}{J. Weneser},
  \bibinfo{title}{{Effect of Transmission Through the Earth on Neutrino
  Oscillations}}, \bibinfo{journal}{Phys. Rev. D} \bibinfo{volume}{35}
  (\bibinfo{year}{1987}) \bibinfo{pages}{528},
  \bibinfo{doi}{\doi{10.1103/PhysRevD.35.528}}.

\bibtype{Article}%
\bibitem{Cribier:1986ak}
\bibinfo{author}{M. Cribier}, \bibinfo{author}{W. Hampel}, \bibinfo{author}{J.
  Rich}, \bibinfo{author}{D. Vignaud}, \bibinfo{title}{{Msw Regeneration of
  Solar $\nu_e$ in the Earth}}, \bibinfo{journal}{Phys. Lett. B}
  \bibinfo{volume}{182} (\bibinfo{year}{1986}) \bibinfo{pages}{89--94},
  \bibinfo{doi}{\doi{10.1016/0370-2693(86)91083-X}}.

\bibtype{Article}%
\bibitem{Capozzi:2018dat}
\bibinfo{author}{Francesco Capozzi}, \bibinfo{author}{Shirley~Weishi Li},
  \bibinfo{author}{Guanying Zhu}, \bibinfo{author}{John~F. Beacom},
  \bibinfo{title}{{DUNE as the Next-Generation Solar Neutrino Experiment}},
  \bibinfo{journal}{Phys. Rev. Lett.} \bibinfo{volume}{123}
  (\bibinfo{number}{13}) (\bibinfo{year}{2019}) \bibinfo{pages}{131803},
  \bibinfo{doi}{\doi{10.1103/PhysRevLett.123.131803}}, \eprint{1808.08232}.

\bibtype{Article}%
\bibitem{DUNE:2024wvj}
\bibinfo{author}{Adam Abed~Abud}, et al. (\bibinfo{collaboration}{DUNE}),
  \bibinfo{title}{{DUNE Phase~II: scientific opportunities, detector concepts,
  technological solutions}}, \bibinfo{journal}{JINST} \bibinfo{volume}{19}
  (\bibinfo{number}{12}) (\bibinfo{year}{2024}) \bibinfo{pages}{P12005},
  \bibinfo{doi}{\doi{10.1088/1748-0221/19/12/P12005}}, \eprint{2408.12725}.

\bibtype{Article}%
\bibitem{Hyper-Kamiokande:2018ofw}
\bibinfo{author}{K. Abe}, et al. (\bibinfo{collaboration}{Hyper-Kamiokande}),
  \bibinfo{title}{{Hyper-Kamiokande Design Report}}  (\bibinfo{year}{2018}),
  \eprint{1805.04163}.

\bibtype{Article}%
\bibitem{Barenboim:2023krl}
\bibinfo{author}{Gabriela Barenboim}, \bibinfo{author}{Pablo
  Mart\'\i{}nez-Mirav\'e}, \bibinfo{author}{Christoph~A. Ternes},
  \bibinfo{author}{Mariam T\'ortola}, \bibinfo{title}{{Neutrino CPT violation
  in the solar sector}}, \bibinfo{journal}{Phys. Rev. D} \bibinfo{volume}{108}
  (\bibinfo{number}{3}) (\bibinfo{year}{2023}) \bibinfo{pages}{035039},
  \bibinfo{doi}{\doi{10.1103/PhysRevD.108.035039}}, \eprint{2305.06384}.

\bibtype{Article}%
\bibitem{SNO:2011hxd}
\bibinfo{author}{B. Aharmim}, et al. (\bibinfo{collaboration}{SNO}),
  \bibinfo{title}{{Combined Analysis of all Three Phases of Solar Neutrino Data
  from the Sudbury Neutrino Observatory}}, \bibinfo{journal}{Phys. Rev. C}
  \bibinfo{volume}{88} (\bibinfo{year}{2013}) \bibinfo{pages}{025501},
  \bibinfo{doi}{\doi{10.1103/PhysRevC.88.025501}}, \eprint{1109.0763}.

\bibtype{Article}%
\bibitem{Super-Kamiokande:1998qwk}
\bibinfo{author}{Y. Fukuda}, et al.
  (\bibinfo{collaboration}{Super-Kamiokande}), \bibinfo{title}{{Measurements of
  the solar neutrino flux from Super-Kamiokande's first 300 days}},
  \bibinfo{journal}{Phys. Rev. Lett.} \bibinfo{volume}{81}
  (\bibinfo{year}{1998}) \bibinfo{pages}{1158--1162},
  \bibinfo{doi}{\doi{10.1103/PhysRevLett.81.1158}}, \bibinfo{note}{[Erratum:
  Phys.Rev.Lett. 81, 4279 (1998)]}, \eprint{hep-ex/9805021}.

\bibtype{Article}%
\bibitem{Super-Kamiokande:1998zvz}
\bibinfo{author}{Y. Fukuda}, et al.
  (\bibinfo{collaboration}{Super-Kamiokande}), \bibinfo{title}{{Measurement of
  the solar neutrino energy spectrum using neutrino electron scattering}},
  \bibinfo{journal}{Phys. Rev. Lett.} \bibinfo{volume}{82}
  (\bibinfo{year}{1999}) \bibinfo{pages}{2430--2434},
  \bibinfo{doi}{\doi{10.1103/PhysRevLett.82.2430}}, \eprint{hep-ex/9812011}.

\bibtype{Article}%
\bibitem{Super-Kamiokande:1998oic}
\bibinfo{author}{Y. Fukuda}, et al.
  (\bibinfo{collaboration}{Super-Kamiokande}), \bibinfo{title}{{Constraints on
  neutrino oscillation parameters from the measurement of day night solar
  neutrino fluxes at Super-Kamiokande}}, \bibinfo{journal}{Phys. Rev. Lett.}
  \bibinfo{volume}{82} (\bibinfo{year}{1999}) \bibinfo{pages}{1810--1814},
  \bibinfo{doi}{\doi{10.1103/PhysRevLett.82.1810}}, \eprint{hep-ex/9812009}.

\bibtype{Article}%
\bibitem{Super-Kamiokande:2001bfk}
\bibinfo{author}{S. Fukuda}, et al.
  (\bibinfo{collaboration}{Super-Kamiokande}), \bibinfo{title}{{Constraints on
  neutrino oscillations using 1258 days of Super-Kamiokande solar neutrino
  data}}, \bibinfo{journal}{Phys. Rev. Lett.} \bibinfo{volume}{86}
  (\bibinfo{year}{2001}) \bibinfo{pages}{5656--5660},
  \bibinfo{doi}{\doi{10.1103/PhysRevLett.86.5656}}, \eprint{hep-ex/0103033}.

\bibtype{Article}%
\bibitem{Super-Kamiokande:2002ujc}
\bibinfo{author}{S. Fukuda}, et al.
  (\bibinfo{collaboration}{Super-Kamiokande}), \bibinfo{title}{{Determination
  of solar neutrino oscillation parameters using 1496 days of Super-Kamiokande
  I data}}, \bibinfo{journal}{Phys. Lett. B} \bibinfo{volume}{539}
  (\bibinfo{year}{2002}) \bibinfo{pages}{179--187},
  \bibinfo{doi}{\doi{10.1016/S0370-2693(02)02090-7}}, \eprint{hep-ex/0205075}.

\bibtype{Article}%
\bibitem{Super-Kamiokande:2005wtt}
\bibinfo{author}{J. Hosaka}, et al.
  (\bibinfo{collaboration}{Super-Kamiokande}), \bibinfo{title}{{Solar neutrino
  measurements in super-Kamiokande-I}}, \bibinfo{journal}{Phys. Rev. D}
  \bibinfo{volume}{73} (\bibinfo{year}{2006}) \bibinfo{pages}{112001},
  \bibinfo{doi}{\doi{10.1103/PhysRevD.73.112001}}, \eprint{hep-ex/0508053}.

\bibtype{Article}%
\bibitem{Super-Kamiokande:2008ecj}
\bibinfo{author}{J.~P. Cravens}, et al.
  (\bibinfo{collaboration}{Super-Kamiokande}), \bibinfo{title}{{Solar neutrino
  measurements in Super-Kamiokande-II}}, \bibinfo{journal}{Phys. Rev. D}
  \bibinfo{volume}{78} (\bibinfo{year}{2008}) \bibinfo{pages}{032002},
  \bibinfo{doi}{\doi{10.1103/PhysRevD.78.032002}}, \eprint{0803.4312}.

\bibtype{Article}%
\bibitem{Super-Kamiokande:2010tar}
\bibinfo{author}{K. Abe}, et al. (\bibinfo{collaboration}{Super-Kamiokande}),
  \bibinfo{title}{{Solar neutrino results in Super-Kamiokande-III}},
  \bibinfo{journal}{Phys. Rev. D} \bibinfo{volume}{83} (\bibinfo{year}{2011})
  \bibinfo{pages}{052010}, \bibinfo{doi}{\doi{10.1103/PhysRevD.83.052010}},
  \eprint{1010.0118}.

\bibtype{Article}%
\bibitem{Super-Kamiokande:2016yck}
\bibinfo{author}{K. Abe}, et al. (\bibinfo{collaboration}{Super-Kamiokande}),
  \bibinfo{title}{{Solar Neutrino Measurements in Super-Kamiokande-IV}},
  \bibinfo{journal}{Phys. Rev. D} \bibinfo{volume}{94} (\bibinfo{number}{5})
  (\bibinfo{year}{2016}) \bibinfo{pages}{052010},
  \bibinfo{doi}{\doi{10.1103/PhysRevD.94.052010}}, \eprint{1606.07538}.

\bibtype{Article}%
\bibitem{Super-Kamiokande:2023jbt}
\bibinfo{author}{K. Abe}, et al. (\bibinfo{collaboration}{Super-Kamiokande}),
  \bibinfo{title}{{Solar neutrino measurements using the full data period of
  Super-Kamiokande-IV}}, \bibinfo{journal}{Phys. Rev. D} \bibinfo{volume}{109}
  (\bibinfo{number}{9}) (\bibinfo{year}{2024}) \bibinfo{pages}{092001},
  \bibinfo{doi}{\doi{10.1103/PhysRevD.109.092001}}, \eprint{2312.12907}.

\bibtype{Article}%
\bibitem{Raffelt:2007nv}
\bibinfo{author}{Georg~G. Raffelt}, \bibinfo{title}{{Supernova neutrino
  observations: What can we learn?}}, \bibinfo{journal}{Nucl. Phys. B Proc.
  Suppl.} \bibinfo{volume}{221} (\bibinfo{year}{2011})
  \bibinfo{pages}{218--229},
  \bibinfo{doi}{\doi{10.1016/j.nuclphysbps.2011.09.006}},
  \eprint{astro-ph/0701677}.

\bibtype{Article}%
\bibitem{Dighe:2008dq}
\bibinfo{author}{Amol Dighe}, \bibinfo{title}{{Physics potential of future
  supernova neutrino observations}}, \bibinfo{journal}{J. Phys. Conf. Ser.}
  \bibinfo{volume}{136} (\bibinfo{year}{2008}) \bibinfo{pages}{022041},
  \bibinfo{doi}{\doi{10.1088/1742-6596/136/2/022041}}, \eprint{0809.2977}.

\bibtype{Article}%
\bibitem{Scholberg:2012id}
\bibinfo{author}{Kate Scholberg}, \bibinfo{title}{{Supernova Neutrino
  Detection}}, \bibinfo{journal}{Ann. Rev. Nucl. Part. Sci.}
  \bibinfo{volume}{62} (\bibinfo{year}{2012}) \bibinfo{pages}{81--103},
  \bibinfo{doi}{\doi{10.1146/annurev-nucl-102711-095006}}, \eprint{1205.6003}.

\bibtype{Article}%
\bibitem{Scholberg:2017czd}
\bibinfo{author}{Kate Scholberg}, \bibinfo{title}{{Supernova Signatures of
  Neutrino Mass Ordering}}, \bibinfo{journal}{J. Phys. G} \bibinfo{volume}{45}
  (\bibinfo{number}{1}) (\bibinfo{year}{2018}) \bibinfo{pages}{014002},
  \bibinfo{doi}{\doi{10.1088/1361-6471/aa97be}}, \eprint{1707.06384}.

\bibtype{Article}%
\bibitem{DeSalas:2018rby}
\bibinfo{author}{P.~F. De~Salas}, \bibinfo{author}{S. Gariazzo},
  \bibinfo{author}{O. Mena}, \bibinfo{author}{C.~A. Ternes},
  \bibinfo{author}{M. T\'ortola}, \bibinfo{title}{{Neutrino Mass Ordering from
  Oscillations and Beyond: 2018 Status and Future Prospects}},
  \bibinfo{journal}{Front. Astron. Space Sci.} \bibinfo{volume}{5}
  (\bibinfo{year}{2018}) \bibinfo{pages}{36},
  \bibinfo{doi}{\doi{10.3389/fspas.2018.00036}}, \eprint{1806.11051}.

\bibtype{Inbook}%
\bibitem{Suliga:2022ica}
\bibinfo{author}{Anna~M. Suliga}, \bibinfo{title}{{Diffuse Supernova Neutrino
  Background}} \bibinfo{year}{2022} pp. \bibinfo{pages}{1--18},
  \bibinfo{doi}{\doi{10.1007/978-981-15-8818-1_129-1}}, \eprint{2207.09632}.

\bibtype{Article}%
\bibitem{Moller:2018kpn}
\bibinfo{author}{Klaes M\o{}ller}, \bibinfo{author}{Anna~M. Suliga},
  \bibinfo{author}{Irene Tamborra}, \bibinfo{author}{Peter~B. Denton},
  \bibinfo{title}{{Measuring the supernova unknowns at the next-generation
  neutrino telescopes through the diffuse neutrino background}},
  \bibinfo{journal}{JCAP} \bibinfo{volume}{05} (\bibinfo{year}{2018})
  \bibinfo{pages}{066}, \bibinfo{doi}{\doi{10.1088/1475-7516/2018/05/066}},
  \eprint{1804.03157}.

\bibtype{Article}%
\bibitem{Gonzalez-Garcia:2000opv}
\bibinfo{author}{M.~C. Gonzalez-Garcia}, \bibinfo{author}{M. Maltoni},
  \bibinfo{author}{Carlos Pena-Garay}, \bibinfo{author}{J.~W.~F. Valle},
  \bibinfo{title}{{Global three neutrino oscillation analysis of neutrino
  data}}, \bibinfo{journal}{Phys. Rev. D} \bibinfo{volume}{63}
  (\bibinfo{year}{2001}) \bibinfo{pages}{033005},
  \bibinfo{doi}{\doi{10.1103/PhysRevD.63.033005}}, \eprint{hep-ph/0009350}.

\bibtype{Article}%
\bibitem{Maltoni:2002ni}
\bibinfo{author}{M. Maltoni}, \bibinfo{author}{T. Schwetz},
  \bibinfo{author}{M.~A. Tortola}, \bibinfo{author}{J.~W.~F. Valle},
  \bibinfo{title}{{Constraining neutrino oscillation parameters with current
  solar and atmospheric data}}, \bibinfo{journal}{Phys. Rev. D}
  \bibinfo{volume}{67} (\bibinfo{year}{2003}) \bibinfo{pages}{013011},
  \bibinfo{doi}{\doi{10.1103/PhysRevD.67.013011}}, \eprint{hep-ph/0207227}.

\bibtype{Article}%
\bibitem{Schwetz:2008er}
\bibinfo{author}{Thomas Schwetz}, \bibinfo{author}{M.~A. Tortola},
  \bibinfo{author}{Jose W.~F. Valle}, \bibinfo{title}{{Three-flavour neutrino
  oscillation update}}, \bibinfo{journal}{New J. Phys.} \bibinfo{volume}{10}
  (\bibinfo{year}{2008}) \bibinfo{pages}{113011},
  \bibinfo{doi}{\doi{10.1088/1367-2630/10/11/113011}}, \eprint{0808.2016}.

\bibtype{Article}%
\bibitem{Gonzalez-Garcia:2010zke}
\bibinfo{author}{M.~C. Gonzalez-Garcia}, \bibinfo{author}{Michele Maltoni},
  \bibinfo{author}{Jordi Salvado}, \bibinfo{title}{{Updated global fit to three
  neutrino mixing: status of the hints of theta13 \ensuremath{>} 0}},
  \bibinfo{journal}{JHEP} \bibinfo{volume}{04} (\bibinfo{year}{2010})
  \bibinfo{pages}{056}, \bibinfo{doi}{\doi{10.1007/JHEP04(2010)056}},
  \eprint{1001.4524}.

\bibtype{Article}%
\bibitem{Forero:2012faj}
\bibinfo{author}{D.~V. Forero}, \bibinfo{author}{M. Tortola},
  \bibinfo{author}{J.~W.~F. Valle}, \bibinfo{title}{{Global status of neutrino
  oscillation parameters after Neutrino-2012}}, \bibinfo{journal}{Phys. Rev. D}
  \bibinfo{volume}{86} (\bibinfo{year}{2012}) \bibinfo{pages}{073012},
  \bibinfo{doi}{\doi{10.1103/PhysRevD.86.073012}}, \eprint{1205.4018}.

\bibtype{Article}%
\bibitem{Forero:2014bxa}
\bibinfo{author}{D.~V. Forero}, \bibinfo{author}{M. Tortola},
  \bibinfo{author}{J.~W.~F. Valle}, \bibinfo{title}{{Neutrino oscillations
  refitted}}, \bibinfo{journal}{Phys. Rev. D} \bibinfo{volume}{90}
  (\bibinfo{number}{9}) (\bibinfo{year}{2014}) \bibinfo{pages}{093006},
  \bibinfo{doi}{\doi{10.1103/PhysRevD.90.093006}}, \eprint{1405.7540}.

\bibtype{Article}%
\bibitem{deSalas:2017kay}
\bibinfo{author}{P.~F. de Salas}, \bibinfo{author}{D.~V. Forero},
  \bibinfo{author}{C.~A. Ternes}, \bibinfo{author}{M. Tortola},
  \bibinfo{author}{J.~W.~F. Valle}, \bibinfo{title}{{Status of neutrino
  oscillations 2018: 3$\sigma$ hint for normal mass ordering and improved CP
  sensitivity}}, \bibinfo{journal}{Phys. Lett. B} \bibinfo{volume}{782}
  (\bibinfo{year}{2018}) \bibinfo{pages}{633--640},
  \bibinfo{doi}{\doi{10.1016/j.physletb.2018.06.019}}, \eprint{1708.01186}.

\bibtype{Article}%
\bibitem{Capozzi:2017ipn}
\bibinfo{author}{Francesco Capozzi}, \bibinfo{author}{Eleonora Di~Valentino},
  \bibinfo{author}{Eligio Lisi}, \bibinfo{author}{Antonio Marrone},
  \bibinfo{author}{Alessandro Melchiorri}, \bibinfo{author}{Antonio Palazzo},
  \bibinfo{title}{{Global constraints on absolute neutrino masses and their
  ordering}}, \bibinfo{journal}{Phys. Rev. D} \bibinfo{volume}{95}
  (\bibinfo{number}{9}) (\bibinfo{year}{2017}) \bibinfo{pages}{096014},
  \bibinfo{doi}{\doi{10.1103/PhysRevD.95.096014}}, \bibinfo{note}{[Addendum:
  Phys.Rev.D 101, 116013 (2020)]}, \eprint{2003.08511}.

\bibtype{Article}%
\bibitem{Esteban:2020cvm}
\bibinfo{author}{Ivan Esteban}, \bibinfo{author}{M.~C. Gonzalez-Garcia},
  \bibinfo{author}{Michele Maltoni}, \bibinfo{author}{Thomas Schwetz},
  \bibinfo{author}{Albert Zhou}, \bibinfo{title}{{The fate of hints: updated
  global analysis of three-flavor neutrino oscillations}},
  \bibinfo{journal}{JHEP} \bibinfo{volume}{09} (\bibinfo{year}{2020})
  \bibinfo{pages}{178}, \bibinfo{doi}{\doi{10.1007/JHEP09(2020)178}},
  \eprint{2007.14792}.

\bibtype{Article}%
\bibitem{Denton:2020exu}
\bibinfo{author}{Peter~B. Denton}, \bibinfo{title}{{A Return To Neutrino
  Normalcy}}  (\bibinfo{year}{2020}), \eprint{2003.04319}.

\bibtype{Article}%
\bibitem{Denton:2021vtf}
\bibinfo{author}{Peter~B. Denton}, \bibinfo{author}{Stephen~J. Parke},
  \bibinfo{title}{{Parameter symmetries of neutrino oscillations in vacuum,
  matter, and approximation schemes}}, \bibinfo{journal}{Phys. Rev. D}
  \bibinfo{volume}{105} (\bibinfo{number}{1}) (\bibinfo{year}{2022})
  \bibinfo{pages}{013002}, \bibinfo{doi}{\doi{10.1103/PhysRevD.105.013002}},
  \eprint{2106.12436}.

\bibtype{Article}%
\bibitem{deGouvea:2000pqg}
\bibinfo{author}{Andre de Gouvea}, \bibinfo{author}{Alexander Friedland},
  \bibinfo{author}{Hitoshi Murayama}, \bibinfo{title}{{The Dark side of the
  solar neutrino parameter space}}, \bibinfo{journal}{Phys. Lett. B}
  \bibinfo{volume}{490} (\bibinfo{year}{2000}) \bibinfo{pages}{125--130},
  \bibinfo{doi}{\doi{10.1016/S0370-2693(00)00989-8}}, \eprint{hep-ph/0002064}.

\bibtype{Article}%
\bibitem{Esteban:2024eli}
\bibinfo{author}{Ivan Esteban}, \bibinfo{author}{M.~C. Gonzalez-Garcia},
  \bibinfo{author}{Michele Maltoni}, \bibinfo{author}{Ivan Martinez-Soler},
  \bibinfo{author}{Jo{\~a}o~Paulo Pinheiro}, \bibinfo{author}{Thomas Schwetz},
  \bibinfo{title}{{NuFit-6.0: updated global analysis of three-flavor neutrino
  oscillations}}, \bibinfo{journal}{JHEP} \bibinfo{volume}{12}
  (\bibinfo{year}{2024}) \bibinfo{pages}{216},
  \bibinfo{doi}{\doi{10.1007/JHEP12(2024)216}}, \eprint{2410.05380}.

\bibtype{Article}%
\bibitem{Capozzi:2025wyn}
\bibinfo{author}{Francesco Capozzi}, \bibinfo{author}{William Giar{\`e}},
  \bibinfo{author}{Eligio Lisi}, \bibinfo{author}{Antonio Marrone},
  \bibinfo{author}{Alessandro Melchiorri}, \bibinfo{author}{Antonio Palazzo},
  \bibinfo{title}{{Neutrino masses and mixing: Entering the era of subpercent
  precision}}, \bibinfo{journal}{Phys. Rev. D} \bibinfo{volume}{111}
  (\bibinfo{number}{9}) (\bibinfo{year}{2025}) \bibinfo{pages}{093006},
  \bibinfo{doi}{\doi{10.1103/PhysRevD.111.093006}}, \eprint{2503.07752}.

\bibtype{Article}%
\bibitem{Esteban:2026phq}
\bibinfo{author}{Ivan Esteban}, \bibinfo{author}{M.~C. Gonzalez-Garcia},
  \bibinfo{author}{Michele Maltoni}, \bibinfo{author}{Ivan Martinez-Soler},
  \bibinfo{author}{Joao~Paulo Pinheiro}, \bibinfo{author}{Thomas Schwetz},
  \bibinfo{title}{{Lessons from the first JUNO results}}
  (\bibinfo{year}{2026}), \eprint{2601.09791}.

\bibtype{Article}%
\bibitem{DUNE:2022aul}
\bibinfo{author}{A. Abed~Abud}, et al. (\bibinfo{collaboration}{DUNE}),
  \bibinfo{title}{{Snowmass Neutrino Frontier: DUNE Physics Summary}}
  (\bibinfo{year}{2022}), \eprint{2203.06100}.

\bibtype{Article}%
\bibitem{eBOSS:2020yzd}
\bibinfo{author}{Shadab Alam}, et al. (\bibinfo{collaboration}{eBOSS}),
  \bibinfo{title}{{Completed SDSS-IV extended Baryon Oscillation Spectroscopic
  Survey: Cosmological implications from two decades of spectroscopic surveys
  at the Apache Point Observatory}}, \bibinfo{journal}{Phys. Rev. D}
  \bibinfo{volume}{103} (\bibinfo{number}{8}) (\bibinfo{year}{2021})
  \bibinfo{pages}{083533}, \bibinfo{doi}{\doi{10.1103/PhysRevD.103.083533}},
  \eprint{2007.08991}.

\bibtype{Article}%
\bibitem{DES:2021wwk}
\bibinfo{author}{T.~M.~C. Abbott}, et al. (\bibinfo{collaboration}{DES}),
  \bibinfo{title}{{Dark Energy Survey Year 3 results: Cosmological constraints
  from galaxy clustering and weak lensing}}, \bibinfo{journal}{Phys. Rev. D}
  \bibinfo{volume}{105} (\bibinfo{number}{2}) (\bibinfo{year}{2022})
  \bibinfo{pages}{023520}, \bibinfo{doi}{\doi{10.1103/PhysRevD.105.023520}},
  \eprint{2105.13549}.

\bibtype{Article}%
\bibitem{DESI:2024hhd}
\bibinfo{author}{A.~G. Adame}, et al. (\bibinfo{collaboration}{DESI}),
  \bibinfo{title}{{DESI 2024 VII: cosmological constraints from the full-shape
  modeling of clustering measurements}}, \bibinfo{journal}{JCAP}
  \bibinfo{volume}{07} (\bibinfo{year}{2025}) \bibinfo{pages}{028},
  \bibinfo{doi}{\doi{10.1088/1475-7516/2025/07/028}}, \eprint{2411.12022}.

\bibtype{Article}%
\bibitem{Long:2014zva}
\bibinfo{author}{Andrew~J. Long}, \bibinfo{author}{Cecilia Lunardini},
  \bibinfo{author}{Eray Sabancilar}, \bibinfo{title}{{Detecting
  non-relativistic cosmic neutrinos by capture on tritium: phenomenology and
  physics potential}}, \bibinfo{journal}{JCAP} \bibinfo{volume}{08}
  (\bibinfo{year}{2014}) \bibinfo{pages}{038},
  \bibinfo{doi}{\doi{10.1088/1475-7516/2014/08/038}}, \eprint{1405.7654}.

\bibtype{Article}%
\bibitem{KamLAND:2013rgu}
\bibinfo{author}{A. Gando}, et al. (\bibinfo{collaboration}{KamLAND}),
  \bibinfo{title}{{Reactor On-Off Antineutrino Measurement with KamLAND}},
  \bibinfo{journal}{Phys. Rev. D} \bibinfo{volume}{88} (\bibinfo{number}{3})
  (\bibinfo{year}{2013}) \bibinfo{pages}{033001},
  \bibinfo{doi}{\doi{10.1103/PhysRevD.88.033001}}, \eprint{1303.4667}.

\bibtype{Article}%
\bibitem{JUNO:2022mxj}
\bibinfo{author}{Angel Abusleme}, et al. (\bibinfo{collaboration}{JUNO}),
  \bibinfo{title}{{Sub-percent precision measurement of neutrino oscillation
  parameters with JUNO}}, \bibinfo{journal}{Chin. Phys. C} \bibinfo{volume}{46}
  (\bibinfo{number}{12}) (\bibinfo{year}{2022}) \bibinfo{pages}{123001},
  \bibinfo{doi}{\doi{10.1088/1674-1137/ac8bc9}}, \eprint{2204.13249}.

\bibtype{Article}%
\bibitem{Denton:2023zwa}
\bibinfo{author}{Peter~B. Denton}, \bibinfo{author}{Julia Gehrlein},
  \bibinfo{title}{{Solar parameters in long-baseline accelerator neutrino
  oscillations}}, \bibinfo{journal}{JHEP} \bibinfo{volume}{06}
  (\bibinfo{year}{2023}) \bibinfo{pages}{090},
  \bibinfo{doi}{\doi{10.1007/JHEP06(2023)090}}, \eprint{2302.08513}.

\bibtype{Article}%
\bibitem{Denton:2023qmd}
\bibinfo{author}{Peter~B. Denton}, \bibinfo{title}{{Probing CP Violation with
  Neutrino Disappearance Alone}}, \bibinfo{journal}{Phys. Rev. Lett.}
  \bibinfo{volume}{133} (\bibinfo{number}{3}) (\bibinfo{year}{2024})
  \bibinfo{pages}{031801}, \bibinfo{doi}{\doi{10.1103/PhysRevLett.133.031801}},
  \eprint{2309.03262}.

\bibtype{Article}%
\bibitem{BOREXINO:2018ohr}
\bibinfo{author}{M. Agostini}, et al. (\bibinfo{collaboration}{BOREXINO}),
  \bibinfo{title}{{Comprehensive measurement of $pp$-chain solar neutrinos}},
  \bibinfo{journal}{Nature} \bibinfo{volume}{562} (\bibinfo{number}{7728})
  (\bibinfo{year}{2018}) \bibinfo{pages}{505--510},
  \bibinfo{doi}{\doi{10.1038/s41586-018-0624-y}}.

\bibtype{Article}%
\bibitem{Albright:2006cw}
\bibinfo{author}{Carl~H. Albright}, \bibinfo{author}{Mu-Chun Chen},
  \bibinfo{title}{{Model Predictions for Neutrino Oscillation Parameters}},
  \bibinfo{journal}{Phys. Rev. D} \bibinfo{volume}{74} (\bibinfo{year}{2006})
  \bibinfo{pages}{113006}, \bibinfo{doi}{\doi{10.1103/PhysRevD.74.113006}},
  \eprint{hep-ph/0608137}.

\bibtype{Article}%
\bibitem{DayaBay:2018yms}
\bibinfo{author}{D. Adey}, et al. (\bibinfo{collaboration}{Daya Bay}),
  \bibinfo{title}{{Measurement of the Electron Antineutrino Oscillation with
  1958 Days of Operation at Daya Bay}}, \bibinfo{journal}{Phys. Rev. Lett.}
  \bibinfo{volume}{121} (\bibinfo{number}{24}) (\bibinfo{year}{2018})
  \bibinfo{pages}{241805}, \bibinfo{doi}{\doi{10.1103/PhysRevLett.121.241805}},
  \eprint{1809.02261}.

\bibtype{Article}%
\bibitem{RENO:2018dro}
\bibinfo{author}{G. Bak}, et al. (\bibinfo{collaboration}{RENO}),
  \bibinfo{title}{{Measurement of Reactor Antineutrino Oscillation Amplitude
  and Frequency at RENO}}, \bibinfo{journal}{Phys. Rev. Lett.}
  \bibinfo{volume}{121} (\bibinfo{number}{20}) (\bibinfo{year}{2018})
  \bibinfo{pages}{201801}, \bibinfo{doi}{\doi{10.1103/PhysRevLett.121.201801}},
  \eprint{1806.00248}.

\bibtype{Article}%
\bibitem{DoubleChooz:2019qbj}
\bibinfo{author}{H. de Kerret}, et al. (\bibinfo{collaboration}{Double Chooz}),
  \bibinfo{title}{{Double Chooz $\theta_{13}$ measurement via total neutron
  capture detection}}, \bibinfo{journal}{Nature Phys.} \bibinfo{volume}{16}
  (\bibinfo{number}{5}) (\bibinfo{year}{2020}) \bibinfo{pages}{558--564},
  \bibinfo{doi}{\doi{10.1038/s41567-020-0831-y}}, \eprint{1901.09445}.

\bibtype{Article}%
\bibitem{Balantekin:2008zm}
\bibinfo{author}{A.~B. Balantekin}, \bibinfo{author}{D. Yilmaz},
  \bibinfo{title}{{Contrasting solar and reactor neutrinos with a non-zero
  value of theta(13)}}, \bibinfo{journal}{J. Phys. G} \bibinfo{volume}{35}
  (\bibinfo{year}{2008}) \bibinfo{pages}{075007},
  \bibinfo{doi}{\doi{10.1088/0954-3899/35/7/075007}}, \eprint{0804.3345}.

\bibtype{Article}%
\bibitem{Fogli:2011qn}
\bibinfo{author}{G.~L. Fogli}, \bibinfo{author}{E. Lisi}, \bibinfo{author}{A.
  Marrone}, \bibinfo{author}{A. Palazzo}, \bibinfo{author}{A.~M. Rotunno},
  \bibinfo{title}{{Evidence of $\theta_{13}$\ensuremath{>}0 from global
  neutrino data analysis}}, \bibinfo{journal}{Phys. Rev. D}
  \bibinfo{volume}{84} (\bibinfo{year}{2011}) \bibinfo{pages}{053007},
  \bibinfo{doi}{\doi{10.1103/PhysRevD.84.053007}}, \eprint{1106.6028}.

\bibtype{Article}%
\bibitem{Denton:2020uda}
\bibinfo{author}{Peter~B. Denton}, \bibinfo{author}{Julia Gehrlein},
  \bibinfo{author}{Rebekah Pestes}, \bibinfo{title}{{$CP$ -Violating Neutrino
  Nonstandard Interactions in Long-Baseline-Accelerator Data}},
  \bibinfo{journal}{Phys. Rev. Lett.} \bibinfo{volume}{126}
  (\bibinfo{number}{5}) (\bibinfo{year}{2021}) \bibinfo{pages}{051801},
  \bibinfo{doi}{\doi{10.1103/PhysRevLett.126.051801}}, \eprint{2008.01110}.

\bibtype{Article}%
\bibitem{Kaplan:1993ej}
\bibinfo{author}{David~B. Kaplan}, \bibinfo{author}{Martin Schmaltz},
  \bibinfo{title}{{Flavor unification and discrete nonAbelian symmetries}},
  \bibinfo{journal}{Phys. Rev. D} \bibinfo{volume}{49} (\bibinfo{year}{1994})
  \bibinfo{pages}{3741--3750}, \bibinfo{doi}{\doi{10.1103/PhysRevD.49.3741}},
  \eprint{hep-ph/9311281}.

\bibtype{Article}%
\bibitem{Ishimori:2010au}
\bibinfo{author}{Hajime Ishimori}, \bibinfo{author}{Tatsuo Kobayashi},
  \bibinfo{author}{Hiroshi Ohki}, \bibinfo{author}{Yusuke Shimizu},
  \bibinfo{author}{Hiroshi Okada}, \bibinfo{author}{Morimitsu Tanimoto},
  \bibinfo{title}{{Non-Abelian Discrete Symmetries in Particle Physics}},
  \bibinfo{journal}{Prog. Theor. Phys. Suppl.} \bibinfo{volume}{183}
  (\bibinfo{year}{2010}) \bibinfo{pages}{1--163},
  \bibinfo{doi}{\doi{10.1143/PTPS.183.1}}, \eprint{1003.3552}.

\bibtype{Article}%
\bibitem{Feruglio:2019ybq}
\bibinfo{author}{Ferruccio Feruglio}, \bibinfo{author}{Andrea Romanino},
  \bibinfo{title}{{Lepton flavor symmetries}}, \bibinfo{journal}{Rev. Mod.
  Phys.} \bibinfo{volume}{93} (\bibinfo{number}{1}) (\bibinfo{year}{2021})
  \bibinfo{pages}{015007}, \bibinfo{doi}{\doi{10.1103/RevModPhys.93.015007}},
  \eprint{1912.06028}.

\bibtype{Article}%
\bibitem{Almumin:2022rml}
\bibinfo{author}{Yahya Almumin}, \bibinfo{author}{Mu-Chun Chen},
  \bibinfo{author}{Murong Cheng}, \bibinfo{author}{Victor Knapp-Perez},
  \bibinfo{author}{Yulun Li}, \bibinfo{author}{Adreja Mondol},
  \bibinfo{author}{Saul Ramos-Sanchez}, \bibinfo{author}{Michael Ratz},
  \bibinfo{author}{Shreya Shukla}, \bibinfo{title}{{Neutrino Flavor Model
  Building and the Origins of Flavor and CP Violation}},
  \bibinfo{journal}{Universe} \bibinfo{volume}{9} (\bibinfo{number}{12})
  (\bibinfo{year}{2023}) \bibinfo{pages}{512},
  \bibinfo{doi}{\doi{10.3390/universe9120512}}, \eprint{2204.08668}.

\bibtype{Article}%
\bibitem{Chauhan:2023faf}
\bibinfo{author}{Garv Chauhan}, \bibinfo{author}{P.~S.~Bhupal Dev},
  \bibinfo{author}{Ievgen Dubovyk}, \bibinfo{author}{Bartosz Dziewit},
  \bibinfo{author}{Wojciech Flieger}, \bibinfo{author}{Krzysztof Grzanka},
  \bibinfo{author}{Janusz Gluza}, \bibinfo{author}{Biswajit Karmakar},
  \bibinfo{author}{Szymon Zi\k{e}ba}, \bibinfo{title}{{Phenomenology of lepton
  masses and mixing with discrete flavor symmetries}}, \bibinfo{journal}{Prog.
  Part. Nucl. Phys.} \bibinfo{volume}{138} (\bibinfo{year}{2024})
  \bibinfo{pages}{104126}, \bibinfo{doi}{\doi{10.1016/j.ppnp.2024.104126}},
  \eprint{2310.20681}.

\bibtype{Article}%
\bibitem{Denton:2023hkx}
\bibinfo{author}{Peter~B. Denton}, \bibinfo{author}{Julia Gehrlein},
  \bibinfo{title}{{Survey of neutrino flavor predictions and the neutrinoless
  double beta decay funnel}}, \bibinfo{journal}{Phys. Rev. D}
  \bibinfo{volume}{109} (\bibinfo{number}{5}) (\bibinfo{year}{2024})
  \bibinfo{pages}{055028}, \bibinfo{doi}{\doi{10.1103/PhysRevD.109.055028}},
  \eprint{2308.09737}.

\bibtype{Article}%
\bibitem{Harrison:2002er}
\bibinfo{author}{P.~F. Harrison}, \bibinfo{author}{D.~H. Perkins},
  \bibinfo{author}{W.~G. Scott}, \bibinfo{title}{{Tri-bimaximal mixing and the
  neutrino oscillation data}}, \bibinfo{journal}{Phys. Lett. B}
  \bibinfo{volume}{530} (\bibinfo{year}{2002}) \bibinfo{pages}{167},
  \bibinfo{doi}{\doi{10.1016/S0370-2693(02)01336-9}}, \eprint{hep-ph/0202074}.

\bibtype{Article}%
\bibitem{Harrison:2002kp}
\bibinfo{author}{P.~F. Harrison}, \bibinfo{author}{W.~G. Scott},
  \bibinfo{title}{{Symmetries and generalizations of tri - bimaximal neutrino
  mixing}}, \bibinfo{journal}{Phys. Lett. B} \bibinfo{volume}{535}
  (\bibinfo{year}{2002}) \bibinfo{pages}{163--169},
  \bibinfo{doi}{\doi{10.1016/S0370-2693(02)01753-7}}, \eprint{hep-ph/0203209}.

\bibtype{Article}%
\bibitem{Harrison:2003aw}
\bibinfo{author}{P.~F. Harrison}, \bibinfo{author}{W.~G. Scott},
  \bibinfo{title}{{Permutation symmetry, tri - bimaximal neutrino mixing and
  the S3 group characters}}, \bibinfo{journal}{Phys. Lett. B}
  \bibinfo{volume}{557} (\bibinfo{year}{2003}) \bibinfo{pages}{76},
  \bibinfo{doi}{\doi{10.1016/S0370-2693(03)00183-7}}, \eprint{hep-ph/0302025}.

\bibtype{Article}%
\bibitem{King:2007pr}
\bibinfo{author}{S.~F. King}, \bibinfo{title}{{Parametrizing the lepton mixing
  matrix in terms of deviations from tri-bimaximal mixing}},
  \bibinfo{journal}{Phys. Lett. B} \bibinfo{volume}{659} (\bibinfo{year}{2008})
  \bibinfo{pages}{244--251},
  \bibinfo{doi}{\doi{10.1016/j.physletb.2007.10.078}}, \eprint{0710.0530}.

\bibtype{Article}%
\bibitem{Pakvasa:2007zj}
\bibinfo{author}{Sandip Pakvasa}, \bibinfo{author}{Werner Rodejohann},
  \bibinfo{author}{Thomas~J. Weiler}, \bibinfo{title}{{Unitary parametrization
  of perturbations to tribimaximal neutrino mixing}}, \bibinfo{journal}{Phys.
  Rev. Lett.} \bibinfo{volume}{100} (\bibinfo{year}{2008})
  \bibinfo{pages}{111801}, \bibinfo{doi}{\doi{10.1103/PhysRevLett.100.111801}},
  \eprint{0711.0052}.

\bibtype{Article}%
\bibitem{King:2012vj}
\bibinfo{author}{S.~F. King}, \bibinfo{title}{{Tri-bimaximal-Cabibbo Mixing}},
  \bibinfo{journal}{Phys. Lett. B} \bibinfo{volume}{718} (\bibinfo{year}{2012})
  \bibinfo{pages}{136--142},
  \bibinfo{doi}{\doi{10.1016/j.physletb.2012.10.028}}, \eprint{1205.0506}.

\bibtype{Article}%
\bibitem{Giunti:2002ye}
\bibinfo{author}{Carlo Giunti}, \bibinfo{author}{Morimitsu Tanimoto},
  \bibinfo{title}{{Deviation of neutrino mixing from bimaximal}},
  \bibinfo{journal}{Phys. Rev. D} \bibinfo{volume}{66} (\bibinfo{year}{2002})
  \bibinfo{pages}{053013}, \bibinfo{doi}{\doi{10.1103/PhysRevD.66.053013}},
  \eprint{hep-ph/0207096}.

\bibtype{Article}%
\bibitem{Minakata:2004xt}
\bibinfo{author}{Hisakazu Minakata}, \bibinfo{author}{Alexei~Yu. Smirnov},
  \bibinfo{title}{{Neutrino mixing and quark-lepton complementarity}},
  \bibinfo{journal}{Phys. Rev. D} \bibinfo{volume}{70} (\bibinfo{year}{2004})
  \bibinfo{pages}{073009}, \bibinfo{doi}{\doi{10.1103/PhysRevD.70.073009}},
  \eprint{hep-ph/0405088}.

\bibtype{Article}%
\bibitem{Datta:2005ci}
\bibinfo{author}{AseshKrishna Datta}, \bibinfo{author}{Lisa Everett},
  \bibinfo{author}{Pierre Ramond}, \bibinfo{title}{{Cabibbo haze in lepton
  mixing}}, \bibinfo{journal}{Phys. Lett. B} \bibinfo{volume}{620}
  (\bibinfo{year}{2005}) \bibinfo{pages}{42--51},
  \bibinfo{doi}{\doi{10.1016/j.physletb.2005.05.075}}, \eprint{hep-ph/0503222}.

\bibtype{Article}%
\bibitem{Everett:2005ku}
\bibinfo{author}{Lisa~L. Everett}, \bibinfo{title}{{Viewing lepton mixing
  through the Cabibbo haze}}, \bibinfo{journal}{Phys. Rev. D}
  \bibinfo{volume}{73} (\bibinfo{year}{2006}) \bibinfo{pages}{013011},
  \bibinfo{doi}{\doi{10.1103/PhysRevD.73.013011}}, \eprint{hep-ph/0510256}.

\bibtype{Article}%
\bibitem{King:2013eh}
\bibinfo{author}{Stephen~F. King}, \bibinfo{author}{Christoph Luhn},
  \bibinfo{title}{{Neutrino Mass and Mixing with Discrete Symmetry}},
  \bibinfo{journal}{Rept. Prog. Phys.} \bibinfo{volume}{76}
  (\bibinfo{year}{2013}) \bibinfo{pages}{056201},
  \bibinfo{doi}{\doi{10.1088/0034-4885/76/5/056201}}, \eprint{1301.1340}.

\bibtype{Article}%
\bibitem{King:2014nza}
\bibinfo{author}{Stephen~F. King}, \bibinfo{author}{Alexander Merle},
  \bibinfo{author}{Stefano Morisi}, \bibinfo{author}{Yusuke Shimizu},
  \bibinfo{author}{Morimitsu Tanimoto}, \bibinfo{title}{{Neutrino Mass and
  Mixing: from Theory to Experiment}}, \bibinfo{journal}{New J. Phys.}
  \bibinfo{volume}{16} (\bibinfo{year}{2014}) \bibinfo{pages}{045018},
  \bibinfo{doi}{\doi{10.1088/1367-2630/16/4/045018}}, \eprint{1402.4271}.

\bibtype{Article}%
\bibitem{Petcov:2017ggy}
\bibinfo{author}{S.~T. Petcov}, \bibinfo{title}{{Discrete Flavour Symmetries,
  Neutrino Mixing and Leptonic CP Violation}}, \bibinfo{journal}{Eur. Phys. J.
  C} \bibinfo{volume}{78} (\bibinfo{number}{9}) (\bibinfo{year}{2018})
  \bibinfo{pages}{709}, \bibinfo{doi}{\doi{10.1140/epjc/s10052-018-6158-5}},
  \eprint{1711.10806}.

\bibtype{Article}%
\bibitem{Merle:2006du}
\bibinfo{author}{Alexander Merle}, \bibinfo{author}{Werner Rodejohann},
  \bibinfo{title}{{The Elements of the neutrino mass matrix: Allowed ranges and
  implications of texture zeros}}, \bibinfo{journal}{Phys. Rev. D}
  \bibinfo{volume}{73} (\bibinfo{year}{2006}) \bibinfo{pages}{073012},
  \bibinfo{doi}{\doi{10.1103/PhysRevD.73.073012}}, \eprint{hep-ph/0603111}.

\bibtype{Article}%
\bibitem{Lashin:2011dn}
\bibinfo{author}{E.~I. Lashin}, \bibinfo{author}{N. Chamoun},
  \bibinfo{title}{{The One-zero Textures of Majorana Neutrino Mass Matrix and
  Current Experimental Tests}}, \bibinfo{journal}{Phys. Rev. D}
  \bibinfo{volume}{85} (\bibinfo{year}{2012}) \bibinfo{pages}{113011},
  \bibinfo{doi}{\doi{10.1103/PhysRevD.85.113011}}, \eprint{1108.4010}.

\bibtype{Article}%
\bibitem{Dev:2006qe}
\bibinfo{author}{S. Dev}, \bibinfo{author}{Sanjeev Kumar},
  \bibinfo{author}{Surender Verma}, \bibinfo{author}{Shivani Gupta},
  \bibinfo{title}{{Phenomenology of two-texture zero neutrino mass matrices}},
  \bibinfo{journal}{Phys. Rev. D} \bibinfo{volume}{76} (\bibinfo{year}{2007})
  \bibinfo{pages}{013002}, \bibinfo{doi}{\doi{10.1103/PhysRevD.76.013002}},
  \eprint{hep-ph/0612102}.

\bibtype{Article}%
\bibitem{Verma:2020gpl}
\bibinfo{author}{Surender Verma}, \bibinfo{author}{Monal Kashav},
  \bibinfo{title}{{Ramifications of texture one-zero neutrino mass model in
  coherence with the latest neutrino data}}, \bibinfo{journal}{Mod. Phys. Lett.
  A} \bibinfo{volume}{35} (\bibinfo{number}{20}) (\bibinfo{year}{2020})
  \bibinfo{pages}{2050165}, \bibinfo{doi}{\doi{10.1142/S0217732320501655}}.

\bibtype{Article}%
\bibitem{Xing:2003jf}
\bibinfo{author}{Zhi-zhong Xing}, \bibinfo{title}{{Vanishing effective mass of
  the neutrinoless double beta decay?}}, \bibinfo{journal}{Phys. Rev. D}
  \bibinfo{volume}{68} (\bibinfo{year}{2003}) \bibinfo{pages}{053002},
  \bibinfo{doi}{\doi{10.1103/PhysRevD.68.053002}}, \eprint{hep-ph/0305195}.

\bibtype{Article}%
\bibitem{Xing:2003ic}
\bibinfo{author}{Zhi-zhong Xing}, \bibinfo{title}{{Implications of generalized
  Frampton-Glashow-Yanagida ansaetze on neutrino masses and lepton flavor
  mixing}}, \bibinfo{journal}{Phys. Rev. D} \bibinfo{volume}{69}
  (\bibinfo{year}{2004}) \bibinfo{pages}{013006},
  \bibinfo{doi}{\doi{10.1103/PhysRevD.69.013006}}, \eprint{hep-ph/0307007}.

\bibtype{Article}%
\bibitem{Grimus:2004hf}
\bibinfo{author}{Walter Grimus}, \bibinfo{author}{Anjan~S. Joshipura},
  \bibinfo{author}{Luis Lavoura}, \bibinfo{author}{Morimitsu Tanimoto},
  \bibinfo{title}{{Symmetry realization of texture zeros}},
  \bibinfo{journal}{Eur. Phys. J. C} \bibinfo{volume}{36}
  (\bibinfo{year}{2004}) \bibinfo{pages}{227--232},
  \bibinfo{doi}{\doi{10.1140/epjc/s2004-01896-y}}, \eprint{hep-ph/0405016}.

\bibtype{Article}%
\bibitem{BenTov:2011tj}
\bibinfo{author}{Yoni BenTov}, \bibinfo{author}{A. Zee},
  \bibinfo{title}{{Neutrino Mass Matrices with $M_{ee} = 0$}},
  \bibinfo{journal}{Phys. Rev. D} \bibinfo{volume}{84} (\bibinfo{year}{2011})
  \bibinfo{pages}{073012}, \bibinfo{doi}{\doi{10.1103/PhysRevD.84.073012}},
  \eprint{1103.2616}.

\bibtype{Article}%
\bibitem{Liu:2012axa}
\bibinfo{author}{Xue-wen Liu}, \bibinfo{author}{Shun Zhou},
  \bibinfo{title}{{Texture Zeros for Dirac Neutrinos and Current Experimental
  Tests}}, \bibinfo{journal}{Int. J. Mod. Phys. A} \bibinfo{volume}{28}
  (\bibinfo{year}{2013}) \bibinfo{pages}{1350040},
  \bibinfo{doi}{\doi{10.1142/S0217751X13500401}}, \eprint{1211.0472}.

\bibtype{Article}%
\bibitem{Frampton:2002yf}
\bibinfo{author}{Paul~H. Frampton}, \bibinfo{author}{Sheldon~L. Glashow},
  \bibinfo{author}{Danny Marfatia}, \bibinfo{title}{{Zeroes of the neutrino
  mass matrix}}, \bibinfo{journal}{Phys. Lett. B} \bibinfo{volume}{536}
  (\bibinfo{year}{2002}) \bibinfo{pages}{79--82},
  \bibinfo{doi}{\doi{10.1016/S0370-2693(02)01817-8}}, \eprint{hep-ph/0201008}.

\bibtype{Article}%
\bibitem{Guo:2002ei}
\bibinfo{author}{Wan-lei Guo}, \bibinfo{author}{Zhi-zhong Xing},
  \bibinfo{title}{{Implications of the KamLAND measurement on the lepton flavor
  mixing matrix and the neutrino mass matrix}}, \bibinfo{journal}{Phys. Rev. D}
  \bibinfo{volume}{67} (\bibinfo{year}{2003}) \bibinfo{pages}{053002},
  \bibinfo{doi}{\doi{10.1103/PhysRevD.67.053002}}, \eprint{hep-ph/0212142}.

\bibtype{Article}%
\bibitem{Xing:2002ta}
\bibinfo{author}{Zhi-zhong Xing}, \bibinfo{title}{{Texture zeros and Majorana
  phases of the neutrino mass matrix}}, \bibinfo{journal}{Phys. Lett. B}
  \bibinfo{volume}{530} (\bibinfo{year}{2002}) \bibinfo{pages}{159--166},
  \bibinfo{doi}{\doi{10.1016/S0370-2693(02)01354-0}}, \eprint{hep-ph/0201151}.

\bibtype{Article}%
\bibitem{Desai:2002sz}
\bibinfo{author}{Bipin~R. Desai}, \bibinfo{author}{D.~P. Roy},
  \bibinfo{author}{Alexander~R. Vaucher}, \bibinfo{title}{{Three neutrino mass
  matrices with two texture zeros}}, \bibinfo{journal}{Mod. Phys. Lett. A}
  \bibinfo{volume}{18} (\bibinfo{year}{2003}) \bibinfo{pages}{1355--1366},
  \bibinfo{doi}{\doi{10.1142/S0217732303011071}}, \eprint{hep-ph/0209035}.

\bibtype{Article}%
\bibitem{Xing:2002ap}
\bibinfo{author}{Zhi-zhong Xing}, \bibinfo{title}{{A Full determination of the
  neutrino mass spectrum from two zero textures of the neutrino mass matrix}},
  \bibinfo{journal}{Phys. Lett. B} \bibinfo{volume}{539} (\bibinfo{year}{2002})
  \bibinfo{pages}{85--90}, \bibinfo{doi}{\doi{10.1016/S0370-2693(02)02062-2}},
  \eprint{hep-ph/0205032}.

\bibtype{Article}%
\bibitem{Dev:2006if}
\bibinfo{author}{S. Dev}, \bibinfo{author}{Sanjeev Kumar},
  \bibinfo{title}{{Neutrino Parameter Space for a Vanishing ee Element in the
  Neutrino Mass Matrix}}, \bibinfo{journal}{Mod. Phys. Lett. A}
  \bibinfo{volume}{22} (\bibinfo{year}{2007}) \bibinfo{pages}{1401--1410},
  \bibinfo{doi}{\doi{10.1142/S0217732307021767}}, \eprint{hep-ph/0607048}.

\bibtype{Article}%
\bibitem{Dev:2006xu}
\bibinfo{author}{S. Dev}, \bibinfo{author}{Sanjeev Kumar},
  \bibinfo{author}{Surender Verma}, \bibinfo{author}{Shivani Gupta},
  \bibinfo{title}{{Phenomenological implications of a class of neutrino mass
  matrices}}, \bibinfo{journal}{Nucl. Phys. B} \bibinfo{volume}{784}
  (\bibinfo{year}{2007}) \bibinfo{pages}{103--117},
  \bibinfo{doi}{\doi{10.1016/j.nuclphysb.2007.06.030}},
  \eprint{hep-ph/0611313}.

\bibtype{Article}%
\bibitem{Fritzsch:2011qv}
\bibinfo{author}{Harald Fritzsch}, \bibinfo{author}{Zhi-zhong Xing},
  \bibinfo{author}{Shun Zhou}, \bibinfo{title}{{Two-zero Textures of the
  Majorana Neutrino Mass Matrix and Current Experimental Tests}},
  \bibinfo{journal}{JHEP} \bibinfo{volume}{09} (\bibinfo{year}{2011})
  \bibinfo{pages}{083}, \bibinfo{doi}{\doi{10.1007/JHEP09(2011)083}},
  \eprint{1108.4534}.

\bibtype{Article}%
\bibitem{Honda:2003pg}
\bibinfo{author}{Mizue Honda}, \bibinfo{author}{Satoru Kaneko},
  \bibinfo{author}{Morimitsu Tanimoto}, \bibinfo{title}{{Prediction and its
  stability in neutrino mass matrix with two zeros}}, \bibinfo{journal}{JHEP}
  \bibinfo{volume}{09} (\bibinfo{year}{2003}) \bibinfo{pages}{028},
  \bibinfo{doi}{\doi{10.1088/1126-6708/2003/09/028}}, \eprint{hep-ph/0303227}.

\bibtype{Article}%
\bibitem{Hollik:2017get}
\bibinfo{author}{Wolfgang~G. Hollik}, \bibinfo{author}{Ulises~J.
  Saldana-Salazar}, \bibinfo{title}{{Texture zeros and hierarchical masses from
  flavour (mis)alignment}}, \bibinfo{journal}{Nucl. Phys. B}
  \bibinfo{volume}{928} (\bibinfo{year}{2018}) \bibinfo{pages}{535--554},
  \bibinfo{doi}{\doi{10.1016/j.nuclphysb.2018.01.030}}, \eprint{1712.05387}.

\bibtype{Inbook}%
\bibitem{Feruglio:2017spp}
\bibinfo{author}{Ferruccio Feruglio}, \bibinfo{title}{{Are neutrino masses
  modular forms?}} \bibinfo{year}{2019} pp. \bibinfo{pages}{227--266},
  \bibinfo{doi}{\doi{10.1142/9789813238053_0012}}, \eprint{1706.08749}.

\bibtype{Article}%
\bibitem{Penedo:2018nmg}
\bibinfo{author}{J.~T. Penedo}, \bibinfo{author}{S.~T. Petcov},
  \bibinfo{title}{{Lepton Masses and Mixing from Modular $S_4$ Symmetry}},
  \bibinfo{journal}{Nucl. Phys. B} \bibinfo{volume}{939} (\bibinfo{year}{2019})
  \bibinfo{pages}{292--307},
  \bibinfo{doi}{\doi{10.1016/j.nuclphysb.2018.12.016}}, \eprint{1806.11040}.

\bibtype{Article}%
\bibitem{Kobayashi:2018vbk}
\bibinfo{author}{Tatsuo Kobayashi}, \bibinfo{author}{Kentaro Tanaka},
  \bibinfo{author}{Takuya~H. Tatsuishi}, \bibinfo{title}{{Neutrino mixing from
  finite modular groups}}, \bibinfo{journal}{Phys. Rev. D} \bibinfo{volume}{98}
  (\bibinfo{number}{1}) (\bibinfo{year}{2018}) \bibinfo{pages}{016004},
  \bibinfo{doi}{\doi{10.1103/PhysRevD.98.016004}}, \eprint{1803.10391}.

\bibtype{Article}%
\bibitem{Gehrlein:2020jnr}
\bibinfo{author}{J. Gehrlein}, \bibinfo{author}{M. Spinrath},
  \bibinfo{title}{{Leptonic Sum Rules from Flavour Models with Modular
  Symmetries}}, \bibinfo{journal}{JHEP} \bibinfo{volume}{03}
  (\bibinfo{year}{2021}) \bibinfo{pages}{177},
  \bibinfo{doi}{\doi{10.1007/JHEP03(2021)177}}, \eprint{2012.04131}.

\bibtype{Article}%
\bibitem{Kikuchi:2021ogn}
\bibinfo{author}{Shota Kikuchi}, \bibinfo{author}{Tatsuo Kobayashi},
  \bibinfo{author}{Hikaru Uchida}, \bibinfo{title}{{Modular flavor symmetries
  of three-generation modes on magnetized toroidal orbifolds}},
  \bibinfo{journal}{Phys. Rev. D} \bibinfo{volume}{104} (\bibinfo{number}{6})
  (\bibinfo{year}{2021}) \bibinfo{pages}{065008},
  \bibinfo{doi}{\doi{10.1103/PhysRevD.104.065008}}, \eprint{2101.00826}.

\bibtype{Article}%
\bibitem{Liu:2021gwa}
\bibinfo{author}{Xiang-Gan Liu}, \bibinfo{author}{Gui-Jun Ding},
  \bibinfo{title}{{Modular flavor symmetry and vector-valued modular forms}},
  \bibinfo{journal}{JHEP} \bibinfo{volume}{03} (\bibinfo{year}{2022})
  \bibinfo{pages}{123}, \bibinfo{doi}{\doi{10.1007/JHEP03(2022)123}},
  \eprint{2112.14761}.

\bibtype{Article}%
\bibitem{Kobayashi:2023zzc}
\bibinfo{author}{Tatsuo Kobayashi}, \bibinfo{author}{Morimitsu Tanimoto},
  \bibinfo{title}{{Modular flavor symmetric models}}, \bibinfo{journal}{Int. J.
  Mod. Phys. A} \bibinfo{volume}{39} (\bibinfo{number}{09n10})
  (\bibinfo{year}{2024}) \bibinfo{pages}{2441012},
  \bibinfo{doi}{\doi{10.1142/S0217751X24410124}}, \eprint{2307.03384}.

\bibtype{Article}%
\bibitem{Wolfenstein:1981rk}
\bibinfo{author}{Lincoln Wolfenstein}, \bibinfo{title}{{CP Properties of
  Majorana Neutrinos and Double beta Decay}}, \bibinfo{journal}{Phys. Lett. B}
  \bibinfo{volume}{107} (\bibinfo{year}{1981}) \bibinfo{pages}{77--79},
  \bibinfo{doi}{\doi{10.1016/0370-2693(81)91151-5}}.

\bibtype{Article}%
\bibitem{Kayser:1984ge}
\bibinfo{author}{Boris Kayser}, \bibinfo{title}{{CPT, CP, and c Phases and
  their Effects in Majorana Particle Processes}}, \bibinfo{journal}{Phys. Rev.
  D} \bibinfo{volume}{30} (\bibinfo{year}{1984}) \bibinfo{pages}{1023},
  \bibinfo{doi}{\doi{10.1103/PhysRevD.30.1023}}.

\bibtype{Article}%
\bibitem{Bilenky:1984fg}
\bibinfo{author}{Samoil~M. Bilenky}, \bibinfo{author}{N.~P. Nedelcheva},
  \bibinfo{author}{S.~T. Petcov}, \bibinfo{title}{{Some Implications of the
  {CP} Invariance for Mixing of Majorana Neutrinos}}, \bibinfo{journal}{Nucl.
  Phys. B} \bibinfo{volume}{247} (\bibinfo{year}{1984})
  \bibinfo{pages}{61--69}, \bibinfo{doi}{\doi{10.1016/0550-3213(84)90372-9}}.

\bibtype{Article}%
\bibitem{Branco:1986gr}
\bibinfo{author}{G.~C. Branco}, \bibinfo{author}{L. Lavoura},
  \bibinfo{author}{M.~N. Rebelo}, \bibinfo{title}{{Majorana Neutrinos and {CP}
  Violation in the Leptonic Sector}}, \bibinfo{journal}{Phys. Lett. B}
  \bibinfo{volume}{180} (\bibinfo{year}{1986}) \bibinfo{pages}{264--268},
  \bibinfo{doi}{\doi{10.1016/0370-2693(86)90307-2}}.

\bibtype{Article}%
\bibitem{Feruglio:2012cw}
\bibinfo{author}{Ferruccio Feruglio}, \bibinfo{author}{Claudia Hagedorn},
  \bibinfo{author}{Robert Ziegler}, \bibinfo{title}{{Lepton Mixing Parameters
  from Discrete and CP Symmetries}}, \bibinfo{journal}{JHEP}
  \bibinfo{volume}{07} (\bibinfo{year}{2013}) \bibinfo{pages}{027},
  \bibinfo{doi}{\doi{10.1007/JHEP07(2013)027}}, \eprint{1211.5560}.

\bibtype{Article}%
\bibitem{Holthausen:2012dk}
\bibinfo{author}{Martin Holthausen}, \bibinfo{author}{Manfred Lindner},
  \bibinfo{author}{Michael~A. Schmidt}, \bibinfo{title}{{CP and Discrete
  Flavour Symmetries}}, \bibinfo{journal}{JHEP} \bibinfo{volume}{04}
  (\bibinfo{year}{2013}) \bibinfo{pages}{122},
  \bibinfo{doi}{\doi{10.1007/JHEP04(2013)122}}, \eprint{1211.6953}.

\bibtype{Article}%
\bibitem{Ding:2013nsa}
\bibinfo{author}{Gui-Jun Ding}, \bibinfo{author}{Ye-Ling Zhou},
  \bibinfo{title}{{Predicting lepton flavor mixing from $\Delta$(48) and
  generalized $CP$ symmetries}}, \bibinfo{journal}{Chin. Phys. C}
  \bibinfo{volume}{39} (\bibinfo{number}{2}) (\bibinfo{year}{2015})
  \bibinfo{pages}{021001}, \bibinfo{doi}{\doi{10.1088/1674-1137/39/2/021001}},
  \eprint{1312.5222}.

\bibtype{Article}%
\bibitem{Ding:2014hva}
\bibinfo{author}{Gui-Jun Ding}, \bibinfo{author}{Ye-Ling Zhou},
  \bibinfo{title}{{Lepton mixing parameters from $\Delta(48)$ family symmetry
  and generalised CP}}, \bibinfo{journal}{JHEP} \bibinfo{volume}{06}
  (\bibinfo{year}{2014}) \bibinfo{pages}{023},
  \bibinfo{doi}{\doi{10.1007/JHEP06(2014)023}}, \eprint{1404.0592}.

\bibtype{Article}%
\bibitem{King:2014rwa}
\bibinfo{author}{Stephen~F. King}, \bibinfo{author}{Thomas Neder},
  \bibinfo{title}{{Lepton mixing predictions including Majorana phases from
  \ensuremath{\Delta}(6n$^2$) flavour symmetry and generalised CP}},
  \bibinfo{journal}{Phys. Lett. B} \bibinfo{volume}{736} (\bibinfo{year}{2014})
  \bibinfo{pages}{308--316},
  \bibinfo{doi}{\doi{10.1016/j.physletb.2014.07.043}}, \eprint{1403.1758}.

\bibtype{Article}%
\bibitem{Ding:2014ssa}
\bibinfo{author}{Gui-Jun Ding}, \bibinfo{author}{Stephen~F. King},
  \bibinfo{title}{{Generalized $CP$ and $\Delta(96)$ family symmetry}},
  \bibinfo{journal}{Phys. Rev. D} \bibinfo{volume}{89} (\bibinfo{number}{9})
  (\bibinfo{year}{2014}) \bibinfo{pages}{093020},
  \bibinfo{doi}{\doi{10.1103/PhysRevD.89.093020}}, \eprint{1403.5846}.

\bibtype{Article}%
\bibitem{Hagedorn:2014wha}
\bibinfo{author}{Claudia Hagedorn}, \bibinfo{author}{Aurora Meroni},
  \bibinfo{author}{Emiliano Molinaro}, \bibinfo{title}{{Lepton mixing from
  \ensuremath{\Delta}(3$n^2$) and \ensuremath{\Delta}(6$n^2$) and CP}},
  \bibinfo{journal}{Nucl. Phys. B} \bibinfo{volume}{891} (\bibinfo{year}{2015})
  \bibinfo{pages}{499--557},
  \bibinfo{doi}{\doi{10.1016/j.nuclphysb.2014.12.013}}, \eprint{1408.7118}.

\bibtype{Article}%
\bibitem{Ding:2014ora}
\bibinfo{author}{Gui-Jun Ding}, \bibinfo{author}{Stephen~F. King},
  \bibinfo{author}{Thomas Neder}, \bibinfo{title}{{Generalised CP and
  $\Delta(6n^2)$ family symmetry in semi-direct models of leptons}},
  \bibinfo{journal}{JHEP} \bibinfo{volume}{12} (\bibinfo{year}{2014})
  \bibinfo{pages}{007}, \bibinfo{doi}{\doi{10.1007/JHEP12(2014)007}},
  \eprint{1409.8005}.

\bibtype{Article}%
\bibitem{Ding:2015rwa}
\bibinfo{author}{Gui-Jun Ding}, \bibinfo{author}{Stephen~F. King},
  \bibinfo{title}{{Generalized CP and $\Delta (3n^2)$ Family Symmetry for
  Semi-Direct Predictions of the PMNS Matrix}}, \bibinfo{journal}{Phys. Rev. D}
  \bibinfo{volume}{93} (\bibinfo{year}{2016}) \bibinfo{pages}{025013},
  \bibinfo{doi}{\doi{10.1103/PhysRevD.93.025013}}, \eprint{1510.03188}.

\bibtype{Article}%
\bibitem{Penedo:2018kpc}
\bibinfo{author}{J.~T. Penedo}, \bibinfo{author}{S.~T. Petcov},
  \bibinfo{title}{{The 10$^{-3}$ eV frontier in neutrinoless double beta
  decay}}, \bibinfo{journal}{Phys. Lett. B} \bibinfo{volume}{786}
  (\bibinfo{year}{2018}) \bibinfo{pages}{410--417},
  \bibinfo{doi}{\doi{10.1016/j.physletb.2018.09.059}}, \eprint{1806.03203}.

\bibtype{Article}%
\bibitem{Barry:2010yk}
\bibinfo{author}{James Barry}, \bibinfo{author}{Werner Rodejohann},
  \bibinfo{title}{{Neutrino Mass Sum-rules in Flavor Symmetry Models}},
  \bibinfo{journal}{Nucl. Phys. B} \bibinfo{volume}{842} (\bibinfo{year}{2011})
  \bibinfo{pages}{33--50},
  \bibinfo{doi}{\doi{10.1016/j.nuclphysb.2010.08.015}}, \eprint{1007.5217}.

\bibtype{Article}%
\bibitem{Bazzocchi:2009da}
\bibinfo{author}{Federica Bazzocchi}, \bibinfo{author}{Luca Merlo},
  \bibinfo{author}{Stefano Morisi}, \bibinfo{title}{{Phenomenological
  Consequences of See-Saw in S(4) Based Models}}, \bibinfo{journal}{Phys. Rev.
  D} \bibinfo{volume}{80} (\bibinfo{year}{2009}) \bibinfo{pages}{053003},
  \bibinfo{doi}{\doi{10.1103/PhysRevD.80.053003}}, \eprint{0902.2849}.

\bibtype{Article}%
\bibitem{Ding:2010pc}
\bibinfo{author}{Gui-Jun Ding}, \bibinfo{title}{{SUSY adjoint $SU$(5) grand
  unified model with $S_4$ flavor symmetry}}, \bibinfo{journal}{Nucl. Phys. B}
  \bibinfo{volume}{846} (\bibinfo{year}{2011}) \bibinfo{pages}{394--428},
  \bibinfo{doi}{\doi{10.1016/j.nuclphysb.2011.01.009}}, \eprint{1006.4800}.

\bibtype{Article}%
\bibitem{Ma:2005sha}
\bibinfo{author}{Ernest Ma}, \bibinfo{title}{{Aspects of the tetrahedral
  neutrino mass matrix}}, \bibinfo{journal}{Phys. Rev. D} \bibinfo{volume}{72}
  (\bibinfo{year}{2005}) \bibinfo{pages}{037301},
  \bibinfo{doi}{\doi{10.1103/PhysRevD.72.037301}}, \eprint{hep-ph/0505209}.

\bibtype{Article}%
\bibitem{Ma:2006wm}
\bibinfo{author}{Ernest Ma}, \bibinfo{title}{{Suitability of A(4) as a Family
  Symmetry in Grand Unification}}, \bibinfo{journal}{Mod. Phys. Lett. A}
  \bibinfo{volume}{21} (\bibinfo{year}{2006}) \bibinfo{pages}{2931--2936},
  \bibinfo{doi}{\doi{10.1142/S0217732306022262}}, \eprint{hep-ph/0607190}.

\bibtype{Article}%
\bibitem{Kang:2015xfa}
\bibinfo{author}{Sin~Kyu Kang}, \bibinfo{author}{Morimitsu Tanimoto},
  \bibinfo{title}{{Prediction of Leptonic CP Phase in $A_4$ symmetric model}},
  \bibinfo{journal}{Phys. Rev. D} \bibinfo{volume}{91} (\bibinfo{number}{7})
  (\bibinfo{year}{2015}) \bibinfo{pages}{073010},
  \bibinfo{doi}{\doi{10.1103/PhysRevD.91.073010}}, \eprint{1501.07428}.

\bibtype{Article}%
\bibitem{Honda:2008rs}
\bibinfo{author}{Mizue Honda}, \bibinfo{author}{Morimitsu Tanimoto},
  \bibinfo{title}{{Deviation from tri-bimaximal neutrino mixing in A(4) flavor
  symmetry}}, \bibinfo{journal}{Prog. Theor. Phys.} \bibinfo{volume}{119}
  (\bibinfo{year}{2008}) \bibinfo{pages}{583--598},
  \bibinfo{doi}{\doi{10.1143/PTP.119.583}}, \eprint{0801.0181}.

\bibtype{Article}%
\bibitem{Brahmachari:2008fn}
\bibinfo{author}{Biswajoy Brahmachari}, \bibinfo{author}{Sandhya Choubey},
  \bibinfo{author}{Manimala Mitra}, \bibinfo{title}{{The A(4) flavor symmetry
  and neutrino phenomenology}}, \bibinfo{journal}{Phys. Rev. D}
  \bibinfo{volume}{77} (\bibinfo{year}{2008}) \bibinfo{pages}{073008},
  \bibinfo{doi}{\doi{10.1103/PhysRevD.77.119901}}, \bibinfo{note}{[Erratum:
  Phys.Rev.D 77, 119901 (2008)]}, \eprint{0801.3554}.

\bibtype{Article}%
\bibitem{Altarelli:2005yx}
\bibinfo{author}{Guido Altarelli}, \bibinfo{author}{Ferruccio Feruglio},
  \bibinfo{title}{{Tri-bimaximal neutrino mixing, A(4) and the modular
  symmetry}}, \bibinfo{journal}{Nucl. Phys. B} \bibinfo{volume}{741}
  (\bibinfo{year}{2006}) \bibinfo{pages}{215--235},
  \bibinfo{doi}{\doi{10.1016/j.nuclphysb.2006.02.015}},
  \eprint{hep-ph/0512103}.

\bibtype{Article}%
\bibitem{Chen:2009um}
\bibinfo{author}{Mu-Chun Chen}, \bibinfo{author}{Stephen~F. King},
  \bibinfo{title}{{A4 See-Saw Models and Form Dominance}},
  \bibinfo{journal}{JHEP} \bibinfo{volume}{06} (\bibinfo{year}{2009})
  \bibinfo{pages}{072}, \bibinfo{doi}{\doi{10.1088/1126-6708/2009/06/072}},
  \eprint{0903.0125}.

\bibtype{Article}%
\bibitem{Chen:2009gy}
\bibinfo{author}{Mu-Chun Chen}, \bibinfo{author}{K.~T. Mahanthappa},
  \bibinfo{author}{Felix Yu}, \bibinfo{title}{{A Viable Randall-Sundrum Model
  for Quarks and Leptons with T-prime Family Symmetry}},
  \bibinfo{journal}{Phys. Rev. D} \bibinfo{volume}{81} (\bibinfo{year}{2010})
  \bibinfo{pages}{036004}, \bibinfo{doi}{\doi{10.1103/PhysRevD.81.036004}},
  \eprint{0907.3963}.

\bibtype{Article}%
\bibitem{Cooper:2012bd}
\bibinfo{author}{Iain~K. Cooper}, \bibinfo{author}{Stephen~F. King},
  \bibinfo{author}{Alexander~J. Stuart}, \bibinfo{title}{{A Golden $A_5$ Model
  of Leptons with a Minimal NLO Correction}}, \bibinfo{journal}{Nucl. Phys. B}
  \bibinfo{volume}{875} (\bibinfo{year}{2013}) \bibinfo{pages}{650--677},
  \bibinfo{doi}{\doi{10.1016/j.nuclphysb.2013.07.027}}, \eprint{1212.1066}.

\bibtype{Article}%
\bibitem{Altarelli:2009kr}
\bibinfo{author}{Guido Altarelli}, \bibinfo{author}{Davide Meloni},
  \bibinfo{title}{{A Simplest A4 Model for Tri-Bimaximal Neutrino Mixing}},
  \bibinfo{journal}{J. Phys. G} \bibinfo{volume}{36} (\bibinfo{year}{2009})
  \bibinfo{pages}{085005}, \bibinfo{doi}{\doi{10.1088/0954-3899/36/8/085005}},
  \eprint{0905.0620}.

\bibtype{Article}%
\bibitem{Altarelli:2008bg}
\bibinfo{author}{Guido Altarelli}, \bibinfo{author}{Ferruccio Feruglio},
  \bibinfo{author}{Claudia Hagedorn}, \bibinfo{title}{{A SUSY SU(5) Grand
  Unified Model of Tri-Bimaximal Mixing from A$_4$}}, \bibinfo{journal}{JHEP}
  \bibinfo{volume}{03} (\bibinfo{year}{2008}) \bibinfo{pages}{052},
  \bibinfo{doi}{\doi{10.1088/1126-6708/2008/03/052}}, \eprint{0802.0090}.

\bibtype{Article}%
\bibitem{Hirsch:2008rp}
\bibinfo{author}{M. Hirsch}, \bibinfo{author}{S. Morisi},
  \bibinfo{author}{J.~W.~F. Valle}, \bibinfo{title}{{Tri-bimaximal neutrino
  mixing and neutrinoless double beta decay}}, \bibinfo{journal}{Phys. Rev. D}
  \bibinfo{volume}{78} (\bibinfo{year}{2008}) \bibinfo{pages}{093007},
  \bibinfo{doi}{\doi{10.1103/PhysRevD.78.093007}}, \eprint{0804.1521}.

\bibtype{Article}%
\bibitem{Bazzocchi:2009pv}
\bibinfo{author}{Federica Bazzocchi}, \bibinfo{author}{Luca Merlo},
  \bibinfo{author}{Stefano Morisi}, \bibinfo{title}{{Fermion Masses and Mixings
  in a S(4)-based Model}}, \bibinfo{journal}{Nucl. Phys. B}
  \bibinfo{volume}{816} (\bibinfo{year}{2009}) \bibinfo{pages}{204--226},
  \bibinfo{doi}{\doi{10.1016/j.nuclphysb.2009.03.005}}, \eprint{0901.2086}.

\bibtype{Article}%
\bibitem{Everett:2008et}
\bibinfo{author}{Lisa~L. Everett}, \bibinfo{author}{Alexander~J. Stuart},
  \bibinfo{title}{{Icosahedral (A(5)) Family Symmetry and the Golden Ratio
  Prediction for Solar Neutrino Mixing}}, \bibinfo{journal}{Phys. Rev. D}
  \bibinfo{volume}{79} (\bibinfo{year}{2009}) \bibinfo{pages}{085005},
  \bibinfo{doi}{\doi{10.1103/PhysRevD.79.085005}}, \eprint{0812.1057}.

\bibtype{Article}%
\bibitem{Boucenna:2012qb}
\bibinfo{author}{M.~S. Boucenna}, \bibinfo{author}{S. Morisi},
  \bibinfo{author}{E. Peinado}, \bibinfo{author}{Y. Shimizu},
  \bibinfo{author}{J.~W.~F. Valle}, \bibinfo{title}{{Predictive discrete dark
  matter model and neutrino oscillations}}, \bibinfo{journal}{Phys. Rev. D}
  \bibinfo{volume}{86} (\bibinfo{year}{2012}) \bibinfo{pages}{073008},
  \bibinfo{doi}{\doi{10.1103/PhysRevD.86.073008}}, \eprint{1204.4733}.

\bibtype{Article}%
\bibitem{Mohapatra:2012tb}
\bibinfo{author}{R.~N. Mohapatra}, \bibinfo{author}{C.~C. Nishi},
  \bibinfo{title}{{$S_4$ Flavored CP Symmetry for Neutrinos}},
  \bibinfo{journal}{Phys. Rev. D} \bibinfo{volume}{86} (\bibinfo{year}{2012})
  \bibinfo{pages}{073007}, \bibinfo{doi}{\doi{10.1103/PhysRevD.86.073007}},
  \eprint{1208.2875}.

\bibtype{Article}%
\bibitem{Altarelli:2005yp}
\bibinfo{author}{Guido Altarelli}, \bibinfo{author}{Ferruccio Feruglio},
  \bibinfo{title}{{Tri-bimaximal neutrino mixing from discrete symmetry in
  extra dimensions}}, \bibinfo{journal}{Nucl. Phys. B} \bibinfo{volume}{720}
  (\bibinfo{year}{2005}) \bibinfo{pages}{64--88},
  \bibinfo{doi}{\doi{10.1016/j.nuclphysb.2005.05.005}},
  \eprint{hep-ph/0504165}.

\bibtype{Article}%
\bibitem{Altarelli:2006kg}
\bibinfo{author}{Guido Altarelli}, \bibinfo{author}{Ferruccio Feruglio},
  \bibinfo{author}{Yin Lin}, \bibinfo{title}{{Tri-bimaximal neutrino mixing
  from orbifolding}}, \bibinfo{journal}{Nucl. Phys. B} \bibinfo{volume}{775}
  (\bibinfo{year}{2007}) \bibinfo{pages}{31--44},
  \bibinfo{doi}{\doi{10.1016/j.nuclphysb.2007.03.042}},
  \eprint{hep-ph/0610165}.

\bibtype{Article}%
\bibitem{Ma:2006vq}
\bibinfo{author}{Ernest Ma}, \bibinfo{title}{{Supersymmetric A(4) x Z(3) and
  A(4) realizations of neutrino tribimaximal mixing without and with
  corrections}}, \bibinfo{journal}{Mod. Phys. Lett. A} \bibinfo{volume}{22}
  (\bibinfo{year}{2007}) \bibinfo{pages}{101--106},
  \bibinfo{doi}{\doi{10.1142/S0217732307022505}}, \eprint{hep-ph/0610342}.

\bibtype{Article}%
\bibitem{Bazzocchi:2007na}
\bibinfo{author}{F. Bazzocchi}, \bibinfo{author}{S. Kaneko},
  \bibinfo{author}{S. Morisi}, \bibinfo{title}{{A SUSY A(4) model for fermion
  masses and mixings}}, \bibinfo{journal}{JHEP} \bibinfo{volume}{03}
  (\bibinfo{year}{2008}) \bibinfo{pages}{063},
  \bibinfo{doi}{\doi{10.1088/1126-6708/2008/03/063}}, \eprint{0707.3032}.

\bibtype{Article}%
\bibitem{Bazzocchi:2007au}
\bibinfo{author}{Federica Bazzocchi}, \bibinfo{author}{Stefano Morisi},
  \bibinfo{author}{Marco Picariello}, \bibinfo{title}{{Embedding A(4) into
  left-right flavor symmetry: Tribimaximal neutrino mixing and fermion
  hierarchy}}, \bibinfo{journal}{Phys. Lett. B} \bibinfo{volume}{659}
  (\bibinfo{year}{2008}) \bibinfo{pages}{628--633},
  \bibinfo{doi}{\doi{10.1016/j.physletb.2007.11.083}}, \eprint{0710.2928}.

\bibtype{Article}%
\bibitem{Lin:2008aj}
\bibinfo{author}{Yin Lin}, \bibinfo{title}{{A Predictive A(4) model, Charged
  Lepton Hierarchy and Tri-bimaximal Sum Rule}}, \bibinfo{journal}{Nucl. Phys.
  B} \bibinfo{volume}{813} (\bibinfo{year}{2009}) \bibinfo{pages}{91--105},
  \bibinfo{doi}{\doi{10.1016/j.nuclphysb.2008.12.025}}, \eprint{0804.2867}.

\bibtype{Article}%
\bibitem{Ma:2009wi}
\bibinfo{author}{Ernest Ma}, \bibinfo{title}{{Neutrino Tribimaximal Mixing from
  A(4) Alone}}, \bibinfo{journal}{Mod. Phys. Lett. A} \bibinfo{volume}{25}
  (\bibinfo{year}{2010}) \bibinfo{pages}{2215--2221},
  \bibinfo{doi}{\doi{10.1142/S021773231003361X}}, \eprint{0908.3165}.

\bibtype{Article}%
\bibitem{Ciafaloni:2009qs}
\bibinfo{author}{Paolo Ciafaloni}, \bibinfo{author}{Marco Picariello},
  \bibinfo{author}{Alfredo Urbano}, \bibinfo{author}{Emilio Torrente-Lujan},
  \bibinfo{title}{{Toward minimal renormalizable SUSY SU(5) Grand Unified Model
  with tribimaximal mixing from A(4) Flavor symmetry}}, \bibinfo{journal}{Phys.
  Rev. D} \bibinfo{volume}{81} (\bibinfo{year}{2010}) \bibinfo{pages}{016004},
  \bibinfo{doi}{\doi{10.1103/PhysRevD.81.016004}}, \eprint{0909.2553}.

\bibtype{Article}%
\bibitem{Bazzocchi:2008ej}
\bibinfo{author}{Federica Bazzocchi}, \bibinfo{author}{Stefano Morisi},
  \bibinfo{title}{{S(4) as a natural flavor symmetry for lepton mixing}},
  \bibinfo{journal}{Phys. Rev. D} \bibinfo{volume}{80} (\bibinfo{year}{2009})
  \bibinfo{pages}{096005}, \bibinfo{doi}{\doi{10.1103/PhysRevD.80.096005}},
  \eprint{0811.0345}.

\bibtype{Article}%
\bibitem{Feruglio:2013hia}
\bibinfo{author}{Ferruccio Feruglio}, \bibinfo{author}{Claudia Hagedorn},
  \bibinfo{author}{Robert Ziegler}, \bibinfo{title}{{A realistic pattern of
  lepton mixing and masses from $S_4$ and CP}}, \bibinfo{journal}{Eur. Phys. J.
  C} \bibinfo{volume}{74} (\bibinfo{year}{2014}) \bibinfo{pages}{2753},
  \bibinfo{doi}{\doi{10.1140/epjc/s10052-014-2753-2}}, \eprint{1303.7178}.

\bibtype{Article}%
\bibitem{Chen:2007afa}
\bibinfo{author}{Mu-Chun Chen}, \bibinfo{author}{K.~T. Mahanthappa},
  \bibinfo{title}{{CKM and Tri-bimaximal MNS Matrices in a $SU(5) \times
  ^{(d)}T$ Model}}, \bibinfo{journal}{Phys. Lett. B} \bibinfo{volume}{652}
  (\bibinfo{year}{2007}) \bibinfo{pages}{34--39},
  \bibinfo{doi}{\doi{10.1016/j.physletb.2007.06.064}}, \eprint{0705.0714}.

\bibtype{Article}%
\bibitem{Ding:2008rj}
\bibinfo{author}{Gui-Jun Ding}, \bibinfo{title}{{Fermion Mass Hierarchies and
  Flavor Mixing from T-prime Symmetry}}, \bibinfo{journal}{Phys. Rev. D}
  \bibinfo{volume}{78} (\bibinfo{year}{2008}) \bibinfo{pages}{036011},
  \bibinfo{doi}{\doi{10.1103/PhysRevD.78.036011}}, \eprint{0803.2278}.

\bibtype{Article}%
\bibitem{Chen:2009gf}
\bibinfo{author}{Mu-Chun Chen}, \bibinfo{author}{K.~T. Mahanthappa},
  \bibinfo{title}{{Group Theoretical Origin of CP Violation}},
  \bibinfo{journal}{Phys. Lett. B} \bibinfo{volume}{681} (\bibinfo{year}{2009})
  \bibinfo{pages}{444--447},
  \bibinfo{doi}{\doi{10.1016/j.physletb.2009.10.059}}, \eprint{0904.1721}.

\bibtype{Article}%
\bibitem{Feruglio:2007uu}
\bibinfo{author}{Ferruccio Feruglio}, \bibinfo{author}{Claudia Hagedorn},
  \bibinfo{author}{Yin Lin}, \bibinfo{author}{Luca Merlo},
  \bibinfo{title}{{Tri-bimaximal Neutrino Mixing and Quark Masses from a
  Discrete Flavour Symmetry}}, \bibinfo{journal}{Nucl. Phys. B}
  \bibinfo{volume}{775} (\bibinfo{year}{2007}) \bibinfo{pages}{120--142},
  \bibinfo{doi}{\doi{10.1016/j.nuclphysb.2007.04.002}},
  \bibinfo{note}{[Erratum: Nucl.Phys.B 836, 127--128 (2010)]},
  \eprint{hep-ph/0702194}.

\bibtype{Article}%
\bibitem{Merlo:2011hw}
\bibinfo{author}{Luca Merlo}, \bibinfo{author}{Stefano Rigolin},
  \bibinfo{author}{Bryan Zaldivar}, \bibinfo{title}{{Flavour violation in a
  supersymmetric T' model}}, \bibinfo{journal}{JHEP} \bibinfo{volume}{11}
  (\bibinfo{year}{2011}) \bibinfo{pages}{047},
  \bibinfo{doi}{\doi{10.1007/JHEP11(2011)047}}, \eprint{1108.1795}.

\bibtype{Article}%
\bibitem{Luhn:2012bc}
\bibinfo{author}{Christoph Luhn}, \bibinfo{author}{Krishna~Mohan Parattu},
  \bibinfo{author}{Akin Wingerter}, \bibinfo{title}{{A Minimal Model of
  Neutrino Flavor}}, \bibinfo{journal}{JHEP} \bibinfo{volume}{12}
  (\bibinfo{year}{2012}) \bibinfo{pages}{096},
  \bibinfo{doi}{\doi{10.1007/JHEP12(2012)096}}, \eprint{1210.1197}.

\bibtype{Article}%
\bibitem{Fukuyama:2010mz}
\bibinfo{author}{Takeshi Fukuyama}, \bibinfo{author}{Hiroaki Sugiyama},
  \bibinfo{author}{Koji Tsumura}, \bibinfo{title}{{Phenomenology in the Higgs
  Triplet Model With the $A_4$ Symmetry}}, \bibinfo{journal}{Phys. Rev. D}
  \bibinfo{volume}{82} (\bibinfo{year}{2010}) \bibinfo{pages}{036004},
  \bibinfo{doi}{\doi{10.1103/PhysRevD.82.036004}}, \eprint{1005.5338}.

\bibtype{Article}%
\bibitem{Ding:2013eca}
\bibinfo{author}{Gui-Jun Ding}, \bibinfo{author}{Ye-Ling Zhou},
  \bibinfo{title}{{Dirac Neutrinos with $S_4$ Flavor Symmetry in Warped Extra
  Dimensions}}, \bibinfo{journal}{Nucl. Phys. B} \bibinfo{volume}{876}
  (\bibinfo{year}{2013}) \bibinfo{pages}{418--452},
  \bibinfo{doi}{\doi{10.1016/j.nuclphysb.2013.08.011}}, \eprint{1304.2645}.

\bibtype{Article}%
\bibitem{Lindner:2010wr}
\bibinfo{author}{Manfred Lindner}, \bibinfo{author}{Alexander Merle},
  \bibinfo{author}{Viviana Niro}, \bibinfo{title}{{Soft $L_e - L_\mu - L_\tau$
  flavour symmetry breaking and sterile neutrino keV Dark Matter}},
  \bibinfo{journal}{JCAP} \bibinfo{volume}{01} (\bibinfo{year}{2011})
  \bibinfo{pages}{034}, \bibinfo{doi}{\doi{10.1088/1475-7516/2011/01/034}},
  \bibinfo{note}{[Erratum: JCAP 07, E01 (2014)]}, \eprint{1011.4950}.

\bibtype{Article}%
\bibitem{Hashimoto:2011tn}
\bibinfo{author}{Kenji Hashimoto}, \bibinfo{author}{Hiroshi Okada},
  \bibinfo{title}{{Lepton Flavor Model and Decaying Dark Matter in The Binary
  Icosahedral Group Symmetry}}  (\bibinfo{year}{2011}), \eprint{1110.3640}.

\bibtype{Article}%
\bibitem{Ding:2011cm}
\bibinfo{author}{Gui-Jun Ding}, \bibinfo{author}{Lisa~L. Everett},
  \bibinfo{author}{Alexander~J. Stuart}, \bibinfo{title}{{Golden Ratio Neutrino
  Mixing and $A_5$ Flavor Symmetry}}, \bibinfo{journal}{Nucl. Phys. B}
  \bibinfo{volume}{857} (\bibinfo{year}{2012}) \bibinfo{pages}{219--253},
  \bibinfo{doi}{\doi{10.1016/j.nuclphysb.2011.12.004}}, \eprint{1110.1688}.

\bibtype{Article}%
\bibitem{Morisi:2007ft}
\bibinfo{author}{Stefano Morisi}, \bibinfo{author}{Marco Picariello},
  \bibinfo{author}{Emilio Torrente-Lujan}, \bibinfo{title}{{Model for fermion
  masses and lepton mixing in SO(10) x A(4)}}, \bibinfo{journal}{Phys. Rev. D}
  \bibinfo{volume}{75} (\bibinfo{year}{2007}) \bibinfo{pages}{075015},
  \bibinfo{doi}{\doi{10.1103/PhysRevD.75.075015}}, \eprint{hep-ph/0702034}.

\bibtype{Article}%
\bibitem{Adhikary:2008au}
\bibinfo{author}{Biswajit Adhikary}, \bibinfo{author}{Ambar Ghosal},
  \bibinfo{title}{{Nonzero U(e3), CP violation and leptogenesis in a see-saw
  type softly broken A(4) symmetric model}}, \bibinfo{journal}{Phys. Rev. D}
  \bibinfo{volume}{78} (\bibinfo{year}{2008}) \bibinfo{pages}{073007},
  \bibinfo{doi}{\doi{10.1103/PhysRevD.78.073007}}, \eprint{0803.3582}.

\bibtype{Article}%
\bibitem{Lin:2009bw}
\bibinfo{author}{Yin Lin}, \bibinfo{title}{{Tri-bimaximal Neutrino Mixing from
  A(4) and $\theta_{13}$ \textasciitilde{} theta(C)}}, \bibinfo{journal}{Nucl.
  Phys. B} \bibinfo{volume}{824} (\bibinfo{year}{2010})
  \bibinfo{pages}{95--110},
  \bibinfo{doi}{\doi{10.1016/j.nuclphysb.2009.08.018}}, \eprint{0905.3534}.

\bibtype{Article}%
\bibitem{Csaki:2008qq}
\bibinfo{author}{Csaba Csaki}, \bibinfo{author}{Cedric Delaunay},
  \bibinfo{author}{Christophe Grojean}, \bibinfo{author}{Yuval Grossman},
  \bibinfo{title}{{A Model of Lepton Masses from a Warped Extra Dimension}},
  \bibinfo{journal}{JHEP} \bibinfo{volume}{10} (\bibinfo{year}{2008})
  \bibinfo{pages}{055}, \bibinfo{doi}{\doi{10.1088/1126-6708/2008/10/055}},
  \eprint{0806.0356}.

\bibtype{Article}%
\bibitem{Hagedorn:2009jy}
\bibinfo{author}{C. Hagedorn}, \bibinfo{author}{E. Molinaro},
  \bibinfo{author}{S.~T. Petcov}, \bibinfo{title}{{Majorana Phases and
  Leptogenesis in See-Saw Models with A(4) Symmetry}}, \bibinfo{journal}{JHEP}
  \bibinfo{volume}{09} (\bibinfo{year}{2009}) \bibinfo{pages}{115},
  \bibinfo{doi}{\doi{10.1088/1126-6708/2009/09/115}}, \eprint{0908.0240}.

\bibtype{Article}%
\bibitem{Burrows:2009pi}
\bibinfo{author}{T.~J. Burrows}, \bibinfo{author}{S.~F. King},
  \bibinfo{title}{{A(4) Family Symmetry from SU(5) SUSY GUTs in 6d}},
  \bibinfo{journal}{Nucl. Phys. B} \bibinfo{volume}{835} (\bibinfo{year}{2010})
  \bibinfo{pages}{174--196},
  \bibinfo{doi}{\doi{10.1016/j.nuclphysb.2010.04.002}}, \eprint{0909.1433}.

\bibtype{Article}%
\bibitem{Ding:2009gh}
\bibinfo{author}{Gui-Jun Ding}, \bibinfo{author}{Jia-Feng Liu},
  \bibinfo{title}{{Lepton Flavor Violation in Models with A(4) and S(4) Flavor
  Symmetries}}, \bibinfo{journal}{JHEP} \bibinfo{volume}{05}
  (\bibinfo{year}{2010}) \bibinfo{pages}{029},
  \bibinfo{doi}{\doi{10.1007/JHEP05(2010)029}}, \eprint{0911.4799}.

\bibtype{Article}%
\bibitem{Mitra:2009jj}
\bibinfo{author}{Manimala Mitra}, \bibinfo{title}{{Spontaneous R-Parity
  Violation, A(4) Flavor Symmetry and Tribimaximal Mixing}},
  \bibinfo{journal}{JHEP} \bibinfo{volume}{11} (\bibinfo{year}{2010})
  \bibinfo{pages}{026}, \bibinfo{doi}{\doi{10.1007/JHEP11(2010)026}},
  \eprint{0912.5291}.

\bibtype{Article}%
\bibitem{delAguila:2010vg}
\bibinfo{author}{Francisco del Aguila}, \bibinfo{author}{Adrian Carmona},
  \bibinfo{author}{Jose Santiago}, \bibinfo{title}{{Neutrino Masses from an A4
  Symmetry in Holographic Composite Higgs Models}}, \bibinfo{journal}{JHEP}
  \bibinfo{volume}{08} (\bibinfo{year}{2010}) \bibinfo{pages}{127},
  \bibinfo{doi}{\doi{10.1007/JHEP08(2010)127}}, \eprint{1001.5151}.

\bibtype{Article}%
\bibitem{Burrows:2010wz}
\bibinfo{author}{T.~J. Burrows}, \bibinfo{author}{S.~F. King},
  \bibinfo{title}{{$A_4$ x SU(5) SUSY GUT of Flavour in 8d}},
  \bibinfo{journal}{Nucl. Phys. B} \bibinfo{volume}{842} (\bibinfo{year}{2011})
  \bibinfo{pages}{107--121},
  \bibinfo{doi}{\doi{10.1016/j.nuclphysb.2010.08.018}}, \eprint{1007.2310}.

\bibtype{Article}%
\bibitem{Ahn:2014zja}
\bibinfo{author}{Y.~H. Ahn}, \bibinfo{author}{Paolo Gondolo},
  \bibinfo{title}{{Towards a realistic model of quarks and leptons, leptonic
  $CP$ violation, and neutrinoless $\beta\beta$-decay}},
  \bibinfo{journal}{Phys. Rev. D} \bibinfo{volume}{91} (\bibinfo{year}{2015})
  \bibinfo{pages}{013007}, \bibinfo{doi}{\doi{10.1103/PhysRevD.91.013007}},
  \eprint{1402.0150}.

\bibtype{Article}%
\bibitem{Karmakar:2014dva}
\bibinfo{author}{Biswajit Karmakar}, \bibinfo{author}{Arunansu Sil},
  \bibinfo{title}{{Nonzero $\theta_{13}$ and leptogenesis in a type-I seesaw
  model with $A_4$ symmetry}}, \bibinfo{journal}{Phys. Rev. D}
  \bibinfo{volume}{91} (\bibinfo{year}{2015}) \bibinfo{pages}{013004},
  \bibinfo{doi}{\doi{10.1103/PhysRevD.91.013004}}, \eprint{1407.5826}.

\bibtype{Article}%
\bibitem{Ahn:2014gva}
\bibinfo{author}{Y.~H. Ahn}, \bibinfo{title}{{Flavored Peccei-Quinn symmetry}},
  \bibinfo{journal}{Phys. Rev. D} \bibinfo{volume}{91} (\bibinfo{year}{2015})
  \bibinfo{pages}{056005}, \bibinfo{doi}{\doi{10.1103/PhysRevD.91.056005}},
  \eprint{1410.1634}.

\bibtype{Article}%
\bibitem{He:2006dk}
\bibinfo{author}{Xiao-Gang He}, \bibinfo{author}{Yong-Yeon Keum},
  \bibinfo{author}{Raymond~R. Volkas}, \bibinfo{title}{{A(4) flavor symmetry
  breaking scheme for understanding quark and neutrino mixing angles}},
  \bibinfo{journal}{JHEP} \bibinfo{volume}{04} (\bibinfo{year}{2006})
  \bibinfo{pages}{039}, \bibinfo{doi}{\doi{10.1088/1126-6708/2006/04/039}},
  \eprint{hep-ph/0601001}.

\bibtype{Article}%
\bibitem{Berger:2009tt}
\bibinfo{author}{Joshua Berger}, \bibinfo{author}{Yuval Grossman},
  \bibinfo{title}{{Model of leptons from SO(3) ---\ensuremath{>} A(4)}},
  \bibinfo{journal}{JHEP} \bibinfo{volume}{02} (\bibinfo{year}{2010})
  \bibinfo{pages}{071}, \bibinfo{doi}{\doi{10.1007/JHEP02(2010)071}},
  \eprint{0910.4392}.

\bibtype{Article}%
\bibitem{Kadosh:2010rm}
\bibinfo{author}{A. Kadosh}, \bibinfo{author}{E. Pallante}, \bibinfo{title}{{An
  A(4) flavor model for quarks and leptons in warped geometry}},
  \bibinfo{journal}{JHEP} \bibinfo{volume}{08} (\bibinfo{year}{2010})
  \bibinfo{pages}{115}, \bibinfo{doi}{\doi{10.1007/JHEP08(2010)115}},
  \eprint{1004.0321}.

\bibtype{Article}%
\bibitem{Lavoura:2012cv}
\bibinfo{author}{L. Lavoura}, \bibinfo{author}{S. Morisi},
  \bibinfo{author}{J.~W.~F. Valle}, \bibinfo{title}{{Accidental Stability of
  Dark Matter}}, \bibinfo{journal}{JHEP} \bibinfo{volume}{02}
  (\bibinfo{year}{2013}) \bibinfo{pages}{118},
  \bibinfo{doi}{\doi{10.1007/JHEP02(2013)118}}, \eprint{1205.3442}.

\bibtype{Article}%
\bibitem{King:2012in}
\bibinfo{author}{Stephen~F. King}, \bibinfo{author}{Christoph Luhn},
  \bibinfo{author}{Alexander~J. Stuart}, \bibinfo{title}{{A Grand Delta(96) x
  SU(5) Flavour Model}}, \bibinfo{journal}{Nucl. Phys. B} \bibinfo{volume}{867}
  (\bibinfo{year}{2013}) \bibinfo{pages}{203--235},
  \bibinfo{doi}{\doi{10.1016/j.nuclphysb.2012.09.021}}, \eprint{1207.5741}.

\bibtype{Article}%
\bibitem{Adulpravitchai:2009gi}
\bibinfo{author}{Adisorn Adulpravitchai}, \bibinfo{author}{Manfred Lindner},
  \bibinfo{author}{Alexander Merle}, \bibinfo{title}{{Confronting Flavour
  Symmetries and extended Scalar Sectors with Lepton Flavour Violation
  Bounds}}, \bibinfo{journal}{Phys. Rev. D} \bibinfo{volume}{80}
  (\bibinfo{year}{2009}) \bibinfo{pages}{055031},
  \bibinfo{doi}{\doi{10.1103/PhysRevD.80.055031}}, \eprint{0907.2147}.

\bibtype{Article}%
\bibitem{Dorame:2011eb}
\bibinfo{author}{L. Dorame}, \bibinfo{author}{D. Meloni}, \bibinfo{author}{S.
  Morisi}, \bibinfo{author}{E. Peinado}, \bibinfo{author}{J.~W.~F. Valle},
  \bibinfo{title}{{Constraining Neutrinoless Double Beta Decay}},
  \bibinfo{journal}{Nucl. Phys. B} \bibinfo{volume}{861} (\bibinfo{year}{2012})
  \bibinfo{pages}{259--270},
  \bibinfo{doi}{\doi{10.1016/j.nuclphysb.2012.04.003}}, \eprint{1111.5614}.

\bibtype{Article}%
\bibitem{Dorame:2012zv}
\bibinfo{author}{L. Dorame}, \bibinfo{author}{S. Morisi}, \bibinfo{author}{E.
  Peinado}, \bibinfo{author}{J.~W.~F. Valle}, \bibinfo{author}{Alma~D. Rojas},
  \bibinfo{title}{{A new neutrino mass sum rule from inverse seesaw}},
  \bibinfo{journal}{Phys. Rev. D} \bibinfo{volume}{86} (\bibinfo{year}{2012})
  \bibinfo{pages}{056001}, \bibinfo{doi}{\doi{10.1103/PhysRevD.86.056001}},
  \eprint{1203.0155}.

\bibtype{Article}%
\bibitem{King:2013psa}
\bibinfo{author}{Stephen~F. King}, \bibinfo{author}{Alexander Merle},
  \bibinfo{author}{Alexander~J. Stuart}, \bibinfo{title}{{The Power of Neutrino
  Mass Sum Rules for Neutrinoless Double Beta Decay Experiments}},
  \bibinfo{journal}{JHEP} \bibinfo{volume}{12} (\bibinfo{year}{2013})
  \bibinfo{pages}{005}, \bibinfo{doi}{\doi{10.1007/JHEP12(2013)005}},
  \eprint{1307.2901}.

\bibtype{Article}%
\bibitem{Gehrlein:2017ryu}
\bibinfo{author}{Julia Gehrlein}, \bibinfo{author}{Martin Spinrath},
  \bibinfo{title}{{Neutrino Mass Sum Rules and Symmetries of the Mass Matrix}},
  \bibinfo{journal}{Eur. Phys. J. C} \bibinfo{volume}{77} (\bibinfo{number}{5})
  (\bibinfo{year}{2017}) \bibinfo{pages}{281},
  \bibinfo{doi}{\doi{10.1140/epjc/s10052-017-4817-6}}, \eprint{1704.02371}.

\bibtype{Article}%
\bibitem{Froggatt:1978nt}
\bibinfo{author}{C.~D. Froggatt}, \bibinfo{author}{Holger~Bech Nielsen},
  \bibinfo{title}{{Hierarchy of Quark Masses, Cabibbo Angles and CP
  Violation}}, \bibinfo{journal}{Nucl. Phys. B} \bibinfo{volume}{147}
  (\bibinfo{year}{1979}) \bibinfo{pages}{277--298},
  \bibinfo{doi}{\doi{10.1016/0550-3213(79)90316-X}}.

\bibtype{Article}%
\bibitem{Planck:2018vyg}
\bibinfo{author}{N. Aghanim}, et al. (\bibinfo{collaboration}{Planck}),
  \bibinfo{title}{{Planck 2018 results. VI. Cosmological parameters}},
  \bibinfo{journal}{Astron. Astrophys.} \bibinfo{volume}{641}
  (\bibinfo{year}{2020}) \bibinfo{pages}{A6},
  \bibinfo{doi}{\doi{10.1051/0004-6361/201833910}}, \bibinfo{note}{[Erratum:
  Astron.Astrophys. 652, C4 (2021)]}, \eprint{1807.06209}.

\bibtype{Article}%
\bibitem{DESI:2024mwx}
\bibinfo{author}{A.~G. Adame}, et al. (\bibinfo{collaboration}{DESI}),
  \bibinfo{title}{{DESI 2024 VI: cosmological constraints from the measurements
  of baryon acoustic oscillations}}, \bibinfo{journal}{JCAP}
  \bibinfo{volume}{02} (\bibinfo{year}{2025}) \bibinfo{pages}{021},
  \bibinfo{doi}{\doi{10.1088/1475-7516/2025/02/021}}, \eprint{2404.03002}.

\bibtype{Article}%
\bibitem{Craig:2024tky}
\bibinfo{author}{Nathaniel Craig}, \bibinfo{author}{Daniel Green},
  \bibinfo{author}{Joel Meyers}, \bibinfo{author}{Surjeet Rajendran},
  \bibinfo{title}{{No \ensuremath{\nu}s is Good News}}, \bibinfo{journal}{JHEP}
  \bibinfo{volume}{09} (\bibinfo{year}{2024}) \bibinfo{pages}{097},
  \bibinfo{doi}{\doi{10.1007/JHEP09(2024)097}}, \eprint{2405.00836}.

\bibtype{Article}%
\bibitem{Jiang:2024viw}
\bibinfo{author}{Jun-Qian Jiang}, \bibinfo{author}{William Giar{\`e}},
  \bibinfo{author}{Stefano Gariazzo}, \bibinfo{author}{Maria~Giovanna
  Dainotti}, \bibinfo{author}{Eleonora Di~Valentino}, \bibinfo{author}{Olga
  Mena}, \bibinfo{author}{Davide Pedrotti}, \bibinfo{author}{Simony~Santos da
  Costa}, \bibinfo{author}{Sunny Vagnozzi}, \bibinfo{title}{{Neutrino cosmology
  after DESI: tightest mass upper limits, preference for the normal ordering,
  and tension with terrestrial observations}}, \bibinfo{journal}{JCAP}
  \bibinfo{volume}{01} (\bibinfo{year}{2025}) \bibinfo{pages}{153},
  \bibinfo{doi}{\doi{10.1088/1475-7516/2025/01/153}}, \eprint{2407.18047}.

\bibtype{Article}%
\bibitem{Loverde:2024nfi}
\bibinfo{author}{Marilena Loverde}, \bibinfo{author}{Zachary~J. Weiner},
  \bibinfo{title}{{Massive neutrinos and cosmic composition}},
  \bibinfo{journal}{JCAP} \bibinfo{volume}{12} (\bibinfo{year}{2024})
  \bibinfo{pages}{048}, \bibinfo{doi}{\doi{10.1088/1475-7516/2024/12/048}},
  \eprint{2410.00090}.

\bibtype{Article}%
\bibitem{PTOLEMY:2018jst}
\bibinfo{author}{E. Baracchini}, et al. (\bibinfo{collaboration}{PTOLEMY}),
  \bibinfo{title}{{PTOLEMY: A Proposal for Thermal Relic Detection of Massive
  Neutrinos and Directional Detection of MeV Dark Matter}}
  (\bibinfo{year}{2018}), \eprint{1808.01892}.

\bibtype{Article}%
\bibitem{PTOLEMY:2019hkd}
\bibinfo{author}{M.~G. Betti}, et al. (\bibinfo{collaboration}{PTOLEMY}),
  \bibinfo{title}{{Neutrino physics with the PTOLEMY project: active neutrino
  properties and the light sterile case}}, \bibinfo{journal}{JCAP}
  \bibinfo{volume}{07} (\bibinfo{year}{2019}) \bibinfo{pages}{047},
  \bibinfo{doi}{\doi{10.1088/1475-7516/2019/07/047}}, \eprint{1902.05508}.

\bibtype{Article}%
\bibitem{Cheipesh:2021fmg}
\bibinfo{author}{Yevheniia Cheipesh}, \bibinfo{author}{Vadim Cheianov},
  \bibinfo{author}{Alexey Boyarsky}, \bibinfo{title}{{Navigating the pitfalls
  of relic neutrino detection}}, \bibinfo{journal}{Phys. Rev. D}
  \bibinfo{volume}{104} (\bibinfo{number}{11}) (\bibinfo{year}{2021})
  \bibinfo{pages}{116004}, \bibinfo{doi}{\doi{10.1103/PhysRevD.104.116004}},
  \eprint{2101.10069}.

\bibtype{Article}%
\bibitem{PTOLEMY:2022ldz}
\bibinfo{author}{A. Apponi}, et al. (\bibinfo{collaboration}{PTOLEMY}),
  \bibinfo{title}{{Heisenberg\textquoteright{}s uncertainty principle in the
  PTOLEMY project: A theory update}}, \bibinfo{journal}{Phys. Rev. D}
  \bibinfo{volume}{106} (\bibinfo{number}{5}) (\bibinfo{year}{2022})
  \bibinfo{pages}{053002}, \bibinfo{doi}{\doi{10.1103/PhysRevD.106.053002}},
  \eprint{2203.11228}.

\bibtype{Article}%
\bibitem{Bauer:2022lri}
\bibinfo{author}{Martin Bauer}, \bibinfo{author}{Jack~D. Shergold},
  \bibinfo{title}{{Limits on the cosmic neutrino background}},
  \bibinfo{journal}{JCAP} \bibinfo{volume}{01} (\bibinfo{year}{2023})
  \bibinfo{pages}{003}, \bibinfo{doi}{\doi{10.1088/1475-7516/2023/01/003}},
  \eprint{2207.12413}.

\bibtype{Article}%
\bibitem{Furry:1939qr}
\bibinfo{author}{W.~H. Furry}, \bibinfo{title}{{On transition probabilities in
  double beta-disintegration}}, \bibinfo{journal}{Phys. Rev.}
  \bibinfo{volume}{56} (\bibinfo{year}{1939}) \bibinfo{pages}{1184--1193},
  \bibinfo{doi}{\doi{10.1103/PhysRev.56.1184}}.

\bibtype{Article}%
\bibitem{Schechter:1981bd}
\bibinfo{author}{J. Schechter}, \bibinfo{author}{J.~W.~F. Valle},
  \bibinfo{title}{{Neutrinoless Double beta Decay in SU(2) x U(1) Theories}},
  \bibinfo{journal}{Phys. Rev. D} \bibinfo{volume}{25} (\bibinfo{year}{1982})
  \bibinfo{pages}{2951}, \bibinfo{doi}{\doi{10.1103/PhysRevD.25.2951}}.

\bibtype{Article}%
\bibitem{KamLAND-Zen:2022tow}
\bibinfo{author}{S. Abe}, et al. (\bibinfo{collaboration}{KamLAND-Zen}),
  \bibinfo{title}{{Search for the Majorana Nature of Neutrinos in the Inverted
  Mass Ordering Region with KamLAND-Zen}}, \bibinfo{journal}{Phys. Rev. Lett.}
  \bibinfo{volume}{130} (\bibinfo{number}{5}) (\bibinfo{year}{2023})
  \bibinfo{pages}{051801}, \bibinfo{doi}{\doi{10.1103/PhysRevLett.130.051801}},
  \eprint{2203.02139}.

\bibtype{Article}%
\bibitem{Katrin:2024cdt}
\bibinfo{author}{Max Aker}, et al. (\bibinfo{collaboration}{KATRIN}),
  \bibinfo{title}{{Direct neutrino-mass measurement based on 259 days of KATRIN
  data}}, \bibinfo{journal}{Science} \bibinfo{volume}{388}
  (\bibinfo{number}{6743}) (\bibinfo{year}{2025}) \bibinfo{pages}{adq9592},
  \bibinfo{doi}{\doi{10.1126/science.adq9592}}, \eprint{2406.13516}.

\bibtype{Article}%
\bibitem{KATRIN:2021dfa}
\bibinfo{author}{M. Aker}, et al. (\bibinfo{collaboration}{KATRIN}),
  \bibinfo{title}{{The design, construction, and commissioning of the KATRIN
  experiment}}, \bibinfo{journal}{JINST} \bibinfo{volume}{16}
  (\bibinfo{number}{08}) (\bibinfo{year}{2021}) \bibinfo{pages}{T08015},
  \bibinfo{doi}{\doi{10.1088/1748-0221/16/08/T08015}}, \eprint{2103.04755}.

\bibtype{Article}%
\bibitem{KATRIN:2001ttj}
\bibinfo{author}{A. Osipowicz}, et al. (\bibinfo{collaboration}{KATRIN}),
  \bibinfo{title}{{KATRIN: A Next generation tritium beta decay experiment with
  sub-eV sensitivity for the electron neutrino mass. Letter of intent}}
  (\bibinfo{year}{2001}), \eprint{hep-ex/0109033}.

\bibtype{Article}%
\bibitem{Project8:2017nal}
\bibinfo{author}{Ali Ashtari~Esfahani}, et al. (\bibinfo{collaboration}{Project
  8}), \bibinfo{title}{{Determining the neutrino mass with cyclotron radiation
  emission spectroscopy\textemdash{}Project 8}}, \bibinfo{journal}{J. Phys. G}
  \bibinfo{volume}{44} (\bibinfo{number}{5}) (\bibinfo{year}{2017})
  \bibinfo{pages}{054004}, \bibinfo{doi}{\doi{10.1088/1361-6471/aa5b4f}},
  \eprint{1703.02037}.

\bibtype{Article}%
\bibitem{Gastaldo:2013wha}
\bibinfo{author}{L. Gastaldo}, et al., \bibinfo{title}{{The Electron Capture
  $^{163}$Ho Experiment ECHo: an overview}}, \bibinfo{journal}{J. Low Temp.
  Phys.} \bibinfo{volume}{176} (\bibinfo{number}{5-6}) (\bibinfo{year}{2014})
  \bibinfo{pages}{876--884}, \bibinfo{doi}{\doi{10.1007/s10909-014-1187-4}},
  \eprint{1309.5214}.

\bibtype{Article}%
\bibitem{Zatsepin:1968kt}
\bibinfo{author}{G.~T. Zatsepin}, \bibinfo{title}{{On the possibility of
  determining the upper limit of the neutrino mass by means of the flight
  time}}, \bibinfo{journal}{Pisma Zh. Eksp. Teor. Fiz.} \bibinfo{volume}{8}
  (\bibinfo{year}{1968}) \bibinfo{pages}{333--334}.

\bibtype{Article}%
\bibitem{Loredo:2001rx}
\bibinfo{author}{Thomas~J. Loredo}, \bibinfo{author}{Don~Q. Lamb},
  \bibinfo{title}{{Bayesian analysis of neutrinos observed from supernova
  SN-1987A}}, \bibinfo{journal}{Phys. Rev. D} \bibinfo{volume}{65}
  (\bibinfo{year}{2002}) \bibinfo{pages}{063002},
  \bibinfo{doi}{\doi{10.1103/PhysRevD.65.063002}}, \eprint{astro-ph/0107260}.

\bibtype{Article}%
\bibitem{Nardi:2003pr}
\bibinfo{author}{Enrico Nardi}, \bibinfo{author}{Jorge~I. Zuluaga},
  \bibinfo{title}{{Exploring the sub-eV neutrino mass range with supernova
  neutrinos}}, \bibinfo{journal}{Phys. Rev. D} \bibinfo{volume}{69}
  (\bibinfo{year}{2004}) \bibinfo{pages}{103002},
  \bibinfo{doi}{\doi{10.1103/PhysRevD.69.103002}}, \eprint{astro-ph/0306384}.

\bibtype{Article}%
\bibitem{Nardi:2004zg}
\bibinfo{author}{Enrico Nardi}, \bibinfo{author}{Jorge~I. Zuluaga},
  \bibinfo{title}{{Constraints on neutrino masses from a galactic supernova
  neutrino signal at present and future detectors}}, \bibinfo{journal}{Nucl.
  Phys. B} \bibinfo{volume}{731} (\bibinfo{year}{2005})
  \bibinfo{pages}{140--163},
  \bibinfo{doi}{\doi{10.1016/j.nuclphysb.2005.10.009}},
  \eprint{hep-ph/0412104}.

\bibtype{Article}%
\bibitem{Pagliaroli:2010ik}
\bibinfo{author}{G. Pagliaroli}, \bibinfo{author}{F. Rossi-Torres},
  \bibinfo{author}{F. Vissani}, \bibinfo{title}{{Neutrino mass bound in the
  standard scenario for supernova electronic antineutrino emission}},
  \bibinfo{journal}{Astropart. Phys.} \bibinfo{volume}{33}
  (\bibinfo{year}{2010}) \bibinfo{pages}{287--291},
  \bibinfo{doi}{\doi{10.1016/j.astropartphys.2010.02.007}}, \eprint{1002.3349}.

\bibtype{Article}%
\bibitem{Lu:2014zma}
\bibinfo{author}{Jia-Shu Lu}, \bibinfo{author}{Jun Cao},
  \bibinfo{author}{Yu-Feng Li}, \bibinfo{author}{Shun Zhou},
  \bibinfo{title}{{Constraining Absolute Neutrino Masses via Detection of
  Galactic Supernova Neutrinos at JUNO}}, \bibinfo{journal}{JCAP}
  \bibinfo{volume}{05} (\bibinfo{year}{2015}) \bibinfo{pages}{044},
  \bibinfo{doi}{\doi{10.1088/1475-7516/2015/05/044}}, \eprint{1412.7418}.

\bibtype{Article}%
\bibitem{Hansen:2019giq}
\bibinfo{author}{Rasmus S.~L. Hansen}, \bibinfo{author}{Manfred Lindner},
  \bibinfo{author}{Oliver Scholer}, \bibinfo{title}{{Timing the neutrino signal
  of a Galactic supernova}}, \bibinfo{journal}{Phys. Rev. D}
  \bibinfo{volume}{101} (\bibinfo{number}{12}) (\bibinfo{year}{2020})
  \bibinfo{pages}{123018}, \bibinfo{doi}{\doi{10.1103/PhysRevD.101.123018}},
  \eprint{1904.11461}.

\bibtype{Article}%
\bibitem{Pompa:2022cxc}
\bibinfo{author}{Federica Pompa}, \bibinfo{author}{Francesco Capozzi},
  \bibinfo{author}{Olga Mena}, \bibinfo{author}{Michel Sorel},
  \bibinfo{title}{{Absolute \ensuremath{\nu} Mass Measurement with the DUNE
  Experiment}}, \bibinfo{journal}{Phys. Rev. Lett.} \bibinfo{volume}{129}
  (\bibinfo{number}{12}) (\bibinfo{year}{2022}) \bibinfo{pages}{121802},
  \bibinfo{doi}{\doi{10.1103/PhysRevLett.129.121802}}, \eprint{2203.00024}.

\bibtype{Article}%
\bibitem{Pitik:2022jjh}
\bibinfo{author}{Tetyana Pitik}, \bibinfo{author}{Daniel~J. Heimsoth},
  \bibinfo{author}{Anna~M. Suliga}, \bibinfo{author}{A.~Baha Balantekin},
  \bibinfo{title}{{Exploiting stellar explosion induced by the QCD phase
  transition in large-scale neutrino detectors}}, \bibinfo{journal}{Phys. Rev.
  D} \bibinfo{volume}{106} (\bibinfo{number}{10}) (\bibinfo{year}{2022})
  \bibinfo{pages}{103007}, \bibinfo{doi}{\doi{10.1103/PhysRevD.106.103007}},
  \eprint{2208.14469}.

\bibtype{Article}%
\bibitem{Brdar:2022vfr}
\bibinfo{author}{Vedran Brdar}, \bibinfo{author}{Xun-Jie Xu},
  \bibinfo{title}{{Timing and multi-channel: novel method for determining the
  neutrino mass ordering from supernovae}}, \bibinfo{journal}{JCAP}
  \bibinfo{volume}{08} (\bibinfo{year}{2022}) \bibinfo{pages}{067},
  \bibinfo{doi}{\doi{10.1088/1475-7516/2022/08/067}}, \eprint{2204.13135}.

\bibtype{Article}%
\bibitem{Denton:2024mlb}
\bibinfo{author}{Peter~B. Denton}, \bibinfo{author}{Yves Kini},
  \bibinfo{title}{{Individual neutrino masses from a supernova}},
  \bibinfo{journal}{Phys. Rev. D} \bibinfo{volume}{111} (\bibinfo{number}{10})
  (\bibinfo{year}{2025}) \bibinfo{pages}{103006},
  \bibinfo{doi}{\doi{10.1103/PhysRevD.111.103006}}, \eprint{2411.13634}.

\bibtype{Article}%
\bibitem{Mirizzi:2015eza}
\bibinfo{author}{Alessandro Mirizzi}, \bibinfo{author}{Irene Tamborra},
  \bibinfo{author}{Hans-Thomas Janka}, \bibinfo{author}{Ninetta Saviano},
  \bibinfo{author}{Kate Scholberg}, \bibinfo{author}{Robert Bollig},
  \bibinfo{author}{Lorenz Hudepohl}, \bibinfo{author}{Sovan Chakraborty},
  \bibinfo{title}{{Supernova Neutrinos: Production, Oscillations and
  Detection}}, \bibinfo{journal}{Riv. Nuovo Cim.} \bibinfo{volume}{39}
  (\bibinfo{number}{1-2}) (\bibinfo{year}{2016}) \bibinfo{pages}{1--112},
  \bibinfo{doi}{\doi{10.1393/ncr/i2016-10120-8}}, \eprint{1508.00785}.

\bibtype{Article}%
\bibitem{Sekiguchi:2010ja}
\bibinfo{author}{Yuichiro Sekiguchi}, \bibinfo{author}{Masaru Shibata},
  \bibinfo{title}{{Formation of black hole and accretion disk in a massive
  high-entropy stellar core collapse}}, \bibinfo{journal}{Astrophys. J.}
  \bibinfo{volume}{737} (\bibinfo{year}{2011}) \bibinfo{pages}{6},
  \bibinfo{doi}{\doi{10.1088/0004-637X/737/1/6}}, \eprint{1009.5303}.

\bibtype{Article}%
\bibitem{Gullin:2021hfv}
\bibinfo{author}{Samuel Gullin}, \bibinfo{author}{Evan~P.
  O\textquoteright{}Connor}, \bibinfo{author}{Jia-Shian Wang},
  \bibinfo{author}{Jeff Tseng}, \bibinfo{title}{{Neutrino Echos following Black
  Hole Formation in Core-collapse Supernovae}}, \bibinfo{journal}{Astrophys.
  J.} \bibinfo{volume}{926} (\bibinfo{number}{2}) (\bibinfo{year}{2022})
  \bibinfo{pages}{212}, \bibinfo{doi}{\doi{10.3847/1538-4357/ac4420}},
  \eprint{2109.13242}.

\bibtype{Article}%
\bibitem{Sagert:2008ka}
\bibinfo{author}{I. Sagert}, \bibinfo{author}{T. Fischer}, \bibinfo{author}{M.
  Hempel}, \bibinfo{author}{G. Pagliara}, \bibinfo{author}{J.
  Schaffner-Bielich}, \bibinfo{author}{A. Mezzacappa}, \bibinfo{author}{F.~K.
  Thielemann}, \bibinfo{author}{M. Liebendorfer}, \bibinfo{title}{{Signals of
  the QCD phase transition in core-collapse supernovae}},
  \bibinfo{journal}{Phys. Rev. Lett.} \bibinfo{volume}{102}
  (\bibinfo{year}{2009}) \bibinfo{pages}{081101},
  \bibinfo{doi}{\doi{10.1103/PhysRevLett.102.081101}}, \eprint{0809.4225}.

\bibtype{Article}%
\bibitem{Fischer:2010wp}
\bibinfo{author}{T. Fischer}, \bibinfo{author}{I. Sagert}, \bibinfo{author}{G.
  Pagliara}, \bibinfo{author}{M. Hempel}, \bibinfo{author}{J.
  Schaffner-Bielich}, \bibinfo{author}{T. Rauscher}, \bibinfo{author}{F.~K.
  Thielemann}, \bibinfo{author}{R. Kappeli}, \bibinfo{author}{G.
  Martinez-Pinedo}, \bibinfo{author}{M. Liebendorfer},
  \bibinfo{title}{{Core-collapse supernova explosions triggered by a
  quark-hadron phase transition during the early post-bounce phase}},
  \bibinfo{journal}{Astrophys. J. Suppl.} \bibinfo{volume}{194}
  (\bibinfo{year}{2011}) \bibinfo{pages}{39},
  \bibinfo{doi}{\doi{10.1088/0067-0049/194/2/39}}, \eprint{1011.3409}.

\bibtype{Article}%
\bibitem{Fischer:2017lag}
\bibinfo{author}{Tobias Fischer}, \bibinfo{author}{Niels-Uwe~F. Bastian},
  \bibinfo{author}{Meng-Ru Wu}, \bibinfo{author}{Petr Baklanov},
  \bibinfo{author}{Elena Sorokina}, \bibinfo{author}{Sergei Blinnikov},
  \bibinfo{author}{Stefan Typel}, \bibinfo{author}{Thomas Kl\"ahn},
  \bibinfo{author}{David~B. Blaschke}, \bibinfo{title}{{Quark deconfinement as
  a supernova explosion engine for massive blue supergiant stars}},
  \bibinfo{journal}{Nature Astron.} \bibinfo{volume}{2} (\bibinfo{number}{12})
  (\bibinfo{year}{2018}) \bibinfo{pages}{980--986},
  \bibinfo{doi}{\doi{10.1038/s41550-018-0583-0}}, \eprint{1712.08788}.

\bibtype{Article}%
\bibitem{Zha:2021fbi}
\bibinfo{author}{Shuai Zha}, \bibinfo{author}{Evan~P. O'Connor},
  \bibinfo{author}{Andr\'e da Silva~Schneider}, \bibinfo{title}{{Progenitor
  Dependence of Hadron-quark Phase Transition in Failing Core-collapse
  Supernovae}}, \bibinfo{journal}{Astrophys. J.} \bibinfo{volume}{911}
  (\bibinfo{number}{2}) (\bibinfo{year}{2021}) \bibinfo{pages}{74},
  \bibinfo{doi}{\doi{10.3847/1538-4357/abec4c}}, \eprint{2103.02268}.

\bibtype{Article}%
\bibitem{Fischer:2021tvv}
\bibinfo{author}{Tobias Fischer}, \bibinfo{title}{{QCD phase transition drives
  supernova explosion of a very massive star: Dependence on metallicity of the
  progenitor star}}, \bibinfo{journal}{Eur. Phys. J. A} \bibinfo{volume}{57}
  (\bibinfo{number}{9}) (\bibinfo{year}{2021}) \bibinfo{pages}{270},
  \bibinfo{doi}{\doi{10.1140/epja/s10050-021-00571-z}}, \eprint{2108.00196}.

\bibtype{Article}%
\bibitem{Kuroda:2021eiv}
\bibinfo{author}{Takami Kuroda}, \bibinfo{author}{Tobias Fischer},
  \bibinfo{author}{Tomoya Takiwaki}, \bibinfo{author}{Kei Kotake},
  \bibinfo{title}{{Core-collapse Supernova Simulations and the Formation of
  Neutron Stars, Hybrid Stars, and Black Holes}}, \bibinfo{journal}{Astrophys.
  J.} \bibinfo{volume}{924} (\bibinfo{number}{1}) (\bibinfo{year}{2022})
  \bibinfo{pages}{38}, \bibinfo{doi}{\doi{10.3847/1538-4357/ac31a8}},
  \eprint{2109.01508}.

\bibtype{Inbook}%
\bibitem{Bauswein:2022vtq}
\bibinfo{author}{Andreas Bauswein}, \bibinfo{author}{David Blaschke},
  \bibinfo{author}{Tobias Fischer}, \bibinfo{title}{{Effects of a strong phase
  transition on supernova explosions, compact stars and their mergers}}
  \bibinfo{year}{2022} \bibinfo{doi}{\doi{10.1142/9789811220944_0008}},
  \eprint{2203.17188}.

\bibtype{Article}%
\bibitem{Lin:2022lck}
\bibinfo{author}{Zidu Lin}, \bibinfo{author}{Shuai Zha},
  \bibinfo{author}{Evan~P. O'Connor}, \bibinfo{author}{Andrew~W. Steiner},
  \bibinfo{title}{{Detectability of neutrino-signal fluctuations induced by the
  hadron-quark phase transition in failing core-collapse supernovae}},
  \bibinfo{journal}{Phys. Rev. D} \bibinfo{volume}{109} (\bibinfo{number}{2})
  (\bibinfo{year}{2024}) \bibinfo{pages}{023005},
  \bibinfo{doi}{\doi{10.1103/PhysRevD.109.023005}}, \eprint{2203.05141}.

\end{thebibliography*}
\let\addcontentsline\oldaddcontentsline

\end{fmffile}

\end{document}